\documentclass[11pt, english]{article}
\usepackage[T1]{fontenc}
\usepackage[latin9]{inputenc}
\usepackage[margin=1in,bottom=1in,top=1in]{geometry}
\usepackage{lscape}
\usepackage{setspace}
\usepackage{amsmath}
\usepackage{amsthm}
\usepackage{bm}
\usepackage{amssymb}
\usepackage{stmaryrd}
\usepackage{graphicx}
\usepackage{float}
\usepackage{setspace}
\usepackage{rotating}
\usepackage{subcaption}
\usepackage{ragged2e}
\usepackage{esint}
\usepackage{ulem}
\usepackage{babel}
\usepackage{longtable}
\usepackage{booktabs}
\usepackage{caption,booktabs,array}

\usepackage[round]{natbib}
\usepackage[hypertexnames=false]{hyperref}

\usepackage{colortbl}
\usepackage{color}
\definecolor{darkblue}{rgb}{0,0,.6}
\hypersetup{
colorlinks=true,
linkcolor=darkblue,
filecolor=darkblue,
urlcolor=darkblue,
citecolor=darkblue,
}
\usepackage[toc,page]{appendix}

\usepackage[sc]{mathpazo}
\usepackage{colortbl}

\usepackage{multirow}

\newtheorem{assumption}{Assumption}

\newtheorem{remark}{Remark}
\newtheorem{example}{Example}
\newtheorem{theorem}{Theorem}
\newtheorem{corollary}{Corollary}
\newtheorem{lemma}{Lemma}

\usepackage{dsfont}
\usepackage{pifont}
\usepackage{algorithm}
\usepackage{algpseudocode}

\begin{document}
\title{Unconditional Quantile Partial Effects via Conditional Quantile Regression\thanks{Alejo: IECON-Universidad de la Republica, Montevideo, Uruguay. E-mail: javier.alejo@fcea.edu.uy; 
Galvao: Department of Economics, Michigan State University, E-mail: agalvao@msu.edu;
Martinez-Iriarte: Department of Economics, UC Santa Cruz. E-mail:
jmart425@ucsc.edu; Montes-Rojas: IIEP-BAIRES-CONICET and Universidad de Buenos Aires. E-mail:
gabriel.montes@economicas.uba.ar. We thank the Editor, Xiaohong Chen, the Associate Editor, anonymous referees, and Michael Jansson, Michael Leung, Jessie Li, Blaise Melly, Yixiao Sun, Takuya Ura, and seminar participants at UC Irvine, MSU, NY Camp Econometrics XVII, CFE-CMStatistics 2022, Econometric Society European Meeting 2023, SEA 93\textsuperscript{rd} Annual Meeting, and 1\textsuperscript{st} Workshop on Quantile Regression in Rome for very helpful and constructive comments. Computer programs to replicate the numerical analyses are available from the authors. All remaining errors are our own.}}
\author{Javier Alejo,
Antonio F. Galvao, 
Julian Martinez-Iriarte, and
Gabriel Montes-Rojas}
\maketitle

\begin{abstract}
This paper develops a semi-parametric procedure for estimation of unconditional quantile partial effects using quantile regression coefficients. The estimator is based on an identification result showing that, for continuous covariates, unconditional quantile effects are a weighted average of conditional ones at particular quantile levels that depend on the covariates. We propose a two-step estimator for the unconditional effects where in the first step one estimates a structural quantile regression model, and in the second step a nonparametric regression is applied to the first step coefficients. We establish the asymptotic properties of the estimator, say consistency and asymptotic normality. Monte Carlo simulations show numerical evidence that the estimator has very good finite sample performance and is robust to the selection of bandwidth and kernel. To illustrate the proposed method, we study the canonical application of the Engel's curve, i.e. food expenditures as a share of income.
\vspace{7mm}

 \vspace{3mm}


\end{abstract}

\normalem


\onehalfspacing
\newpage

\section{Introduction}

Conditional quantile regression (CQR) is a general approach to estimate conditional quantile partial effects (CQPE), i.e., the effect of a covariate variable of interest  (\textit{ceteris paribus}) on the conditional quantile distribution of the outcome.\footnote{ See, \textit{e.g.}, \cite{KoenkerBassett78, KoenkerHallock01,Koenker05, Koenkeretal20} for comprehensive analyses of CQR.} CQR is a useful way to represent heterogeneity using a set of parameters to characterize the entire conditional distribution of an outcome variable given a list of observable covariates. 

Recently, unconditional quantile regression (UQR), proposed  by \cite{FirpoFortinLemieux09}, has attracted interest in both applied and theoretical literatures. UQR is an important tool for practitioners since it provides a method to evaluate the impact of changes in the distribution of the explanatory variables on  quantiles of the unconditional (marginal) distribution of the outcome variable. This method allows researchers to investigate important heterogeneity in the variable of interest. Naturally, UQR leads to the unconditional quantile partial effect (UQPE), which refers to the effect of a covariate (\textit{ceteris paribus}) on the unconditional quantile distribution of the outcome variable.\footnote{Conditional here means that one is conditioning on a set of observable variables, while partial means that one is looking at the effect of one particular covariate controlling for the rest of the covariates. On the other hand, Conditional Quantile Regression (CQR) and Unconditional Quantile Regression (UQR) refer to two regression methodologies to estimate CQPE and UQPE, respectively. Sometimes the acronym of the method is informally interchanged with the parameter of interest, and, therefore, can be somewhat confusing if read lightly.}

\cite{FirpoFortinLemieux09} propose several ways to estimate the UQPE. The most popular approach is the recentered influence function (RIF) regression method, commonly referred to as RIF regression. It is a two-step procedure, where in the first stage one estimates the RIF, and in the second step, a standard OLS regression of the RIF on covariates estimates the UQPE. While the method is appealing due to its simplicity, it relies on ability of the researcher to specify a regression equation for the influence function, a relatively abstract object. \citet[p.959]{FirpoFortinLemieux09} also show an important theoretical result connecting CQPE and UQPE that, when considering a continuous covariate, the UQPE can be expressed as an unconditional weighted average of the CQPE. Derivation of this result relies on a function that matches the conditional quantile whose values are the closest to the unconditional quantile. 

This paper builds on this relationship between CQPE and UQPE and suggests an alternative method for estimation of UQPE using simple CQR methods. The procedure is based on the identification result of UQPE that explores information contained in the CQPE. Since estimation of conditional density function is a high dimension object, we first slightly modify the result in \citet{FirpoFortinLemieux09} to show that the UQPE can be written as a \textit{conditional} average of the CQPE effects (evaluated where unconditional and conditional quantiles are equal), given the outcome variable (evaluated at the unconditional quantile). Hence, by starting with the common assumption of linearity of the quantile process, a simple reweighting of the CQR coefficients using a ratio of conditional and unconditional density functions delivers the UQPE.\footnote{In this paper we focus on a first stage quantile regression, but the model can be extended to non-linear models. An alternative is to employ distribution regression of \cite{ChernozhukovFernandezValMelly13}.} Thus, a useful by-product of the CQR analysis is the ability to express UQPE, for a given quantile $\tau\in(0,1)$, as a function of CQR.\footnote{Other procedures where statistics of interest are based on a combination of CQR coefficients are the following: \cite{BeraGalvaoMontesPark16} propose to estimate a unique representative CQPE based on an asymmetric Laplace framework, and \cite{yingying2021} considers a general weighted average quantile derivative using the CQR coefficients for a \emph{fixed} quantile level.} 

We propose a new two-step semi-parametric estimator based on this identification result that employs CQR coefficients to estimate the UQPE. The practical implementation is simple and makes use of the usual practice in empirical research of estimating the CQR process, that is, many quantiles, to explore heterogeneity in conditional effects of a certain covariate. In the first step one uses standard linear QR methods to estimate CQPE from the conditional model of interest over a grid of $\eta$-quantiles, 
and also estimates the unconditional $\tau$-quantile of the outcome of interest. Then, for a given value of the covariates, a matching function can be applied to select coefficients such that unconditional and conditional quantiles are equal. This matching function estimator is analogous to the QR estimator of the conditional CDF proposed by \cite{ChernozhukovFernandezValMelly13} in the context of counterfactual distributions.
In the second step, one employs a nonparametric regression of the matched CQR coefficients on the outcome, and evaluates this at the unconditional $\tau$-quantile. This is a one-dimensional (reverse) regression: the regressor is the outcome.
Mild sufficient conditions are provided for the two-step estimator to have desired asymptotic properties, namely, consistency and asymptotic normality. We derive the convergence rate of the estimator and show that, as expected, it converges at a standard nonparametric rate.

The proposed method offers important advantages over available techniques to compute UQPE. First, an important motivation and advantage of this paper is to compare and contrast the  conditional and unconditional quantile effects. Along these lines, the proposed methods do not require additional modeling assumptions to obtain the UQPE in addition to the CQPE conditions. Since UQPE is nonparametrically identified, the modeling assumptions for both cases would be the same, hence, while the results could still be affected by misspecification of the CQR model, potential differences between UQPE and CQPE results are not be driven by modeling choices.

Second, our framework provides the researcher with a simple structural conditional quantile model to study, which facilitates the understanding of identification. The conditional QR model in the first step allows for a simple and intuitive modelling framework. It is familiar to model the main output variable as a function of the covariates using CQR. On the other hand, in the second stage of RIF-OLS and RIF-Logit, one needs to model the effect of covariates on the conditional average of the RIF, which is difficult to conceptualize. There is, thus, a risk of misspecifying the regression in the second stage. In our proposed method, covariates enter the structural CQR in standard way, i.e., the dependent variable of interest is a function of covariates, in the first step. This approach may be simpler for the researcher since the CQR could be specified using the previous literature or the economic theory. Naturally, the price to be paid is the linearity of the CQR in the first step.

Finally, there have been considerable extensions on CQR modeling, such as, for example, models for panel data, endogeneity, and large number of covariates (see, e.g., \cite{Koenkeretal20}). Extensions of the proposed UQPE methods to encompass these situations might be simpler, relative to other existing methods, and are left for future research.

\textbf{Related Literature.} Although the literature on applications of UQR methods is extensive, the literature on theoretical developments is relatively small. \citet{Rothe10,Rothe12} generalize the method of \cite{FirpoFortinLemieux09}, for other recent developments, see, e.g., \cite{InoueLiXu21}, \cite{SasakiUraZhang22}, \cite{mms2022}, \cite{yixiao2020}, and \cite{MartinezIriarte21}. 
For a comprehensive survey on counterfactual distributions and decomposition methods, see \cite{FortinLemieuxFirpo11}.
Moreover, in a related work, \cite{ChernozhukovFernandezValMelly13} propose to use distributional regression to compute conditional CDF and estimate general counterfactual distributions. It is possible to obtain the UQPE from this methodology by differentiating the estimated CDF, and then integrating the covariates. 
Another branch of the literature related to this paper uses a combination of standard CQR with simulation exercises to evaluate distributional effects, such as UQPE. While it is feasible to calculate the unconditional distribution of an outcome variable using CQR (see, e.g., \cite{AutorKatzKearney05}, \cite{MachadoMata05}, \cite{Melly05}, and \cite{ChernozhukovFernandezValMelly13}), this task is not obvious, at least compared to the ordinary least-squares (OLS) for the conditional mean. Since an analogue of the law of iterated expectations does not hold in the case of quantiles (\cite{deCastroCostaGalvaoZubelli23}), the CQR analysis cannot be directly employed to analyze unconditional quantiles \cite[see the discussion in][]{FortinLemieuxFirpo11}.

The remaining of the paper is organized as follows. Section \ref{UQPEviaCQPE} presents the main result that motivates the UQPE estimator based on CQR. Section \ref{estimator} proposes an estimator for the UQPE and Section \ref{asymptotictheory} derives its asymptotic properties. Section \ref{MonteCarlo} studies the estimator's finite sample performance using Monte Carlo experiments. Section \ref{empirical} provides an empirical application. Section \ref{conclusion} concludes.

\section{Quantile Partial Effects}\label{UQPEviaCQPE}

In this section we introduce the unconditional quantile partial effect (UQPE), and the conditional quantile partial effect (CQPE). Some developments appear in \cite{FirpoFortinLemieux09} and we reproduce them here for completeness. The relationship between UQPE and CQPE is the foundation for the estimator we propose in the next section.  In the following, the \emph{unconditional} quantiles of $Y$ are indexed by $\tau\in(0,1)$, while the \emph{conditional} quantiles $Y$ given $X$ are indexed by $\eta\in(0,1)$.

\subsection{UQPE in terms of CQPE}

Consider a general model $Y=r(X, U)$, where $X=(X_1, X_2')'$. Here, $Y$ is the dependent variable, $X_1$ is the target variable of interest and is a scalar, $X_2$ is a $(d-1)\times 1$ vector consisting of other observable covariates, and $U$ consists of unobservables. A leading example is the simple linear model $Y = \beta_{0} + \beta_{1}X_{1}+X_{2}'\beta_{2}+U,$
such that the conditional $\eta$-quantile of $Y$ given $(X_{1},X_{2})$ is
\begin{equation}\label{eq:linear qr model}
Q_Y[\eta | X_{1}, X_{2}]=\beta_{0}(\eta) + \beta_{1}(\eta)X_{1}+X_{2}'\beta_{2}(\eta).
\end{equation}
The typical object of study of the standard conditional quantile regression (CQR) is the conditional quantile partial effect (CQPE) defined as
\begin{align}\label{eq:cqpe0}
CQPE_{X_1}(\eta, x) :&= \frac{\partial Q_Y[\eta|X_1 = z, X_2=x_2]}{\partial z}\bigg|_{z=x_1},
\end{align}
and corresponds to the marginal effect of $X_1$ on the conditional quantiles of the outcome when $X_1=x_1$ and $X_2=x_2$. In the case of model \eqref{eq:linear qr model}, $CQPE_{X_1}(\eta, x)=\beta_{1}(\eta)$ and estimation of this parameter follows from standard quantile regression methods. 

To define an unconditional counterpart to $CQPE_{X_1}(\eta, x)$, we follow \cite{FirpoFortinLemieux09}. To that end, consider the counterfactual outcome
\begin{align*}
Y_{\delta,X_1} = r(X_1+\delta, X_2, U),
\end{align*}
where $\delta$ captures a small location change in the variable $X_{1}$. $Y_{\delta,X_1}$ is the outcome we would observe if every individual receives an additional quantity $\delta$ of $X_1$, while keeping $X_2$, $U$, and the joint dependence between $X_1, X_2$, and $U$ constant. Let $Q_Z[\tau]$ be the unconditional $\tau$-quantile of the random variable $Z$. The unconditional quantile partial effect (UQPE) is defined as\footnote{The UQPE can be defined for a vector of covariates as in \cite{FirpoFortinLemieux09} resulting in a vector of UQPEs. \cite{mms2022} provide an interpretation of the linear combination of UQPEs as a compensated counterfactual change.} 
\begin{align}\label{eq:uqpe_0}
UQPE_{X_1}(\tau) :&= \lim_{\delta\to 0}\frac{Q_{Y_{\delta,X_1}}[\tau]-Q_Y[\tau]}{\delta}.
\end{align}
The $UQPE_{X_1}(\tau)$ is the marginal effect of a location shift in $X_1$ on the unconditional $\tau$-quantile of the outcome. 

The interpretation of $CQPE_{X_1}(\eta, x)$ and $UQPE_{X_1}(\tau)$ are different. The $CQPE_{X_1}(\eta, x)$ amounts to manipulating $X_1$ locally at $x$ and evaluating a local impact on $Y$, i.e., it measures the effect of a marginal change in $X_{1}$ on the $\eta$-conditional quantile of $Y$. In other words, it only concerns observations (\textit{i.e.,}  individuals) with $X=x$. The $UQPE_{X_1}(\tau)$ is obtained by what we may refer to as a global change in $X_1$: everyone (not just those with $X=x$ covariates values) moves from $X_1$ to $X_1+\delta$. Then, it looks at its associated impact on the $\tau$-unconditional quantile of $Y$.\footnote{
\cite{galvaowang2015} study quantile treatment effects with a continuous treatment. When treatment is continuous, taking values $t\in\mathcal T$, a subset of the real line, interest lies in the dose response function $t\mapsto Y(t)$, which is the potential outcome associated with different levels of the treatment. The quantile dose response functions aims at estimating the quantiles of $Y(t)$ for different values of $t$. This is different from here, where we estimates the effect on the quantiles of the \textit{observed} outcome $Y$.
}

\cite{FirpoFortinLemieux09} show that under some mild conditions the following identification result holds:
\begin{align}\label{eq:uqpe}
UQPE_{X_1}(\tau) 
&= -\frac{1}{f_Y(Q_Y[\tau])}  \int \frac{\partial F_{Y| X}(Q_Y[\tau]|z,x_2 )}{\partial z}\bigg|_{z=x_1} dF_{X}(x),
\end{align}
where $F_{X}(x)$ is short for the joint distribution, 
$F_{X_1,X_2}(x_1,x_2)$. In a similar manner,  assuming differentiability of $F_{Y|X}(y|\cdot )$, from \eqref{eq:cqpe0} we have that  
\begin{align}
CQPE_{X_1}(\eta, x) &= -\frac{1}{f_{Y| X}(Q_Y[\eta| X= x]| x)}   \frac{\partial F_{Y| X}(Q_Y[\eta| X= x]|z,x_2 )}{\partial z}\bigg|_{z=x_1}. \label{eq:cqpe}
\end{align}
It is interesting to see that $CQPE_{X_1}(\eta, x)$ in equation \eqref{eq:cqpe} has a similar structure to $UQPE_{X_1}(\tau)$ in \eqref{eq:uqpe}.
Comparing the formulas in \eqref{eq:uqpe} and \eqref{eq:cqpe}, one is able to see that even if the conditional quantile is equal to the corresponding unconditional, that is, $Q_Y[\tau|X = x]=  Q_Y[\tau]$, one is \emph{not} able to recover $UQPE_{X_1}(\tau)$ from $CQPE_{X_1}(\tau, \cdot)$ by simply integrating the latter over $X$. Moreover, it is usually the case that $Q_Y[\tau|X = x]\neq  Q_Y[\tau]$. Thus, first we need to match conditional and unconditional quantiles, and then re-weight them appropriately to recover $UQPE_{X_1}(\tau)$ from $CQPE_{X_1}(\cdot, \cdot)$.

The following matching map, introduced by \citet[p.959]{FirpoFortinLemieux09}, is an important tool to relate the CQPE and UQPE:
\begin{align}\label{eq:xi}
\xi_\tau( x) = \left\{ \eta :  Q_Y[\eta|X = x]=  Q_Y[\tau] \right\}.
\end{align}
The map $\xi_\tau( x):(0,1)\times \mathbb R^d\mapsto(0,1)$ corresponds to the quantile index(es) in the conditional model, $\eta$, that produces the closest match with the unconditional quantiles $\tau$ for different values of $x$. In Section \ref{section:matching} we analyze this map in detail. For now, we assume that $\xi_\tau( x) $ is a singleton. Therefore, we have that, for every $x$,  $Q_Y[\xi_\tau(x)|X = x]=Q_Y[\tau]$. Under this condition, it is simple to formalize the relationship between CQPE and UQPE. Note that the $CQPE_{X_1}$ in equation \eqref{eq:cqpe} evaluated at $\eta=\xi_\tau(x)$ can be written as
\begin{align*}
CQPE_{X_1}(\xi_\tau(x),x)
&=-\frac{1}{f_{Y|X}(Q_Y[\tau] |x)}   \frac{\partial F_{Y|X}(Q_Y[\tau] |z,x_2 )}{\partial z}\bigg|_{z=x_1} .
\end{align*}
Now, rearranging $CQPE_{X_1}(\xi_\tau(x),x)$ above and substituting into equation \eqref{eq:uqpe} yields
\begin{align}\label{eq:cqpe_uqpe}
UQPE_{X_1}(\tau) = \int CQPE_{X_1}(\xi_\tau(x),x) \frac{f_{Y|X}(Q_Y[\tau] |x)}{f_{Y}(Q_Y[\tau])}dF_{X}(x). 
\end{align}

The result in equation \eqref{eq:cqpe_uqpe} appears in Proposition 1(ii) of \citet{FirpoFortinLemieux09}.\footnote{Proposition 1(ii) is stated in \citet[p.959]{FirpoFortinLemieux09} as following: We can also represent $UQPE(\tau)$ as a weighted average of $CQPE(\zeta_{\tau}(x),x))$: $UQPE(\tau)=E[\omega_{\tau}(X)\cdot CQPE(\zeta_{\tau}(X),X)]$, where $\omega_{\tau}(x)\equiv f_{Y|X}(q_{\tau}|x)/f_{Y}(q_{\tau})$, $\zeta_\tau(x) = \left\{ s :  Q_s[Y|X = x]=  q_\tau \right\}$, and $q_{\tau}=\inf_{q}\{q: F_{Y}(q) \geq \tau \}$.} One could think about estimating this unconditional expectation. However, notice that the conditional density $f_{Y|X}$ is a high dimensional object, and might be difficult to estimate in practice. Thus, we rewrite this equation to obtain an alternative representation of the UQPE in terms of CQPE.

Note that, when $F_{X}(x)$ is differentiable, weights in \eqref{eq:cqpe_uqpe} can be rearranged as
\begin{align*}
\frac{f_{Y|X}(Q_Y[\tau] |x)}{f_{Y}(Q_Y[\tau])}f_{X}(x) = \frac{f_{Y,X}(Q_Y[\tau] ,x)}{f_{Y}(Q_Y[\tau])f_{X}(x)}f_{X}(x) = f_{X|Y}(x|Q_Y[\tau]).
\end{align*}
Finally, using this weight equation \eqref{eq:cqpe_uqpe} becomes a \textit{conditional} expectation as
\begin{align}\label{eq:cqpe_uqpe_2}
UQPE_{X_1}(\tau)
= E \bigl[ CQPE_{X_1}(\xi_\tau(X),X)|Y= Q_Y[\tau]\bigr].
\end{align}

Equation \eqref{eq:cqpe_uqpe_2} shows that the UQPE is in fact a \emph{local} weighted average of CQPE effects ``near'' the unconditional $\tau$-quantile of $Y$. As noted above, the $\tau$-th unconditional quantile of interest may be different from the (random) $\xi_\tau(X)$-th conditional quantiles used inside the integral.
The preceding informal discussion is summarized in the next lemma.

\begin{lemma}\label{lemma:reverse}
Let the following assumptions hold: $(i)$ $UQPE_{X_1}(\tau) $ is identified by \eqref{eq:uqpe}; $(ii)$ the matching function defined in \eqref{eq:xi} is a singleton; $(iii)$ $F_{Y| X}(y|x )$ and $Q_Y[\tau| X= x]$ are differentiable with respect to $x_1$; $(iv)$ $f_{Y| X}$, $f_Y$ and $f_X$ are strictly positive. Then \eqref{eq:cqpe_uqpe_2} holds.
\end{lemma}

As mentioned above, sufficient conditions for $(i)$ are laid out in \cite{FirpoFortinLemieux09}. Regarding $(ii)$, see Assumption \ref{assumption_matching} stated below for sufficient conditions for $\xi_\tau( x) $ to be singleton. The rest of the assumptions are customary regularity conditions.\footnote{ \citet{Rothe10} extends the identification of UQPE to nonseparable triangular models with endogenous regressors via a control variable approach.}

\subsection{Description of the procedure}

\begin{figure}
     \centering
     \begin{subfigure}[b]{0.4\textwidth}
         \centering
        \includegraphics[width=8cm]{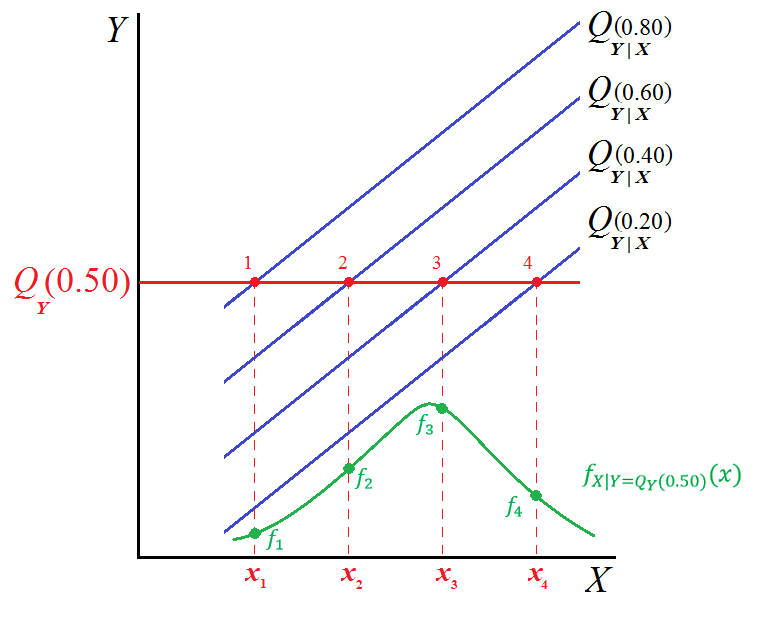}
         \caption{Constant CQPE.}
         \label{fig:betasloc}
     \end{subfigure}
     \begin{subfigure}[b]{0.4\textwidth}
         \centering
         	\includegraphics[width=8cm]{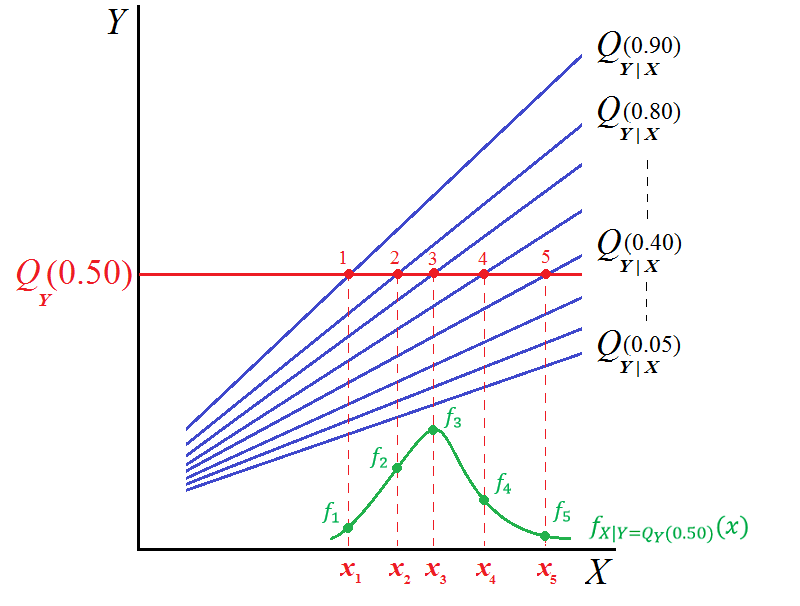}
         \caption{Heterogeneous CQPE.}
         \label{fig:betaslocsca}
     \end{subfigure}
        \caption{Constant vs. Heterogeneous CQPE.}
        \label{fig:graph_1}
\end{figure}

Given the discussion above, it is possible to compute the UQPE from CQPE using equation \eqref{eq:cqpe_uqpe_2}. Consider a linear conditional quantile model as in \eqref{eq:linear qr model}, then $CQPE_{X_1}(\xi_\tau(x),x)=\beta_{1}(\xi_\tau(x))$. This implies that $\eqref{eq:cqpe_uqpe_2}$ becomes a weighted average of matched slopes. Figures \ref{fig:betasloc} and \ref{fig:betaslocsca} 
illustrate how the procedure works in two different linear cases. The figures plot both the unconditional quantile, $Q_Y[\tau]$ (red line) and conditional quantiles, $Q_{Y|X}[\eta]$ (blue lines), as well as the conditional density $f_{X|Y}$ (green curve).
An informal description is the following:
\begin{enumerate}
    \item   Identify the unconditional $\tau$ quantile, $Q_Y[\tau]$, say $Q_Y[0.50]$ as illustrated in the figures for the unconditional median, and drawn in a horizontal (red) line. 
    
    \item Notice that for each $\eta$, the conditional quantiles $Q_{Y|X}[\eta]$ (blue lines) intersect the unconditional quantile $Q_Y[\tau]$ (horizontal red line); in the figures, we illustrate this for $\{x_1,x_2,x_3,x_4\}$ values of $X$ that correspond to $\eta=\xi_\tau(x)$ values $\{0.80,0.60,0.40,0.20\}$ respectively.
    
    \item The UQPE is the weighted average -- with weights given by density $f_{X|Y}(x|Q_Y[\tau])$ (green curve) --  of the intersected slopes on the conditional quantile models.
\end{enumerate}

The figures are useful for analyzing the source of the variation in the UQPE across different unconditional quantiles. For example, in Figure \ref{fig:betasloc}, the CQPE slopes are the same across conditional quantiles $\eta$, which implies that the CQPE is constant across quantiles. Even if the weights change with $\tau$, this is irrelevant, because the slopes are constant.

On the other hand, in Figure \ref{fig:betaslocsca}, the CQPE slopes exhibit some variation across conditional quantiles $\eta$. 
This heterogeneity can be present even if the weights are not a function of $\tau$. The UQPE is then constructed as a weighted average of those. An additional source of variation is given by the potential different conditional densities of $X$ given $Y=Q_Y[\tau]$. The UQPE will then be the based on the different CQPE and the corresponding density weights. Section \ref{estimator} below formalizes the estimator.

\subsection{The matching map}\label{section:matching}
Recall that the matching function selects the quantile that equates the unconditional quantile of $Y$ with the corresponding conditional quantile, which may vary across values of $X$. Equation \eqref{eq:xi} defined the map $\xi_\tau( x):(0,1)\times \mathbb R^d\mapsto(0,1)$ as
\begin{align*}
\xi_\tau( x) = \left\{ \eta :  Q_Y[\eta|X = x]=  Q_Y[\tau] \right\}.
\end{align*}
For a fixed covariate value $X=x$, the map $\tau \mapsto  \xi_\tau(x)$ describes how the unconditional distribution maps on the conditional one. In general, $ \xi_\tau(x)$ may vary across the value of covariates as well. Note that it is entirely possible that $\tau\neq \xi_\tau(x)$.

For the purpose of this paper it is important that $\xi_\tau( x)$ is unique. But, generally, three situations may occur. 
First, $\xi_\tau( x)$ is unique when $F_{Y|X}(y|x)$ is strictly increasing. In this case, there can be at most one $\eta$ that satisfies equation \eqref{eq:xi}. To see this,  note that $Q_Y[\eta|X = x]=  Q_Y[\tau]$ is identical to $\eta=F_{Y|X}(Q_Y[\tau]|x)$, so that $\xi_\tau(x)$ is unique.
Second, $\xi_\tau( x) $ might be an interval. For example if $F_{Y|X}(y|x)$ has a jump discontinuity at $y=Q_Y[\tau]$, but it is otherwise continuous and strictly increasing, then $\xi_\tau( x) = \bigl[\lim_{y\uparrow Q_Y[\tau]}F_{Y|X}(y|x), F_{Y|X}(Q_Y[\tau]|x) \bigr]$. See Figure \ref{fig:interval} below. Third, $\xi_\tau( x)$ might be empty for some $x$. For example, suppose that $F_{Y|X}(y|x)$ is continuous, and aside from a flat interval, it is strictly increasing. If $Q_Y[\tau]$ is in the interior of the interval mapping to the flat interval, then we cannot have $ Q_Y[\eta|X = x]=  Q_Y[\tau]$. This is illustrated in Figure \ref{fig:empty} below.

Now we provide an example to illustrate the how to explicitly write matching function $\xi_\tau(x)$ using a linear QR model.

\begin{figure}
     \centering
     \begin{subfigure}[b]{0.4\textwidth}
         \centering
        \includegraphics[width=7cm]{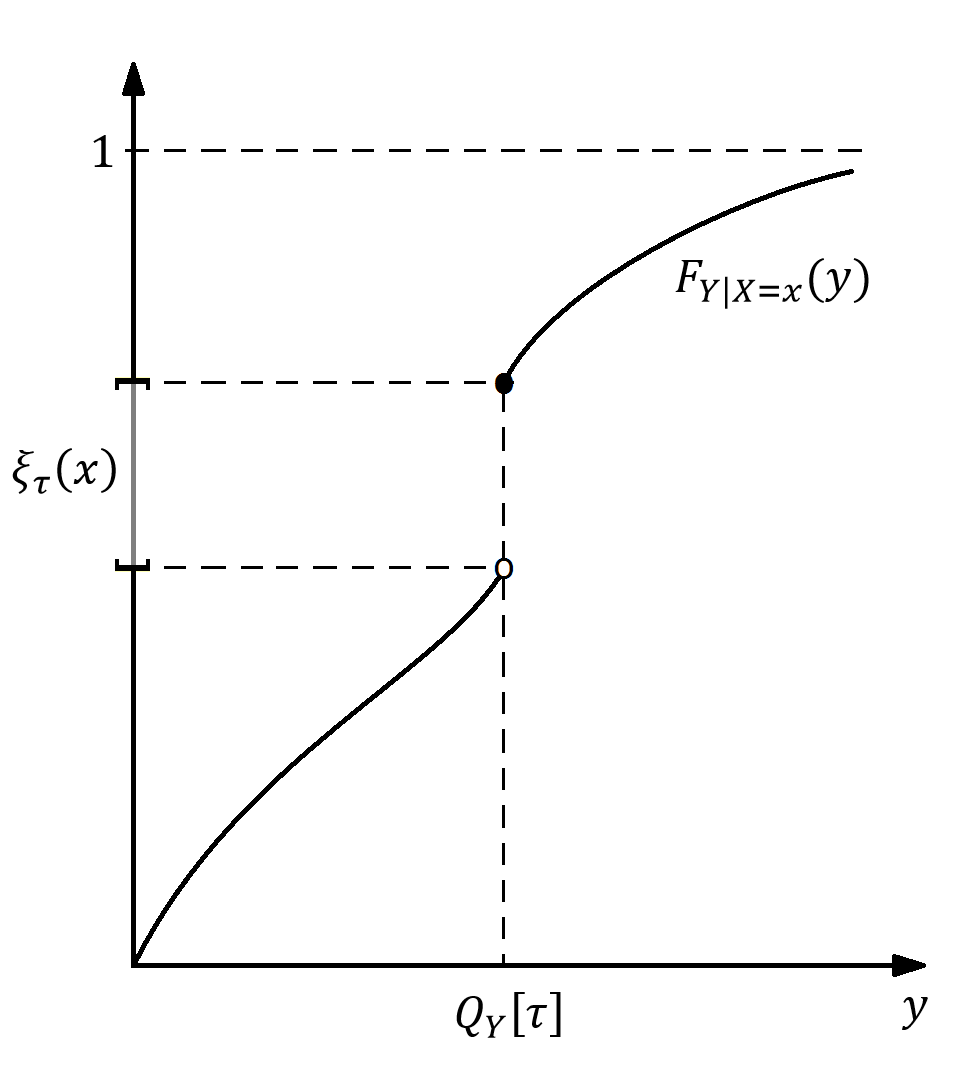}
         \caption{Interval valued $\xi_\tau( x) $.}
         \label{fig:interval}
     \end{subfigure}
     \begin{subfigure}[b]{0.4\textwidth}
         \centering
         	\includegraphics[width=7cm]{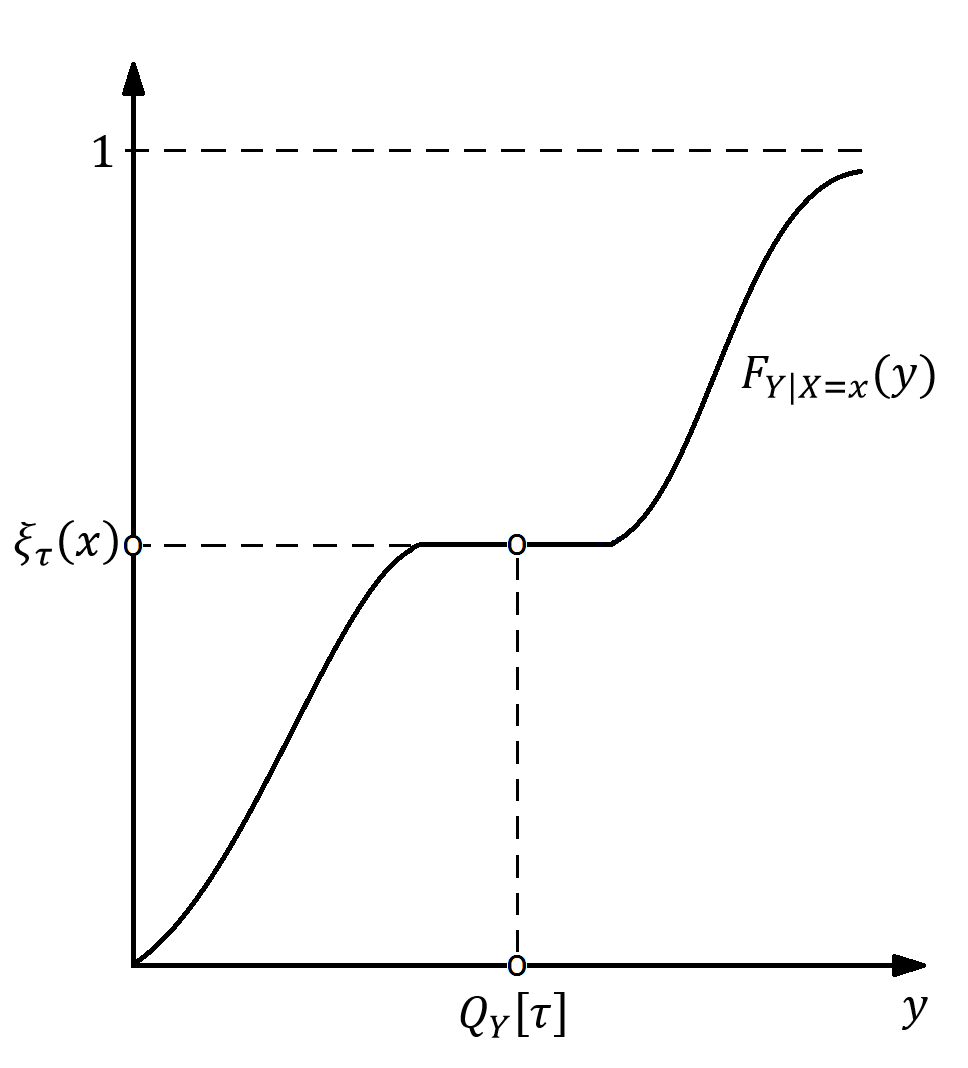}
         \caption{Empty $\xi_\tau( x)$.}
         \label{fig:empty}
     \end{subfigure}
        \caption{Non-uniqueness of $\xi_\tau( x)$.}
        \label{fig:matching}
\end{figure}

\begin{example}\label{example1}
Consider the model $Y=\alpha_0 +  \alpha_1 X_1 + (1+\theta X_1)U$ with $X\perp U$. By standard computations, if $1+\theta x_1>0$, then $Q_{Y}[\eta|X_1=x_1] = \alpha_0 +  \alpha_1 x_1 + (1+\theta x_1)Q_{U}[\eta]$. To find $\xi_\tau( x_1)$ we need the level $\eta$ such that $Q_Y[\eta|X_1 = x_1]=  Q_Y[\tau]$. Thus,
\begin{align}\label{eq:example1}
\xi_\tau( x_1)= F_U\left(\frac{Q_Y[\tau] - \alpha_0 - \alpha_1 x_1}{1+\theta x_1}\right).
\end{align}

\end{example}

\begin{remark}
If the matching function is the identity function: $\xi_\tau(X) = \left\{ \eta :  Q_Y[\eta|X = x]=  Q_Y[\tau] \right\}=\tau,$ Then, by equation \eqref{eq:cqpe_uqpe_2}, $UQPE_{X_1}(\tau)$ can be written as
\begin{align*}
UQPE_{X_1}(\tau)=\int CQPE_{X_1}(\tau, x) \frac{f_{X|Y}(x|Q_Y[\tau])}{f_{X}(x)}f_{X}(x)dx,
\end{align*}
which is the parameter of interest of \cite{yingying2021}: a weighted average quantile derivative. Here the weight is $\frac{f_{X|Y}(x|Q_Y[\tau])}{f_{X}(x)}$. Note that the argument in $CQPE_{X_1}(\cdot, x) $ is $\eta=\tau$, which does not depend on $x$, as opposed to general case when $CQPE_{X_1}(\cdot, x) $ is evaluated at  $\xi_\tau(x)$. If the matching function is not the identity, then our parameter is not covered by the methods of \cite{yingying2021}. 
\end{remark}

Since the estimation of the matching function is crucial for our results, we carry out many Monte Carlo simulations to asses the quality of the estimator proposed below in \eqref{eq:mat_func}.
In both the Monte Carlo simulations and the empirical application we plot the estimated matching function for different quantiles against the values of the of $X_1$. The shape of the matching function can give an indication of the effect of $X_1$ on $Y$, as Example \ref{example1} shows. 

\subsection{$\eta$-heterogeneity vs. $\tau$-heterogeneity}
Here we study how heterogeneity in the conditional effects across $\eta$-quantiles propagates to heterogeneity in unconditional effects across $\tau$-quantiles. We refer to the latter as $\tau$-heterogeneity, and to the former as $\eta$-heterogeneity. More precisely:
\begin{align*}
    \eta\text{-heterogeneity}:=\frac{\partial CQPE_{X_1}(\eta,x)}{\partial \eta},
\end{align*}
and
\begin{align*}
    \tau\text{-heterogeneity}:=\frac{dUQPE_{X_1}(\tau)}{d\tau}.
\end{align*}
While the $\eta\text{-heterogeneity}$ depends on (a fixed) $x$, we keep this dependence implicit. Using the chain rule, we obtain the following marginal change in the unconditional quantile effect with respect to $\tau$:
\begin{align*}
\frac{dUQPE_{X_1}(\tau)}{d\tau}&= \int_{\mathcal X}\frac{\partial  CQPE_{X_1}(\eta,x)}{\partial \eta}\bigg|_{\eta=\xi_\tau(x)} \frac{\partial \xi_\tau(x)}{\partial \tau} f_{X|Y}(x|Q_Y[\tau])dx\\
&+ \frac{ dQ_Y[\tau] }{d\tau} \int_{\mathcal X} CQPE_{X_1}(\xi_\tau(x),x) \frac{df_{X|Y}(x|y)}{dy}\bigg|_{y=Q_Y[\tau]} dx.
\end{align*}
The first term averages across the $\eta$-heterogeneity. In general, even if there is no $\eta$-heterogeneity, we may still have non-zero $\tau$-heterogeneity through the second term. An exception is when $CQPE_{X_1}(\eta,x)=\beta_1$, since in this case, $UQPE_{X_1}(\tau)=\beta_1$ as given in equation \eqref{eq:cqpe_uqpe_2}. This is the case of Example \ref{example1} above, which we discuss in more details now.
\setcounter{example}{0}
\begin{example}[Continued]
Here, $CQPE_{X_1}(\eta,x) =  \beta_1(\eta)=\alpha_1 + \theta Q_U[\eta]$. The $\eta$-heterogeneity is governed by the parameter $\theta$:
\begin{align*}
    \frac{d \beta_1(\eta)}{d\eta}= \theta \frac{ dQ_U[\eta] }{d\eta}.
\end{align*}
 The unconditional effect is  $UQPE_{X_1}(\tau)=E\left [ \beta_1(\xi_\tau(X_1))|Y= Q_Y[\tau]\right]$. It can be shown that 
\begin{align*}
\frac{dUQPE_{X_1}(\tau)}{d\tau}&= \frac{ dQ_Y[\tau] }{d\tau} \int_{\mathcal X_1} \frac{ \theta  }{1+\theta x_1} f_{X_1|Y}(x_1|Q_Y[\tau])dx_1 + \frac{ dQ_Y[\tau] }{d\tau}\int_{\mathcal X_1} \beta_1(\xi_\tau(x_1)) \frac{df_{X_1|Y}(x_1|y)}{dy}\bigg|_{y=Q_Y[\tau]}  dx_1.
\end{align*}
Now, if $\theta=0$, there is no $\eta$-heterogeneity, so the first term in the above expression for the $\tau$-heterogeneity is 0. The second term is also 0, since $\beta_1(\eta)=\alpha_1=\beta_1(\xi_\tau(x_1))$:
\begin{align*}
 \int_{\mathcal X_1} \beta_1(\xi_\tau(x_1)) \frac{f_{X_1|Y}(x_1|y)}{dy}\bigg|_{y=Q_Y[\tau]}  dx_1 &= \frac{ dQ_Y[\tau] }{d\tau} \alpha_1 \int_{\mathcal X_1}  \frac{df_{X_1|Y}(x_1|y)}{dy}\bigg|_{y=Q_Y[\tau]}  dx_1\\
 &=\frac{ dQ_Y[\tau] }{d\tau} \alpha_1\frac{d}{dy} \underbrace{\int_{\mathcal X_1}  f_{X_1|Y}(x_1|y) dx_1}_{=1}  \bigg|_{y=Q_Y[\tau]} \\
 &=0,
\end{align*}
provided we can interchange derivatives with integration. Thus, no $\eta$-heterogeneity implies no $\tau$-heterogeneity.
\end{example}

\section{Estimator}\label{estimator}

In this section we describe a two-step estimator of $UQPE_{X_1}(\tau)$, which is based on the conditional expectation in equation \eqref{eq:cqpe_uqpe_2}. The asymptotic properties are discussed in later sections.

Assume first  that 
\begin{equation}\label{eq:qr structural}
Q_Y[\eta|X_1 = x_1, X_2=x_2] =x_1\beta_1(\eta) + x_2'\beta_2(\eta)=x'\beta(\eta),
\end{equation}
where $\beta = (\beta_1, \beta_2')'$. Note that $x_2$ has to include a constant for correct specification. 
In this paper, we use the conditional quantile function in \eqref{eq:qr structural} to estimate the UQPE. Using this quantile regression model has advantages. First, it allows the researcher to directly model the outcome variable $Y$ as a function of observable covariates $X$, instead of modeling the recentered influence function. This is important because it may be simpler to relate the variable of interest directly from the economic theory or existing literature, than modeling the influence function. Second, practical estimation of \eqref{eq:qr structural} is simple, as we discuss below.

Under \eqref{eq:qr structural}, $CQPE_{X_1}(\xi_\tau(x),x)=\beta_1(\xi_\tau(x))$. Equation \eqref{eq:cqpe_uqpe_2} then has the convenient form
\begin{align}\label{eq:cqpe_uqpe_3}
UQPE_{X_1}(\tau)=E\left [ \beta_1(\xi_\tau(X))|Y= Q_Y[\tau]\right].
\end{align}
Our proposed estimator is a nonparametric regression of $\left\{\beta_1(\xi_\tau(x_i))\right\}_{i=1}^n$ on $\left\{y_i\right\}_{i=1}^n$ evaluated at $Q_Y[\tau]$. To implement this method in practice we are required to estimate $\beta_1(\xi_\tau(x))$ and $Q_Y[\tau]$.

To estimate $\beta_1( \xi_\tau(x_i))$ we first use CQR methods, and estimate $\beta(\eta)$ for a grid of $m$ values of $\eta$'s given by $\mathcal{H}_m=\{\epsilon<\eta_1< \cdots<\eta_j<\cdots<\eta_m<1-\epsilon\}$, $\epsilon\in(0,\tfrac{1}{2})$. In the standard linear case we have that for a given value of $\eta_j,$ and a sample $\left\{ y_i, x_i\right\}_{i=1}^n$, we simply apply standard quantile regression methods as  
\begin{align}\label{eq:qr}
(\hat \beta_1(\eta_j), \hat \beta_2(\eta_j)')'=\hat{\beta}(\eta_j)= \arg\min_{ b}\frac{1}{n}\sum_{i=1}^n \rho_{\eta_j}(y_i-x_i'  b),
\end{align}
where $\rho_\tau(u)=u(\tau-1[u<0])$ is the \cite{KoenkerBassett78} check function. 
We also estimate the unconditional quantile $Q_Y[\tau]$ by
\begin{align}\label{eq:uq}
\hat Q_Y[\tau] = \arg\min_{q}\frac{1}{n}\sum_{i=1}^n \rho_{\tau}(y_i-q).
\end{align}

To find the matched coefficients $\hat\beta_1(\hat\xi_\tau(x_i))$, we employ the two previous estimates as following. Let 
\begin{equation}\label{eq:mat_func}
    \hat\xi_\tau(x_i)=\left\{ 
    \begin{array}{ccc}
        \epsilon& \text{if} & \hat Q_Y[\tau]< x_i'\,\hat{\beta}(\eta_1);\\
        \eta_j\in\mathcal{H}_m & \text{if} & \left\{x_i'\,\hat{\beta}(\eta_{j}) \leq \hat Q_Y[\tau] < x_i'\hat{\beta}(\eta_{j+1})\right\} \mbox{ for } j=1,...,m-1;\\
       \eta_m\in\mathcal{H}_m & \text{if} &  x_i'\,\hat{\beta}(\eta_m)\leq \hat Q_Y[\tau],\\
    \end{array}
    \right.
\end{equation}
for $i=1,...,n$. We note that the estimation of the matching function procedure in equation \eqref{eq:mat_func} is analogous to the estimation of the conditional CDF $F_{Y|X}(Q_Y[\tau]|x_i)$ in equation (3.7) of \citet[p.2219]{ChernozhukovFernandezValMelly13} using $\hat{Q}_Y[\tau]$ instead of ${Q}_Y[\tau]$.\footnote{The above estimator relies on monotonicity of CQR such that there is only one match. In practice, this needs to be checked in small samples as multiple crossings may occur if $x_i$ is very different from $\bar x$. Then an algorithm could be implemented such as taking the average of the selected $\beta_1$ or a rearrangement of estimated quantiles (see, for instance, \cite{ChernozhukovFernandezGalichon10} for a discussion about quantile crossings). 
} Alternatively, one could employ distribution regression to estimate this matching function.

Finally, to estimate the $UQPE_{X_1}(\tau)$, we use a Nadaraya-Watson type-estimator, using the preliminary estimators:
\begin{align}\label{eq:nw_est}
\hat E\left [ \hat \beta_1(\hat \xi_\tau(X))|Y= \hat Q_Y[\tau]\right] = \frac{\sum_{i=1}^nK_h(y_i-\hat Q_Y[\tau])\cdot \hat \beta_1(\hat \xi_\tau(x_i))}{\sum_{i=1}^nK_h(y_i-\hat Q_Y[\tau])},
\end{align}
where $K_h$ is the rescaled kernel $K_h(u) := \frac{1}{h}K\left( \frac{u}{h} \right)$.
The estimator in \eqref{eq:nw_est} avoids the curse of dimensionality because it is a nonparametric regression on just one regressor: $Y$. Indeed, the dimension of $X$ enters in the CQR estimation and in the matching function. 



Equation \eqref{eq:nw_est} highlights the main benefit of our proposed approach: obtaining the unconditional effect is an easy follow-up from the conditional effects. If the researcher, as is usually the case, has estimated a grid of CQR coefficients, then, after they are ``matched'' according to \eqref{eq:mat_func}, they can be averaged following \eqref{eq:nw_est} to yield the unconditional effect for the desired quantile level. Moreover, notice that the second nonparametric step in equation \eqref{eq:nw_est} does not suffer from the curse of dimensionality, in the sense that it is a reverse regression where the one-dimensional outcome is the regressor.\footnote{The RIF-nonparametric approach is potentially subject to the curse of dimensionality since multiple regressors may enter the nonparametric second step.}

\begin{remark}
An alternative approach to estimating $UQPE_{X_1}(\tau)$ based on \eqref{eq:cqpe_uqpe_3} is a linear regression of $\hat \beta_1(\hat \xi_\tau(X))$ on a constant and $Y$. The predicted fit at $Y=\hat Q_Y[\tau]$ is an easy-to-compute approximation to $UQPE_{X_1}(\tau)$. Yet another option is to do a local linear regression. This estimator may help reduce the bias in lower and higher quantiles. The estimator is $\hat a_{\tau,0} + \hat a_{\tau,1} \hat Q_Y[\tau]$, where $(\hat a_{\tau,0} , \hat a_{\tau,1} )'$ solve
\begin{align*}
(\hat a_{\tau,0} , \hat a_{\tau,1} )' = \arg\min_{a_{\tau,0}, a_{\tau,1}}\sum_{i=1}^n K_h(y_i-\hat Q_Y[\tau])\left[
\beta_1(\hat \xi_\tau(x_i)) - a_{\tau,0}  - a_{\tau,1}\left(\frac{y_i-\hat Q_Y[\tau]}{h} \right) 
\right]^2.
\end{align*}
A study of the properties of this estimator in this particular setting is left for future research. 
\end{remark}

Next we provide a concise algorithm to compute the UQPE estimator for a given $\tau$. 
\begin{algorithm}
\caption{UQPE Computation}\label{alg:cap}
\begin{algorithmic}
\item \qquad \textbf{Data:} $\{y_i,x_i\}_{i=1}^n$ with $y_i$ a scalar and $x_i\in \mathbb R^d$ for $d\geq 1.$ 
\item \qquad \textbf{Input:} $\tau\in(0,1)$, $\epsilon\in (0,\tfrac{1}{2})$, $m\in\mathbb N$, bandwidth $h$, and kernel function $K_h$.
\end{algorithmic}
\begin{enumerate}
    \item Construct the grid $$\mathcal{H}_m=\{\epsilon<\eta_1< \cdots<\eta_j<\cdots<\eta_m<1-\epsilon\}.$$
    \item Estimate the QR coefficients $\hat{\beta}(\eta_j)$ for $j=1,2,...,m$ as in \eqref{eq:qr} and $\hat Q_Y[\tau]$ as in \eqref{eq:uq}.
    \item For each observation $i=1,2,...,n$, construct $x_i'\hat{\beta}(\eta_j)$ for $j=1,2,...,m$.
    \item For each observation $i=1,2,...,n$, compute
    the matching function $\hat{\xi}_\tau(x_i)$ as in \eqref{eq:mat_func} and then compute  $\hat\beta(\hat{\xi}_\tau(x_i))$.
    \item For each observation $i=1,2,...,n$ compute the kernel function $K_h(y_i-\hat Q_Y[\tau])$ and estimate \eqref{eq:nw_est} to compute the UQPE.
\end{enumerate}

\end{algorithm}

Note that this procedure is based on the initial estimate of a QR process for $\eta\in\mathcal{H}_m$, a typical output in QR analysis, and the unconditional quantile of $Y$. Then, for each observation compute the estimated conditional quantile functions $x_i'\hat{\beta}_\tau(\eta)$ for $\eta\in\mathcal{H}_m$ (point 3). The matching function involves a simple binary argument from the results in point 3. Then the algorithm needs to retrieve the corresponding QR coefficient (already estimated in point 2) delivered by the match. Finally compute a weighted average using the kernels as weights (point 5). Importantly, if several values of $\tau$ are to be computed, then points 1, 2 and 3 do not have to be redone.

\section{Asymptotic Theory}\label{asymptotictheory}

This section derives the asymptotic properties of the two-step estimator. First, we study the first step, and establish an asymptotic linear representation and rate of convergence for the conditional quantile regression coefficients as a function of the matched quantiles. Second, we study the asymptotic properties of the nonparametric regression in the second step.

\subsection{Structural QR and Matched Quantiles}

The following assumptions are needed to establish that $\hat \beta_1(\hat \xi_\tau(x)) - \beta_1( \xi_\tau(x))  = O_p(n^{-1/2})$, where $\hat \beta_1(\hat \xi_\tau(x)) $ is computed according to \eqref{eq:mat_func}.

\begin{assumption} \label{assumption_matching}Let $\{y_i,x_i\}_{i=1}^n$ be a random sample of independent and identically distributed (\textit{iid}) observations with $y_i$ a scalar and $x_i\in \mathbb R^d$ that satisfy the following properties:

\begin{enumerate}
\item \label{assumption_matching_linearity}The conditional quantiles are linear: $Q_Y[\eta|X = x] =  x' { \beta}(\eta)$, $\eta\in [\epsilon,1-\epsilon]$, $\epsilon\in(0,\tfrac{1}{2})$, with $X\in \mathbb R^d$ and $E|X|<\infty$.

\item \label{assumption_matching_positive_density}For every $x$ in the support of $X$, $f_{Y|X}(y|x)$ is bounded away from zero.

\item  \label{assumption_matching_conditional_sup}The conditional quantile regression estimators satisfy
\begin{align*}
\hat{ \beta}(\eta)-{ \beta}(\eta) &=   E\left[f_{Y|X}(X ' { \beta}(\eta)|X)XX'\right]^{-1}\frac{1}{n}\sum_{i=1}^n\left ( \eta- \mathds 1\left\{ y_i\leq x_i'  { \beta}(\eta) \right\} \right)x_i+ o_p(n^{-1/2})\\
&=  \frac{1}{n}\sum_{i=1}^n \Psi_{i}(\eta)  + o_p(n^{-1/2}),
\end{align*}
uniformly in $\eta\in [\epsilon,1-\epsilon]$, $\epsilon\in(0,\tfrac{1}{2})$, and $\eta \mapsto E\left[f_{Y|X}(X ' { \beta}(\eta)|X)XX'\right]$ has uniformly bounded derivatives.
\item \label{assumption_matching_unconditional}The unconditional quantile estimator satisfies 
\begin{align*}
\hat Q_Y[\tau]- Q_Y[\tau] &=f_Y(Q_Y[\tau])^{-1}\frac{1}{n}\sum_{i=1}^n \left (\tau -\mathds 1\left\{ y_i\leq Q_Y[\tau]  \right\}  \right) \\
&= \frac{1}{n}\sum_{i=1}^n\psi_{i}(\tau) + o_p(n^{-1/2}).
\end{align*}
\item \label{assumption_matching_approx_zero} The grid of quantiles $\{\epsilon<\eta_1<\ldots <\eta_j<\ldots <\eta_m<1-\epsilon\}$, $\epsilon\in(0,\tfrac{1}{2})$, satisfies $\Delta\eta=o(n^{-1/2})$ as $n\to\infty$ for $\Delta\eta:=\eta_j-\eta_{j-1},\ j=2,...,m$, and $\eta_1=\epsilon$ and $\eta_m=1-\epsilon$ for a small $\epsilon>0$.
\end{enumerate}
\end{assumption}

Assumption \ref{assumption_matching}.\ref{assumption_matching_linearity} imposes linearity of the quantile process.\footnote{There is a large set of examples using linear specification of quantiles over the entire conditional distribution as, among others, treatment effects, endogeneity, high-dimensional, stochastic dominance, censoring, as well as most of the theoretical papers providing statistical foundations for the quantile regression process (see, e.g., among many others, \citet{GutenbrunnerJureckova92}, \citet{KoenkerMachado99}, \citet{KoenkerXiao02}, \citet{Knight02}, \citet{ChernozhukovVal05}, \citet{AngristChernozhukovFernandezVal06}, \citet{BelloniChernozhukov11}, \citet{Portnoy12}, \citet{VolgushevChaoCheng19}, and \citet{HePanTanZhou22}).} 
Condition \ref{assumption_matching}.\ref{assumption_matching_positive_density} is very standard in the QR literature, see, e.g., \cite{Koenker05}. Assumptions \ref{assumption_matching}.\ref{assumption_matching_linearity} and  \ref{assumption_matching}.\ref{assumption_matching_positive_density}, allow us to write $F_{Y|X}(x'\beta(\eta)|x)=\eta$, so that $x'{ \dot\beta}( \eta) = f_{Y|X}(x'\beta(\eta)|x)^{-1}>0$, where $\dot\beta( \eta)$ is the Jacobian vector: the derivative of the map $\eta\mapsto {\beta}(\eta)$. This quantity appears in the denominator of the influence function of $\hat \xi_\tau(x)$. Moreover, Assumption \ref{assumption_matching}.\ref{assumption_matching_positive_density} states that $f_{Y|X}(y|x)$ is bounded away from zero. This in turn implies that $y\mapsto F_{Y|X}(y|x)$ is strictly increasing.
Assumption \ref{assumption_matching}.\ref{assumption_matching_conditional_sup} is a uniform Bahadur representation for the QR estimator. It is established in Lemma 3 in \cite{kato2019}. See also Theorem 3 in \cite{AngristChernozhukovFernandezVal06}. It implies $\sup_{\eta\in [\epsilon,1-\epsilon] } |\hat{ \beta}(\eta)-{ \beta}(\eta) | = O_p(n^{-1/2})$ and the stochastic equicontinuity of the process $\tau\mapsto \sqrt n (\hat{ \beta}(\eta)-{ \beta}(\eta))$ on $[\epsilon, 1-\epsilon]$.
Condition \ref{assumption_matching}.\ref{assumption_matching_unconditional} is a simple linear representation for the unconditional quantile.
Sufficient conditions for Assumption \ref{assumption_matching}.\ref{assumption_matching_unconditional} are given in \cite{serfling1980}. 
Finally, Assumption \ref{assumption_matching}.\ref{assumption_matching_approx_zero} requires that the grid for the matching function becomes denser as the sample size increases. This condition has appeared in the QR literature. \citet[Remark 3.1 p.2220]{ChernozhukovFernandezValMelly13} provide a similar condition when computing counterfactual distributions.


The next result provides a rate of convergence and a linear representation for $\hat \beta_1(\hat \xi_\tau(x)) - \beta_1( \xi_\tau(x))$, and for $\hat \xi_\tau(x)- \xi_\tau(x)$.

\begin{theorem}\label{thm:matching}
Under Assumption \ref{assumption_matching}, the CQR coefficient of $X_1$ evaluated at the random quantile $\hat \xi_\tau(x)$ can be represented as
\begin{align*}
\hat \beta_1(\hat \xi_\tau(x)) - \beta_1( \xi_\tau(x)) 
&=\hat \beta_1( \xi_\tau(x)) - \beta_1(  \xi_\tau(x)) + \dot\beta_1( \xi_\tau(x) ) (\hat \xi_\tau(x)- \xi_\tau(x) ) + o_p(n^{-1/2}),
\end{align*}
where
\begin{align*}
\hat \xi_\tau(x)- \xi_\tau(x)  
 &= f_{Y|X}(x'\beta(\xi_\tau(x))|x) \left[-x' \left ( \hat{\beta}( \xi_\tau(x))- {\beta}( \xi_\tau(x)) \right) +  \left ( \hat Q_Y[\tau] - Q_Y[\tau] \right)\right] + o_p(n^{-1/2}).
\end{align*}
Moreover, they are both $O_p(n^{-1/2})$ and asymptotically normal.
\end{theorem}


\subsection{Nadaraya-Watson Estimator}

Our parameter of interest given in \eqref{eq:cqpe_uqpe_3} is
\begin{align*}
UQPE_{X_1}(\tau)=E\left [ \beta_1(\xi_\tau(X))|Y= Q_Y[\tau]\right],
\end{align*}
and we propose the following nonparametric regression Nadaraya-Watson-type estimator: 
\begin{align*}
\widehat{UQPE_{X_1}}(\tau)=\hat E\left [ \hat \beta_1(\hat \xi_\tau(X))|Y= \hat Q_Y[\tau]\right] = \frac{\sum_{i=1}^nK_h(y_i-\hat Q_Y[\tau])\cdot \hat \beta_1(\hat \xi_\tau(x_i))}{\sum_{i=1}^nK_h(y_i-\hat Q_Y[\tau])}.
\end{align*}
The unfeasible (oracle) version is denoted by
\begin{align*}
\widetilde{UQPE_{X_1}}(\tau)= \hat E\left [  \beta_1( \xi_\tau(X))|Y=  Q_Y[\tau]\right] =\frac{\sum_{i=1}^nK_h(y_i- Q_Y[\tau])\cdot  \beta_1( \xi_\tau(x_i))}{\sum_{i=1}^nK_h(y_i- Q_Y[\tau])}.
\end{align*}


We maintain the following assumptions.

\begin{assumption}\label{assumption_kernel}
$K(u)$ is a symmetric function around 0 that satisfies: (i) $\int K(u)du=1$; (ii) For $r\geq 2$, $\int u^j K(u)du = 0$ when $j=1,...,r-1$,
and $\int |u|^r K(u)du < \infty$; (iii) $\int K'(u)du=0$ ; (iv) $u^jK(u)\to 0$ as $u\to\pm\infty$ for j=1,...,r+1; (v) $\sup_{u\in\mathbb R}|K'(u)|<\infty$ and $\sup_{u\in\mathbb R}|K''(u)|<\infty$ ; (vi) $\int K'(u)^2du<\infty$ and $\int uK'(u)^2du<\infty$.
\end{assumption}

\begin{remark}
The Gaussian kernel is (ignoring the constants) $K(u) =  e^{-u^2/2}$. It's first derivative is $K'(u) = -u e^{-u^2/2}$. To show that $K'$ is Lipschitz, we need to show that there exist a constant $C$, independent of $u$  and $v$, such that $|K'(u)-K'(v)|\leq C|u-v|$. We can use the mean value theorem, since the second derivative is uniformly bounded. It is given by $K''(u) = -e^{-u^2/2} + u^{2}e^{-u^2/2}$ and $|K''(u)|\leq C$. Therefore, by the mean value theorem, $K'$ is Lipschitz continuous.
\end{remark}

\begin{assumption}\label{assumption_bandwitdh}
As $n\to\infty$, the bandwidth satisfies $h\propto n^{-a}$ with $1/(1+2r)\leq a< 1/2$.
\end{assumption}

\begin{remark}
     If a second order Kernel is chosen,  $r=2$ and Assumption \ref{assumption_bandwitdh} requires that the admissible bandwidth satisfies $h\propto n^{-a}$ for $1/5\leq a< 1/2$.
\end{remark}

\begin{assumption}\label{assumption_density}
(i) The density of $Y$ is $r+1$ times continuously differentiable, with uniformly bounded derivatives; (ii) The joint density $f_{Y,X}(y,x)$ is $r+1$ times continuously differentiable, with uniformly bounded derivatives for every $x$ in the support of $X$.
\end{assumption}

\begin{assumption}\label{assumption_approx}
(i) The remainder in the expression for $\hat \xi_\tau(x)- \xi_\tau(x)$ holds uniformly over $\mathcal X$, the support of $X$; (ii) $\sup_{\eta\in[\epsilon,1-\epsilon]}|\dot \beta_1(\eta)|$, and $(y,x)\mapsto f_{Y|X}(y|x)$.
\end{assumption}

\begin{remark}
    Assumption \ref{assumption_approx}.(i) is implied by Condition D in \citet[p.2224]{ChernozhukovFernandezValMelly13}. It pertains to a uniform central limit theorem for the estimator of the conditional CDF. For primitive conditions, we refer the reader to \cite{ChernozhukovFernandezValMelly13}, where they provide details on this verification.
\end{remark}

\begin{theorem}\label{thm:first_order_equiv}
Let Assumptions \ref{assumption_matching}, \ref{assumption_kernel}, \ref{assumption_density}, \ref{assumption_bandwitdh}, and \ref{assumption_approx}  hold. Then, as $n\rightarrow\infty$,
\begin{align*}
\widehat{UQPE_{X_1}}(\tau) = \widetilde{UQPE_{X_1}}(\tau) + o_p(n^{-1/2}h^{-1/2}).
\end{align*}
\end{theorem}

This theorem states that the preliminary estimators of the CQR slopes, the matched quantiles and the unconditional quantile of $Y$ vanish asymptotically because they converge at a faster rate: $n^{-1/2}$ as opposed to $n^{-1/2}h^{-1/2}$. Moreover, the asymptotic distribution of the unfeasible estimator $\widetilde{UQPE_{X_1}}(\tau)$ is well-known and can be readily established.

The following assumption is customary in order to apply the Lindeberg-Feller Central Limit Theorem.
\begin{assumption}\label{assumption_clt}
(i) For $U_\tau:= \beta_1(\xi_\tau(X)) - E\left [ \beta_1(\xi_\tau(X))|Y\right]$, and $\gamma>0$,
$E[|U_\tau|^{2+\gamma}|Y]<C<\infty$ a.s. for some $C$; (ii) $\int |K(u)|^{2+\gamma}du<\infty$; (iii) The map $y\mapsto E\left [  \beta_1( \xi_\tau(X))|Y=  y\right]$ is $r+1$ times continuously differentiable, with uniformly bounded derivatives; (iv) The map $y\mapsto \sigma^2_\tau(y):=E[U_\tau^{2}|Y=y]$ is continuous.
\end{assumption}

\begin{corollary}\label{cor:clt}
Let Assumptions \ref{assumption_matching}, \ref{assumption_kernel}, \ref{assumption_density}, \ref{assumption_bandwitdh}, \ref{assumption_approx} and \ref{assumption_clt}  hold. Then, as $n\rightarrow\infty$,
\begin{align*}
\sqrt{nh} \left (\widehat{UQPE_{X_1}}(\tau) -  {UQPE_{X_1}}(\tau)\right)\overset{d}{\to}N\left (0,  \sigma^2_\tau(Q_Y[\tau]) f_Y(Q_Y[\tau])^{-1} \int K(u)^2du \right).
\end{align*}
\end{corollary}


\begin{remark}
The practical computation of the asymptotic variance-covariance matrix in Corollary \ref{cor:clt} is difficult due to the presence of preliminary estimators in the nonparametric regression. Thus, in practice, we employ resampling approach for inference.
There is an extensive literature on constructing nonparametric confidence bands for functions. We refer the reader to \cite{HardleBowman88} and \cite{PeterHorowitz13} and references therein for resampling methods. 
\end{remark}

We describe now the implementation of the pairwise bootstrap procedure.
\begin{enumerate}
    \item Estimate $\{\hat\beta(\eta)\}$ for a given grid  $\mathcal{H}_m=\{\eta_1,...,\eta_m\}$ and $\hat{Q}_Y[\tau]$, then compute $\widehat{UQPE}_{X_1}(\tau)$ using the sample $\{y_i,x_i\}_{i=1}^n$.
    \item Compute samples with replacement $\{y_i^{*b},x_i^{*b}\}_{i=1}^n$, for $b=1,...,B$, and estimators $\{\hat\beta^{*b}(\eta)\}$ for $\mathcal{H}_m$, $\hat{Q}^{*b}_Y[\tau]$ and $\widehat{UQPE}^{*b}_{X_1}(\tau)$.
    \item Compute the standard deviation from the bootstrap sample, $$\hat\sigma^*_{UQPE}=\sqrt{\frac 1 B \sum_{b=1}^B \left(\widehat{UQPE}^{*b}_{X_1}(\tau)-\overline{\widehat{UQPE}^{*}_{X_1}(\tau)}\right)^2}$$ where $\overline{\widehat{UQPE}^{*}_{X_1}(\tau)}=\frac 1 B \sum_{b=1}^B \widehat{UQPE}^{*b}_{X_1}(\tau)$. \\
    \item Compute the $1-\alpha$ confidence interval $[\widehat{UQPE}^{*[\alpha/2]}_{X_1}(\tau),\widehat{UQPE}^{*[1-\alpha/2]}_{X_1}(\tau)]$ using the ordered statistics of the bootstrap sample.
\end{enumerate}

\section{Monte Carlo experiments}\label{MonteCarlo}

This section presents several simulation exercises to study the finite sample performance of the proposed estimator. First, we assess the matching function estimator. Second, we evaluate the unconditional quantile partial effect (UQPE) estimation. 

The first data generating process (DGP) we consider is as following:
\begin{align}\label{eq:monte_carlo1}
y_i = 1 + x_i + (1+\theta x_i) u_{i},
\end{align}
where $x_i \sim N(10,1)$ and $u_{i}$ is a random variable with $E(u_{i})=0$, $V(u_{i})=1$ and independent of $x_i$. The distribution of $u_{i}$ is specified below as either standard Gaussian or (standardized) Chi-squared with 1 degree of freedom.
The parameter $\theta$ controls the type of effect of the covariate $x$ on the distribution of $y|x$: when $\theta=0$ the effect is a \textit{location shift}, and if $\theta\neq0$ is a \textit{location-scale shift}. In the former case the conditional quantile regression (CQR) effects are constant across quantiles, while in the latter case they vary. 

Second, we use a DGP with an additional covariate
\begin{align}\label{eq:monte_carlo2}
y_i = 1 + w_i + x_i + (1+\theta x_i) u_{i},
\end{align}
where we consider two cases: (i) $w_i\sim N(10,1)$ (independent of $x_i)$; (ii) $w_i=10+(x_i+N(10,1)-20)/\sqrt{2}$, where we make $w_{i}$ correlated with $x_i$. 

\subsection{Matching function estimator}

The proposed UQPE estimator relies on the estimator of the matching function for the quantiles, $\hat{\xi}_\tau(x)$. This subsection presents simulations exercises for assessing the accuracy of the matching estimator as given in equation \eqref{eq:mat_func}. Recall from Example \ref{example1}, equation \eqref{eq:example1}, that in the simple linear case we have an explicit formula for the population matching function, $\xi_\tau(x)$. Thus, we are able to use simulations to assess the finite sample performance of the estimator.  

\begin{figure}
    \caption{Estimation of matching functions, $u \sim N(0,1)$ and $\theta=0$ (pure location).}
    \label{fig:matching_theta0}
    \centering
    \begin{tabular}{cc}
    \includegraphics[width=70mm]{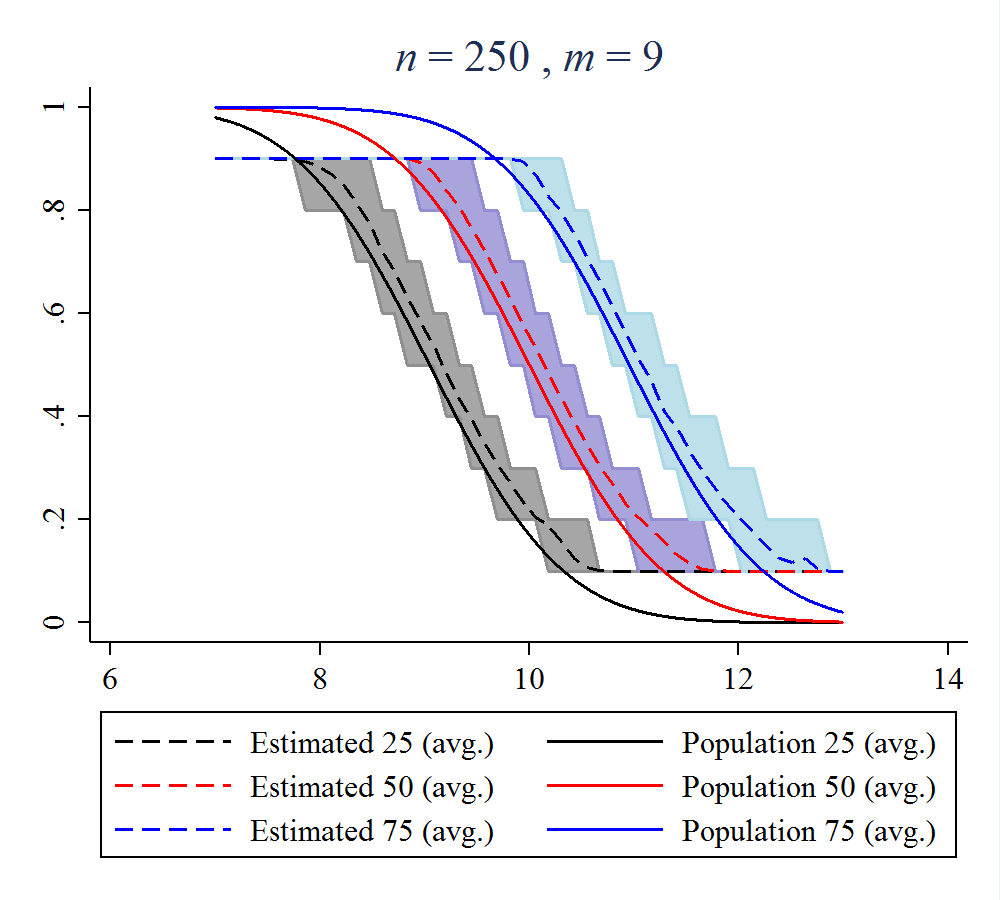}&
    \includegraphics[width=70mm]{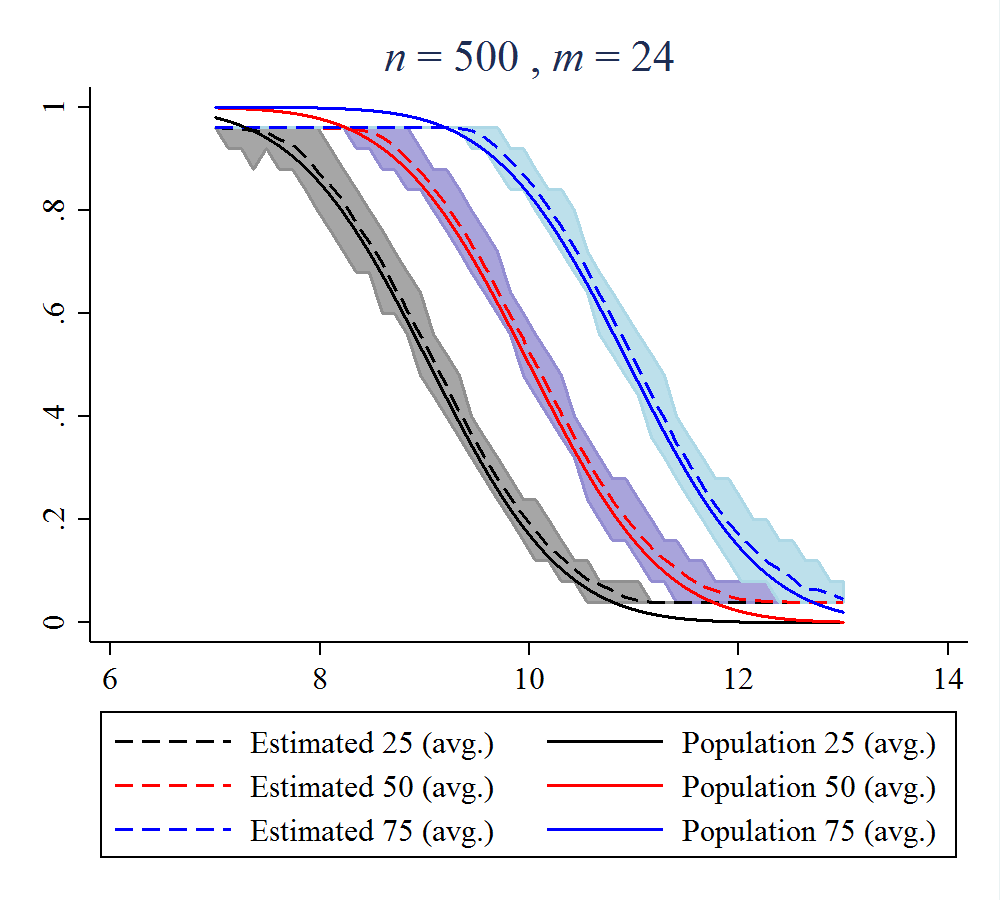}\\
    \includegraphics[width=70mm]{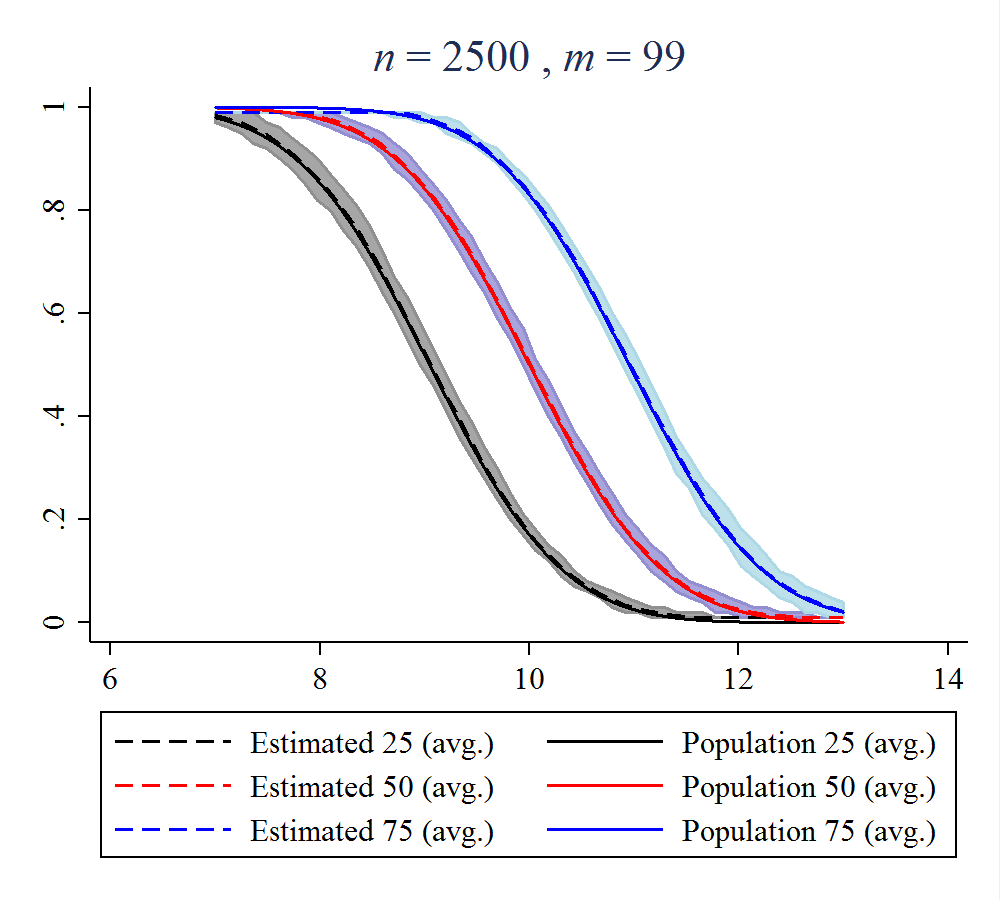}&
    \includegraphics[width=70mm]{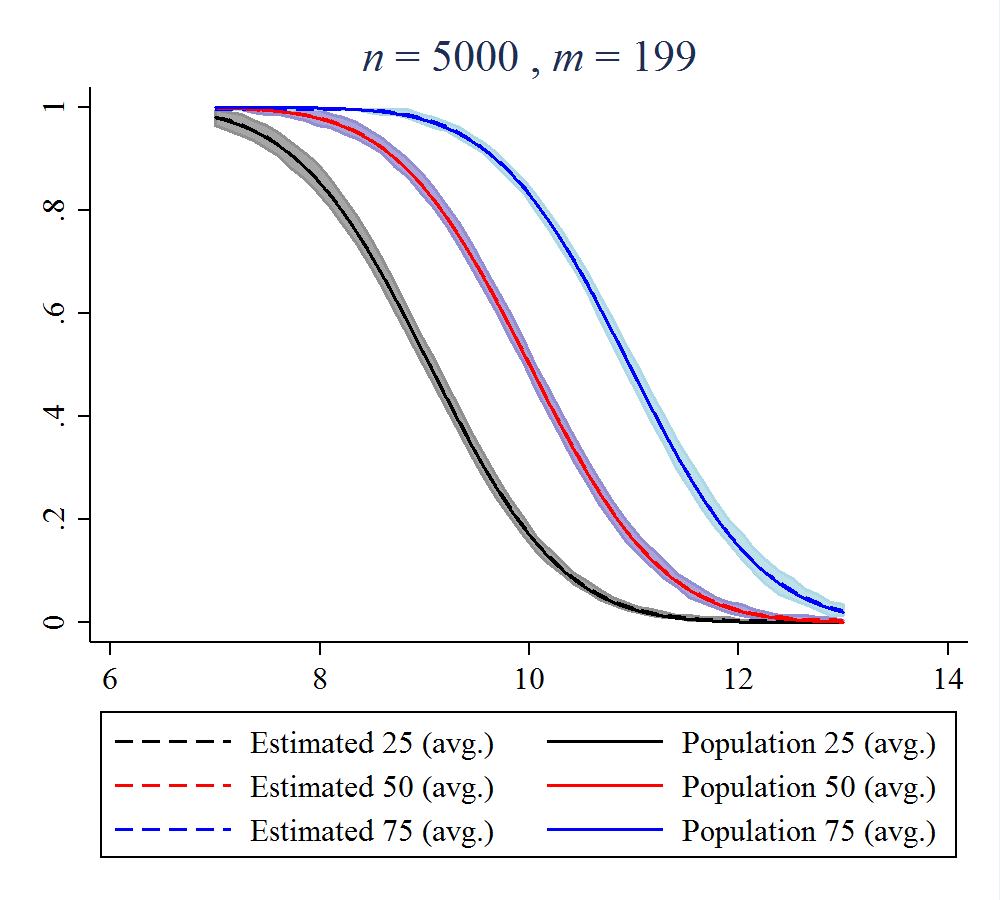}\\
    \end{tabular}

Notes: The shaded areas are 95\% empirical confidence intervals estimated using 100 simulations.

\end{figure}

\begin{figure}
    \caption{Estimation of matching functions, $u \sim N(0,1)$ and  $\theta=1$ (location-scale).}
    \label{fig:matching_theta1}
    \centering
    \begin{tabular}{cc}
    \includegraphics[width=70mm]{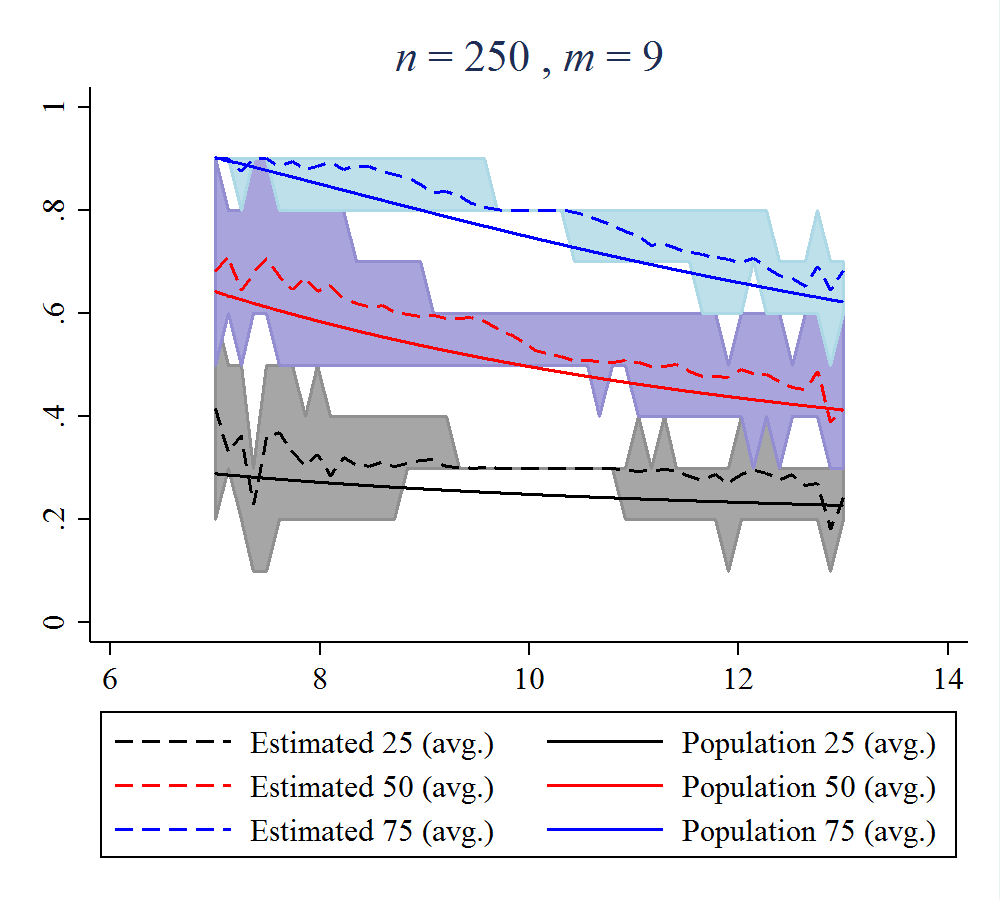}&
    \includegraphics[width=70mm]{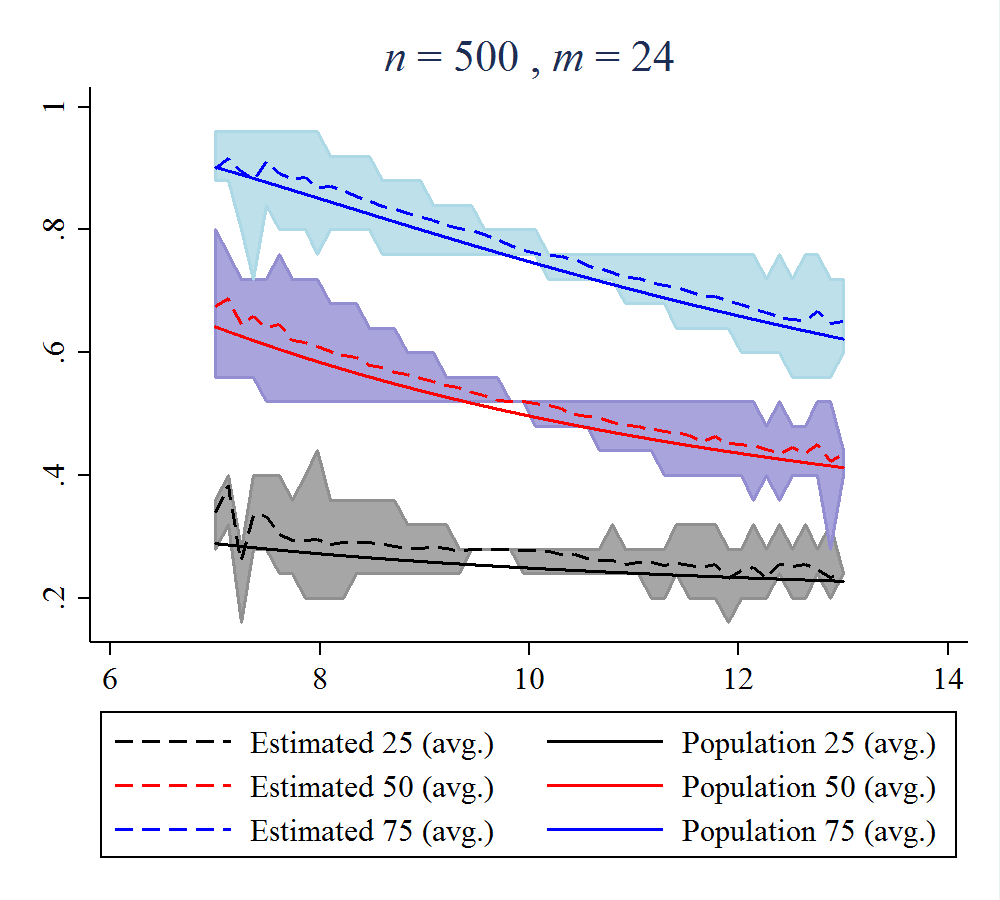}\\
    \includegraphics[width=70mm]{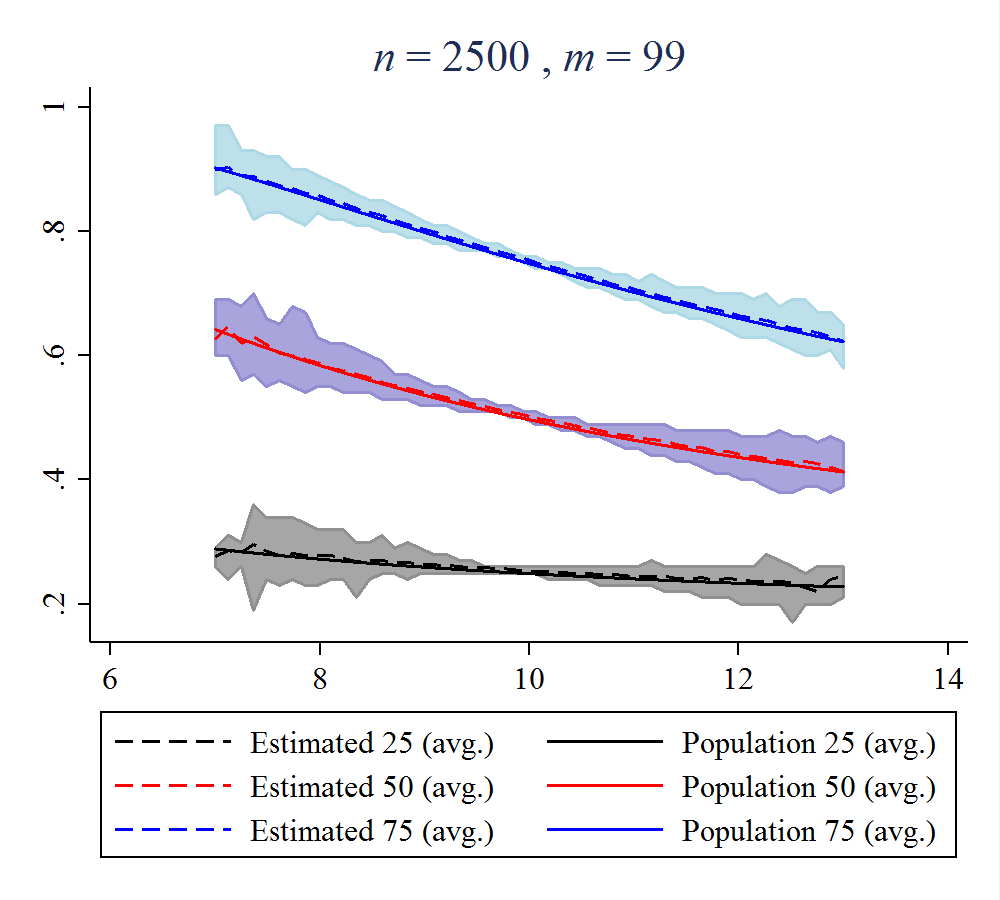}&
    \includegraphics[width=70mm]{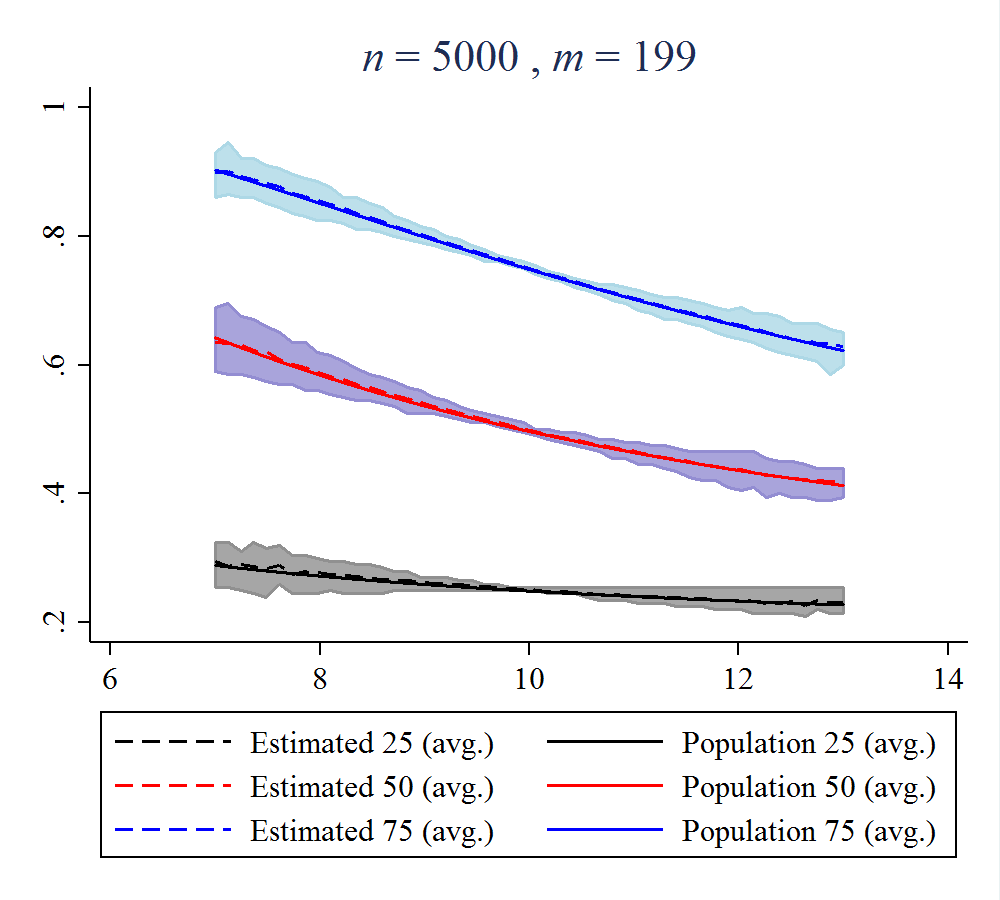}\\
    \end{tabular}

Notes: The shaded areas are 95\% empirical confidence intervals estimated using 100 simulations.

\end{figure}

We consider experiments using DGP model in \eqref{eq:monte_carlo1} for a pure location model, $\theta=0$, as well as a location-scale model, $\theta=1$. We use $x_i \sim N(10,1)$ and $u_{i}\sim N(10,1)$. 
Each experiment has 100 simulations of the DGP with sample sizes $n=\left\{ 250, 500, 2500, 5000 \right\}$, and quantile grid sizes $m=\left\{ 9, 24, 99, 199 \right\}$, respectively. We consider three quantiles $\tau \in \{0.25,0.50,0.75\}$. 
Figure \ref{fig:matching_theta0} reports results for the location case, and Figure \ref{fig:matching_theta1} displays results for the location-scale case. 
In each figure,  we plot the parameter of interest (the true value of the matching function), the estimates (average estimates over the number of simulations), as well as the 95\% empirical confidence interval.\footnote{These are computed as the 2.5-th and 97.5-th empirical percentiles of estimates across simulations.}

Simulation results show evidence that the matching function estimator provides an approximately asymptotically unbiased estimator for both the pure location and location-scale models with a better performance of sample sizes of $n\geq 500$. Point estimates are close to the populations counterparts even for small samples and grids. As sample size and grid increase together, point estimates become very close to the population and confidence intervals shrink. 

\subsection{UQPE estimation}

Now we investigate the finite sample performance of the proposed UQPE estimator as in equation \eqref{eq:nw_est}. In what follows, We label this estimator as Nadaraya-Watson (NW).
For comparison, we also implement the RIF regression model for UQR for each unconditional quantile using the \texttt{rifvar} STATA command \citep{RiosAvila20}. In this case, the UQPE is estimated by OLS with a cubic polynomial model of the RIF for each quantile as a function of $x_i$ (or as a function of $x_i$ and $w_i$ for the model with additional covariates). Although not reported the cubic model overperforms the linear and quadratic implementation, available from the Authors upon request. Finally we also implement the RIF-Logit model suggested by \cite{FirpoFortinLemieux09}.

Each experiment has 1,000 simulations of the DGP with sample sizes $n=\left\{ 250, 500, 2500, 5000 \right\}$, and quantiles grid sizes $m=\left\{ 9, 24, 99, 199 \right\}$, respectively. We consider three quantiles $\tau \in \{0.25,0.50,0.75\}$. Moreover, we use a bandwidth $h_n = 0.9 \hat{\sigma}_y n^{-1/5}$ and the Gaussian kernel function. To evaluate the procedures we report the empirical bias, variance, and mean-squared error (MSE).

Table \ref{table:loc-normal} presents results for the baseline model for the simple location-shift model (i.e. $\theta=0$) and Gaussian covariate and innovation. Both RIF models and NW estimators have a good performance in terms of bias, variance and MSE. These three statistics decrease for both estimators as sample size increases, for all three quantiles. 

Tables \ref{table:locscale-normal} 
and	\ref{table:locscale-chi2} present simulations results for Gaussian and Chi-squared innovations, respectively, 
for the location-scale shift model (i.e. $\theta=1$) with a Gaussian covariate. For all cases we observe that for the proposed NW estimator the bias and variance reduces as $n$ increases. The relative performance to the RIF models varies depending on the simulation exercises, but in most cases either the NW estimator outperforms the RIF one. 

Tables 	\ref{table:locscale-normal-w} and \ref{table:locscale-normal-wcorr} collect simulation results for cases where there is an additional covariate, $w_{i}$. The former case uses an independent additional covariate and in the latter case $w_i$ is correlated with $x_i$. In both cases we use the model with $\theta=1$ and $x_{i} \sim N(10,1)$. The results are also in line with previous ones, highlighting a good performance of the NW estimator in terms of bias, variance, and MSE.

Overall, these simulation results indicate that our proposed method produces a consistent estimator, where both bias and variance reduce as $n$ increases. In some cases, however, the bias improvement applies only for $n\geq500$.

Next, Table \ref{tab:bandkernel} presents simulation exercises where we consider different bandwidth choices. In particular, we use $h_n = 0.9 \hat{\sigma}_y n^{-1/4}$, $h_n = 0.9 \hat{\sigma}_y n^{-1/5}$ and $h_n = 0.9 \hat{\sigma}_y n^{-1/6}$. 
We consider the location-scale model with $\theta=1$ and Gaussian errors. The results show evidence that there are only small differences across bandwidths which suggest that the estimator is robust to these choices. For the empirical researcher this suggests that our proposed estimator can be combined with different nonparametric implementations.

\begin{table}
	\caption{Location model: $\theta=0$, and $x_{i} \sim N(10,1)$, $u_{i} \sim N(0,1)$.}
	\label{table:loc-normal}
	\centering

    \begin{tabular}{cc|ccc|ccc|ccc}
    \toprule
    \multicolumn{2}{c}{\textbf{Estimator}} & \multicolumn{3}{c}{\textbf{RIF-OLS (cubic)}} & \multicolumn{3}{c}{\textbf{RIF-Logit}} & \multicolumn{3}{c}{\textbf{NW}} \\
    \midrule
    \textbf{$\tau$} & \textbf{$n$} & \textbf{Bias} & \textbf{Variance} & \textbf{MSE} & \textbf{Bias} & \textbf{Variance} & \textbf{MSE} & \textbf{Bias} & \textbf{Variance} & \textbf{MSE} \\
    \midrule
    \multirow{4}[2]{*}{25} & 250   & 0.02470 & 0.01559 & 0.01620 & 0.02547 & 0.01736 & 0.01800 & -0.00405 & 0.00458 & 0.00459 \\
          & 500   & 0.02432 & 0.00876 & 0.00935 & 0.01837 & 0.00960 & 0.00994 & 0.00019 & 0.00241 & 0.00241 \\
          & 2500  & 0.01299 & 0.00202 & 0.00219 & 0.00924 & 0.00224 & 0.00233 & -0.00104 & 0.00046 & 0.00046 \\
          & 5000  & 0.00866 & 0.00122 & 0.00129 & 0.00595 & 0.00136 & 0.00140 & -0.00049 & 0.00022 & 0.00022 \\
    \midrule
    \multirow{4}[2]{*}{50} & 250   & 0.04550 & 0.01215 & 0.01422 & 0.03407 & 0.01344 & 0.01460 & -0.00017 & 0.00419 & 0.00419 \\
          & 500   & 0.03899 & 0.00736 & 0.00888 & 0.02971 & 0.00815 & 0.00903 & 0.00222 & 0.00219 & 0.00219 \\
          & 2500  & 0.02319 & 0.00166 & 0.00220 & 0.01701 & 0.00188 & 0.00217 & -0.00050 & 0.00043 & 0.00043 \\
          & 5000  & 0.01764 & 0.00090 & 0.00121 & 0.01287 & 0.00103 & 0.00119 & -0.00005 & 0.00020 & 0.00020 \\
    \midrule
    \multirow{4}[2]{*}{75} & 250   & 0.03803 & 0.01592 & 0.01737 & 0.02125 & 0.01714 & 0.01759 & 0.00242 & 0.00503 & 0.00504 \\
          & 500   & 0.02561 & 0.00846 & 0.00911 & 0.01772 & 0.00920 & 0.00952 & 0.00400 & 0.00261 & 0.00263 \\
          & 2500  & 0.01217 & 0.00219 & 0.00234 & 0.00856 & 0.00242 & 0.00249 & 0.00015 & 0.00047 & 0.00047 \\
          & 5000  & 0.01135 & 0.00120 & 0.00133 & 0.00852 & 0.00136 & 0.00143 & 0.00045 & 0.00023 & 0.00023 \\
    \bottomrule
    \end{tabular}%

Notes: Monte Carlo experiments based on 1000 simulations.
\end{table}

\begin{table}
	\caption{Location-scale shift model: $\theta=1$, and $x_{i} \sim N(10,1)$,  $u_{i} \sim N(0,1)$.}
	\label{table:locscale-normal}
	\centering
    
    \begin{tabular}{cc|ccc|ccc|ccc}
    \toprule
    \multicolumn{2}{c}{\textbf{Estimator}} & \multicolumn{3}{c}{\textbf{RIF-OLS (cubic)}} & \multicolumn{3}{c}{\textbf{RIF-Logit}} & \multicolumn{3}{c}{\textbf{NW}} \\
    \midrule
    \textbf{$\tau$} & \multicolumn{1}{c}{\textbf{$n$}} & \textbf{Bias} & \textbf{Variance} & \textbf{MSE} & \textbf{Bias} & \textbf{Variance} & \textbf{MSE} & \textbf{Bias} & \textbf{Variance} & \textbf{MSE} \\
    \midrule
    \multirow{4}[2]{*}{25} & 250   & -0.00459 & 1.01506 & 1.01508 & 0.00897 & 0.97017 & 0.97025 & 0.14868 & 0.80311 & 0.82522 \\
          & 500   & 0.02808 & 0.47673 & 0.47752 & 0.02558 & 0.46631 & 0.46696 & 0.09142 & 0.42411 & 0.43246 \\
          & 2500  & -0.00302 & 0.09242 & 0.09242 & -0.00503 & 0.09132 & 0.09135 & 0.00379 & 0.08527 & 0.08528 \\
          & 5000  & 0.00355 & 0.04448 & 0.04449 & 0.00239 & 0.04412 & 0.04413 & 0.00602 & 0.04154 & 0.04157 \\
    \midrule
    \multirow{4}[2]{*}{50} & 250   & 0.05602 & 0.86653 & 0.86967 & 0.04100 & 0.81673 & 0.81841 & 0.11159 & 0.68532 & 0.69778 \\
          & 500   & 0.07079 & 0.43029 & 0.43530 & 0.05600 & 0.41888 & 0.42201 & 0.06935 & 0.35727 & 0.36208 \\
          & 2500  & 0.01198 & 0.08004 & 0.08018 & 0.00445 & 0.07946 & 0.07948 & 0.00249 & 0.06978 & 0.06978 \\
          & 5000  & 0.01055 & 0.03963 & 0.03974 & 0.00541 & 0.03937 & 0.03940 & -0.00108 & 0.03341 & 0.03341 \\
    \midrule
    \multirow{4}[2]{*}{75} & 250   & 0.07899 & 1.06619 & 1.07243 & 0.06986 & 0.97638 & 0.98126 & 0.19632 & 0.82485 & 0.86339 \\
          & 500   & 0.04227 & 0.50446 & 0.50625 & 0.03024 & 0.49447 & 0.49538 & 0.09177 & 0.42507 & 0.43349 \\
          & 2500  & 0.03413 & 0.10107 & 0.10223 & 0.02793 & 0.10136 & 0.10214 & 0.02224 & 0.08093 & 0.08143 \\
          & 5000  & 0.02322 & 0.04809 & 0.04863 & 0.01820 & 0.04827 & 0.04860 & 0.01541 & 0.03744 & 0.03768 \\
    \bottomrule
    \end{tabular}%

Notes: Monte Carlo experiments based on 1000 simulations.
\end{table}

\begin{table}
	\caption{Location-scale shift model: $\theta=1$, and $x_{i} \sim N(10,1)$, $u_{i} \sim (\chi^{2}_1-1)/\sqrt 2$.}
	\label{table:locscale-chi2}
	\centering

    \begin{tabular}{cc|ccc|ccc|ccc}
    \toprule
    \multicolumn{2}{c}{\textbf{Estimator}} & \multicolumn{3}{c}{\textbf{RIF-OLS (cubic)}} & \multicolumn{3}{c}{\textbf{RIF-Logit}} & \multicolumn{3}{c}{\textbf{NW}} \\
    \midrule
    \textbf{$\tau$} & \multicolumn{1}{c}{\textbf{$n$}} & \textbf{Bias} & \textbf{Variance} & \textbf{MSE} & \textbf{Bias} & \textbf{Variance} & \textbf{MSE} & \textbf{Bias} & \textbf{Variance} & \textbf{MSE} \\
    \midrule
    \multirow{4}[2]{*}{25} & 250   & 0.38010 & 0.12921 & 0.27369 & 0.29658 & 0.10312 & 0.19108 & 0.04724 & 0.05433 & 0.05656 \\
          & 500   & 0.31075 & 0.05070 & 0.14727 & 0.24543 & 0.04069 & 0.10093 & 0.01919 & 0.02012 & 0.02049 \\
          & 2500  & 0.18007 & 0.00637 & 0.03880 & 0.13126 & 0.00527 & 0.02250 & 0.01149 & 0.00315 & 0.00328 \\
          & 5000  & 0.13945 & 0.00289 & 0.02234 & 0.09669 & 0.00240 & 0.01175 & 0.00796 & 0.00157 & 0.00163 \\
    \midrule
    \multirow{4}[2]{*}{50} & 250   & -0.07688 & 0.22020 & 0.22611 & -0.08321 & 0.20770 & 0.21463 & 0.08695 & 0.34549 & 0.35305 \\
          & 500   & -0.08876 & 0.10207 & 0.10995 & -0.08481 & 0.10242 & 0.10961 & 0.03118 & 0.14885 & 0.14982 \\
          & 2500  & -0.08059 & 0.02002 & 0.02652 & -0.06304 & 0.02133 & 0.02531 & 0.00559 & 0.02382 & 0.02385 \\
          & 5000  & -0.06959 & 0.01075 & 0.01559 & -0.05158 & 0.01150 & 0.01416 & -0.00212 & 0.01240 & 0.01240 \\
    \midrule
    \multirow{4}[2]{*}{75} & 250   & -0.05772 & 1.39024 & 1.39357 & -0.03968 & 1.40400 & 1.40557 & 0.24080 & 1.95959 & 2.01758 \\
          & 500   & -0.08805 & 0.70950 & 0.71725 & -0.06975 & 0.72037 & 0.72524 & 0.04520 & 0.81321 & 0.81526 \\
          & 2500  & -0.02743 & 0.14136 & 0.14211 & -0.02110 & 0.14339 & 0.14384 & 0.01590 & 0.13451 & 0.13477 \\
          & 5000  & -0.03405 & 0.06550 & 0.06666 & -0.02846 & 0.06630 & 0.06711 & -0.00672 & 0.05988 & 0.05992 \\
    \bottomrule
    \end{tabular}%

Notes: Monte Carlo experiments based on 1000 simulations.
\end{table}

\begin{table}[htbp]
  \centering
	\caption{Location-scale shift model with independent covariate: $\theta=1$, $w_i \sim N(10,1) $, and $x_{i} \sim N(10,1)$, $u_{i} \sim N(0,1)$.}
	\label{table:locscale-normal-w}

    \begin{tabular}{cc|ccc|ccc|ccc}
    \toprule
    \multicolumn{2}{c}{\textbf{Estimator}} & \multicolumn{3}{c}{\textbf{RIF-OLS (cubic)}} & \multicolumn{3}{c}{\textbf{RIF-Logit}} & \multicolumn{3}{c}{\textbf{NW}} \\
    \midrule
    \textbf{$\tau$} & \textbf{$n$} & \textbf{Bias} & \textbf{Variance} & \textbf{MSE} & \textbf{Bias} & \textbf{Variance} & \textbf{MSE} & \textbf{Bias} & \textbf{Variance} & \textbf{MSE} \\
    \midrule
    \multirow{4}[2]{*}{25} & 250   & -0.01793 & 0.96720 & 0.96752 & -0.01313 & 0.90922 & 0.90940 & 0.11521 & 0.76408 & 0.77735 \\
          & 500   & 0.02157 & 0.52336 & 0.52383 & 0.01794 & 0.51131 & 0.51163 & 0.08835 & 0.40504 & 0.41285 \\
          & 2500  & 0.00591 & 0.09537 & 0.09541 & 0.00476 & 0.09394 & 0.09396 & 0.01958 & 0.08337 & 0.08375 \\
          & 5000  & -0.00026 & 0.04463 & 0.04463 & -0.00131 & 0.04433 & 0.04433 & 0.00394 & 0.03924 & 0.03926 \\
    \midrule
    \multirow{4}[2]{*}{50} & 250   & 0.00228 & 0.93384 & 0.93384 & -0.02212 & 0.87770 & 0.87819 & 0.07071 & 0.68265 & 0.68765 \\
          & 500   & 0.04824 & 0.41277 & 0.41510 & 0.03867 & 0.39961 & 0.40110 & 0.06020 & 0.32478 & 0.32840 \\
          & 2500  & 0.00705 & 0.08055 & 0.08060 & 0.00002 & 0.07976 & 0.07976 & -0.00757 & 0.07035 & 0.07041 \\
          & 5000  & -0.00166 & 0.03845 & 0.03846 & -0.00712 & 0.03802 & 0.03807 & -0.01254 & 0.03165 & 0.03181 \\
    \midrule
    \multirow{4}[2]{*}{75} & 250   & -0.00581 & 1.01975 & 1.01979 & -0.01927 & 0.98074 & 0.98112 & 0.10028 & 0.86565 & 0.87570 \\
          & 500   & 0.04783 & 0.45838 & 0.46066 & 0.03623 & 0.45086 & 0.45217 & 0.05986 & 0.37305 & 0.37663 \\
          & 2500  & 0.01022 & 0.10191 & 0.10201 & 0.00445 & 0.10232 & 0.10234 & 0.00708 & 0.07831 & 0.07836 \\
          & 5000  & -0.00166 & 0.04808 & 0.04808 & -0.00545 & 0.04845 & 0.04848 & -0.01037 & 0.03879 & 0.03889 \\
    \bottomrule
    \end{tabular}%

Notes: Monte Carlo experiments based on 1000 simulations.
\end{table}

\begin{table}[htbp]
  \centering
	\caption{Location-shift model with correlated covariate: $\theta=1$, $w_i = 10 + (x_i + N(10,1)-20)/\sqrt{2} $, and $x_{i} \sim N(10,1)$, $u_{i} \sim N(0,1)$.}
	\label{table:locscale-normal-wcorr}

    \begin{tabular}{cc|ccc|ccc|ccc}
    \toprule
    \multicolumn{2}{c}{\textbf{Estimator}} & \multicolumn{3}{c}{\textbf{RIF-OLS (cubic)}} & \multicolumn{3}{c}{\textbf{RIF-Logit}} & \multicolumn{3}{c}{\textbf{NW}} \\
    \midrule
    \textbf{$\tau$} & \textbf{$n$} & \textbf{Bias} & \textbf{Variance} & \textbf{MSE} & \textbf{Bias} & \textbf{Variance} & \textbf{MSE} & \textbf{Bias} & \textbf{Variance} & \textbf{MSE} \\
    \midrule
    \multirow{4}[2]{*}{25} & 250   & -0.09324 & 2.05569 & 2.06438 & -0.06573 & 1.96614 & 1.97046 & 0.04637 & 1.61066 & 1.61281 \\
          & 500   & -0.02007 & 0.98728 & 0.98769 & -0.02275 & 0.96771 & 0.96823 & 0.04168 & 0.78219 & 0.78393 \\
          & 2500  & -0.02730 & 0.17002 & 0.17077 & -0.02856 & 0.16783 & 0.16864 & -0.01622 & 0.14786 & 0.14813 \\
          & 5000  & -0.01900 & 0.08653 & 0.08689 & -0.02042 & 0.08588 & 0.08630 & -0.01659 & 0.07614 & 0.07642 \\
    \midrule
    \multirow{4}[2]{*}{50} & 250   & 0.00075 & 1.90333 & 1.90334 & -0.02501 & 1.82672 & 1.82735 & 0.06789 & 1.42734 & 1.43195 \\
          & 500   & 0.03006 & 0.82247 & 0.82337 & 0.01738 & 0.79296 & 0.79326 & 0.02777 & 0.65445 & 0.65522 \\
          & 2500  & -0.01982 & 0.15795 & 0.15834 & -0.02668 & 0.15681 & 0.15752 & -0.02519 & 0.13293 & 0.13356 \\
          & 5000  & -0.00523 & 0.07573 & 0.07576 & -0.01103 & 0.07484 & 0.07497 & -0.01642 & 0.06310 & 0.06337 \\
    \midrule
    \multirow{4}[2]{*}{75} & 250   & 0.00633 & 2.04659 & 2.04663 & -0.01486 & 1.92153 & 1.92175 & 0.12363 & 1.62417 & 1.63946 \\
          & 500   & 0.04492 & 1.03865 & 1.04067 & 0.03612 & 1.02019 & 1.02150 & 0.07378 & 0.78689 & 0.79233 \\
          & 2500  & 0.01423 & 0.19626 & 0.19646 & 0.00868 & 0.19582 & 0.19590 & 0.02220 & 0.15084 & 0.15134 \\
          & 5000  & 0.00299 & 0.09886 & 0.09887 & -0.00072 & 0.09855 & 0.09855 & -0.00242 & 0.07764 & 0.07764 \\
    \bottomrule
    \end{tabular}%

Notes: Monte Carlo experiments based on 1000 simulations.
\end{table}  

\begin{landscape}

\begin{table}[htbp]
  \centering
	\caption{Kernel and bandwidth choice ($\theta=1$ and $u_{i} \sim N(0,1)$).}
  \label{tab:bandkernel}%

    \begin{tabular}{cccccccccccc}
    \toprule
          &       &       & \multicolumn{3}{c}{Bias} & \multicolumn{3}{c}{Variance} & \multicolumn{3}{c}{ECM} \\
\cmidrule{4-12}    Kernel & $\tau$   & $n$     & $n^{-1/4}$ & $n^{-1/5}$ & $n^{-1/6}$ & $n^{-1/4}$ & $n^{-1/5}$ & $n^{-1/6}$ & $n^{-1/4}$ & $n^{-1/5}$ & $n^{-1/6}$ \\
    \midrule
    \multirow{12}[6]{*}{Gauss} & \multirow{4}[2]{*}{25} & 250   & 0.14894 & 0.14868 & 0.14849 & 0.80470 & 0.80311 & 0.80181 & 0.82688 & 0.82522 & 0.82386 \\
          &       & 500   & 0.09184 & 0.09142 & 0.09108 & 0.42566 & 0.42411 & 0.42300 & 0.43409 & 0.43246 & 0.43130 \\
          &       & 2500  & 0.00411 & 0.00379 & 0.00359 & 0.08542 & 0.08527 & 0.08516 & 0.08544 & 0.08528 & 0.08517 \\
          &       & 5000  & 0.00629 & 0.00602 & 0.00585 & 0.04160 & 0.04154 & 0.04150 & 0.04164 & 0.04157 & 0.04153 \\
\cmidrule{2-12}          & \multirow{4}[2]{*}{50} & 250   & 0.11349 & 0.11159 & 0.11028 & 0.68763 & 0.68532 & 0.68358 & 0.70051 & 0.69778 & 0.69574 \\
          &       & 500   & 0.07096 & 0.06935 & 0.06825 & 0.35794 & 0.35727 & 0.35668 & 0.36298 & 0.36208 & 0.36133 \\
          &       & 2500  & 0.00308 & 0.00249 & 0.00201 & 0.06978 & 0.06978 & 0.06975 & 0.06979 & 0.06978 & 0.06975 \\
          &       & 5000  & -0.00071 & -0.00108 & -0.00144 & 0.03340 & 0.03341 & 0.03340 & 0.03340 & 0.03341 & 0.03341 \\
\cmidrule{2-12}          & \multirow{4}[2]{*}{75} & 250   & 0.19834 & 0.19632 & 0.19488 & 0.82528 & 0.82485 & 0.82442 & 0.86462 & 0.86339 & 0.86239 \\
          &       & 500   & 0.09263 & 0.09177 & 0.09123 & 0.42418 & 0.42507 & 0.42552 & 0.43276 & 0.43349 & 0.43384 \\
          &       & 2500  & 0.02237 & 0.02224 & 0.02228 & 0.08078 & 0.08093 & 0.08107 & 0.08128 & 0.08143 & 0.08156 \\
          &       & 5000  & 0.01525 & 0.01541 & 0.01558 & 0.03740 & 0.03744 & 0.03750 & 0.03763 & 0.03768 & 0.03774 \\
    \bottomrule
    \end{tabular}%

Notes: Monte Carlo experiments based on 1000 simulations.

\end{table}%

\end{landscape}

\subsection{UQPE bootstrap inference}
Finally, we evaluate the finite sample performance of the pairwise bootstrap procedure discussed above for practical inference of the proposed NW estimator. For this we report empirical coverages for a nominal coverage of the 95\% confidence intervals for three models considered above with one covariate $x_i\sim N(10,1)$ as in eq. \eqref{eq:monte_carlo1}: DGP 1: $\theta=0$ and $u_i\sim N(0,1)$ (as in Table \ref{table:loc-normal}); DGP 2: $\theta=1$ and $u_i\sim N(0,1)$ (as in Table \ref{table:locscale-normal}); and DGP 3: $\theta=1$ and $u_i\sim (\chi_1^2-1)/\sqrt{2}$ (as in Table \ref{table:locscale-chi2}). In particular, we consider the Gaussian confidence interval, constructed using $\pm 1.96$ times the estimated bootstrap variance, and the percentile bootstrap that uses the 0.025 and 0.975 percentiles of the bootstrapped distribution. 

The results for empirical coverage are presented in Table \ref{table:coverage}. The coverage is accurate for all cases, being close to the nominal 0.95. There is no systematic difference between Gaussian and percentile bootstrap, thus indicating that both work in a similar fashion.

\begin{table}[htbp]
  \centering
  \caption{Empirical coverage of the bootstrap inference}

    \begin{tabular}{cc|cc|cc|cc}
    \toprule
          &       & \multicolumn{2}{c|}{\textbf{DGP 1}} & \multicolumn{2}{c|}{\textbf{DGP 2}} & \multicolumn{2}{c}{\textbf{DGP 3}} \\
    \midrule
    \textbf{$\tau$} & \textbf{$n$} & \textbf{Gaussian} & \textbf{Percentile} & \textbf{Gaussian} & \textbf{Percentile} & \textbf{Gaussian} & \textbf{Percentile} \\
    \midrule
    \multirow{2}[2]{*}{25} & 500   & 0.949 & 0.941 & 0.942 & 0.953 & 0.920 & 0.932 \\
          & 1000  & 0.940 & 0.937 & 0.941 & 0.941 & 0.936 & 0.939 \\
    \midrule
    \multirow{2}[2]{*}{50} & 500   & 0.954 & 0.936 & 0.945 & 0.948 & 0.928 & 0.937 \\
          & 1000  & 0.940 & 0.934 & 0.940 & 0.945 & 0.931 & 0.939 \\
    \midrule
    \multirow{2}[2]{*}{75} & 500   & 0.949 & 0.945 & 0.935 & 0.947 & 0.922 & 0.940 \\
          & 1000  & 0.939 & 0.935 & 0.930 & 0.934 & 0.934 & 0.942 \\
    \bottomrule
    \end{tabular}%
  
  \label{table:coverage}%
  
Notes: Monte Carlo experiments based on 1000 simulations. Bootstrap with 100 replications. Coverage corresponding to 95\% confidence intervals.
\end{table}%

\section{Empirical Application}\label{empirical}

This section illustrates the UQPE estimator with an analysis of Engel's curves. The original concept corresponds to Ernst Engel (1857, cited in \citet{Koenker05}, pp. 78-79) who studied the European working class households consumption in the 19th century. 
Engel curves describe how household expenditures on particular goods and services depend on household income. The analysis of Engel curves has a long history of estimating the expenditure-income relationship. An empirical result commonly referred to as ``Engel's law'' states that the poorer a family is, the larger the budget share it spends on food. Other categories of expenditure present a less robust pattern. Hence, we investigate the hypothesis that food expenditure constitutes a declining share of household income.\footnote{There is a large literature on empirical applications of the Engel's curve, see, e.g., for example, among many others, \citet{Lewbel1997, Lewbel2008}, \cite{BlundellChenKristensen2007}, \cite{Charles09}, \cite{Heffetz11}, \cite{PerezTuglia13}, \citet{ChernozhukovFernandezValKowalski15}, \cite{Atkin20} and \cite{Li21}. We follow a regression approach to model the conditional effects. In this paper, for simplicity, we do not consider endogeneity issues that potentially arise from simultaneous decisions on income and expenditures. }

We apply this framework to household expenditures in Argentina using the national survey of expenditures (Encuesta Nacional de Gasto de los Hogares, known as ENGHO 2017-2018), implemented by the Instituto Nacional de Estad\'istica y Censos (INDEC). The survey was carried out between November 2017 and November 2018. The ENGHO 2017-2018 surveys the households' living conditions in terms of their access to goods and services, as well as their income. The data contains information about household expenditures on different goods and services. About 21,547 households were randomly selected and participated on the survey. We consider both food household expenditures and total non-durable consumption for comparison.\footnote{In particular, the sample has information on: $(i)$ food and non-alcoholic beverages, $(ii)$
alcoholic beverages and tobacco, $(iii)$ clothing and footwear, $(iv)$
housing, water, electricity, gas and other fuels, $(v)$ home equipment and maintenance, $(vi)$ health, $(vii)$ transportation, $(viii)$ communications, $(ix)$ recreation and culture, $(x)$ education, $(xi)$ restaurants and hotels, and $(xii)$ miscellaneous goods and services. Both expenses and income are transformed to represent monthly values. Since the monetary values of each household are expressed in current currency at the time of the survey, an inflation adjustment was made to transform them into constant currency for December of the fourth quarter of 2018 using the Consumer Price Index (CPI) computed by national statistical office, INDEC.} 

We estimate both UQPE and CQPE. The former analysis corresponds to evaluating effect of an increase in income for every household in a uniform pattern  on the  unconditional quantile of food expenditure while focusing on the entire distribution of expenditure. The latter effect corresponds to the study of how expenditure changes when marginally increasing income conditional on income. For comparison, we also provide estimate results for the RIF regression  and RIF-Logit of \cite{FirpoFortinLemieux09}. We use both income and expenditures in logarithm, so that the coefficient estimates can be interpreted as an elasticity. Confidence intervals are computed using 200 bootstrap replications.

\begin{table}[htbp]
  \centering
  \caption{Engel's curve for food expenditures.}
  \label{tab:food-short}%

      \begin{tabular}{rrrrrr}
    \toprule
          & \multicolumn{5}{c}{\textbf{Quantile Partial Effect}} \\
\cmidrule{2-6}          & \multicolumn{1}{c}{\textbf{10}} & \multicolumn{1}{c}{\textbf{25}} & \multicolumn{1}{c}{\textbf{50}} & \multicolumn{1}{c}{\textbf{75}} & \multicolumn{1}{c}{\textbf{90}} \\
    \midrule
    \multicolumn{1}{l}{\textit{Conditional distribution}} &       &       &       &       &  \\
          &       &       &       &       &  \\
    \multicolumn{1}{l}{CQR} & \multicolumn{1}{c}{0.383***} & \multicolumn{1}{c}{0.407***} & \multicolumn{1}{c}{0.408***} & \multicolumn{1}{c}{0.408***} & \multicolumn{1}{c}{0.425***} \\
          & \multicolumn{1}{c}{(0.000571)} & \multicolumn{1}{c}{(0.000422)} & \multicolumn{1}{c}{(0.000246)} & \multicolumn{1}{c}{(0.000278)} & \multicolumn{1}{c}{(0.000336)} \\
    \multicolumn{1}{l}{\textit{Unconditional distribution}} &       &       &       &       &  \\
          &       &       &       &       &  \\
    \multicolumn{1}{l}{RIF-OLS (linear)} & \multicolumn{1}{c}{0.367***} & \multicolumn{1}{c}{0.388***} & \multicolumn{1}{c}{0.427***} & \multicolumn{1}{c}{0.396***} & \multicolumn{1}{c}{0.393***} \\
          & \multicolumn{1}{c}{(0.0285)} & \multicolumn{1}{c}{(0.0170)} & \multicolumn{1}{c}{(0.0139)} & \multicolumn{1}{c}{(0.0130)} & \multicolumn{1}{c}{(0.0181)} \\
    \multicolumn{1}{l}{RIF-OLS (quadratic)} & \multicolumn{1}{c}{0.360***} & \multicolumn{1}{c}{0.383***} & \multicolumn{1}{c}{0.427***} & \multicolumn{1}{c}{0.403***} & \multicolumn{1}{c}{0.406***} \\
          & \multicolumn{1}{c}{(0.0275)} & \multicolumn{1}{c}{(0.0166)} & \multicolumn{1}{c}{(0.0140)} & \multicolumn{1}{c}{(0.0129)} & \multicolumn{1}{c}{(0.0182)} \\
    \multicolumn{1}{l}{RIF-OLS (cubic)} & \multicolumn{1}{c}{0.370***} & \multicolumn{1}{c}{0.394***} & \multicolumn{1}{c}{0.440***} & \multicolumn{1}{c}{0.415***} & \multicolumn{1}{c}{0.412***} \\
          & \multicolumn{1}{c}{(0.0279)} & \multicolumn{1}{c}{(0.0169)} & \multicolumn{1}{c}{(0.0143)} & \multicolumn{1}{c}{(0.0137)} & \multicolumn{1}{c}{(0.0183)} \\
    \multicolumn{1}{l}{RIF-Logit} & \multicolumn{1}{c}{0.327***} & \multicolumn{1}{c}{0.395***} & \multicolumn{1}{c}{0.434***} & \multicolumn{1}{c}{0.403***} & \multicolumn{1}{c}{0.427***} \\
          & \multicolumn{1}{l}{(0.0373)} & \multicolumn{1}{l}{(0.0260)} & \multicolumn{1}{l}{(0.0234)} & \multicolumn{1}{l}{(0.0242)} & \multicolumn{1}{l}{(0.0316)} \\
          &       &       &       &       &  \\
    \multicolumn{1}{l}{NW} & \multicolumn{1}{c}{0.395***} & \multicolumn{1}{c}{0.405***} & \multicolumn{1}{c}{0.408***} & \multicolumn{1}{c}{0.409***} & \multicolumn{1}{c}{0.410***} \\
          & \multicolumn{1}{c}{(0.0166)} & \multicolumn{1}{c}{(0.0111)} & \multicolumn{1}{c}{(0.00851)} & \multicolumn{1}{c}{(0.00810)} & \multicolumn{1}{c}{(0.00868)} \\
          &       &       &       &       &  \\
    \multicolumn{1}{l}{Observations} & \multicolumn{1}{c}{21,017} & \multicolumn{1}{c}{21,017} & \multicolumn{1}{c}{21,017} & \multicolumn{1}{c}{21,012} & \multicolumn{1}{c}{21,017} \\
          &       &       &       &       &  \\
    \bottomrule
    \end{tabular}%

\footnotesize
Notes: The CQR analysis corresponds to a regression of log food expenditures on log income. UQPE estimates the effect of a marginal change in log income on the unconditional distribution of log food expenditures.
Standard errors in parentheses (analytical for CQR, bootstrap with 200 replications for RIF and NW). * indicates significance at 10 \%, ** at 5 \% and *** at 1 \%.	
\end{table}%

\begin{figure}[htbp]
        \caption{Engel's curves for food expenditures}
        \label{fig:food-short}
        \centering
       \includegraphics[width=13cm]{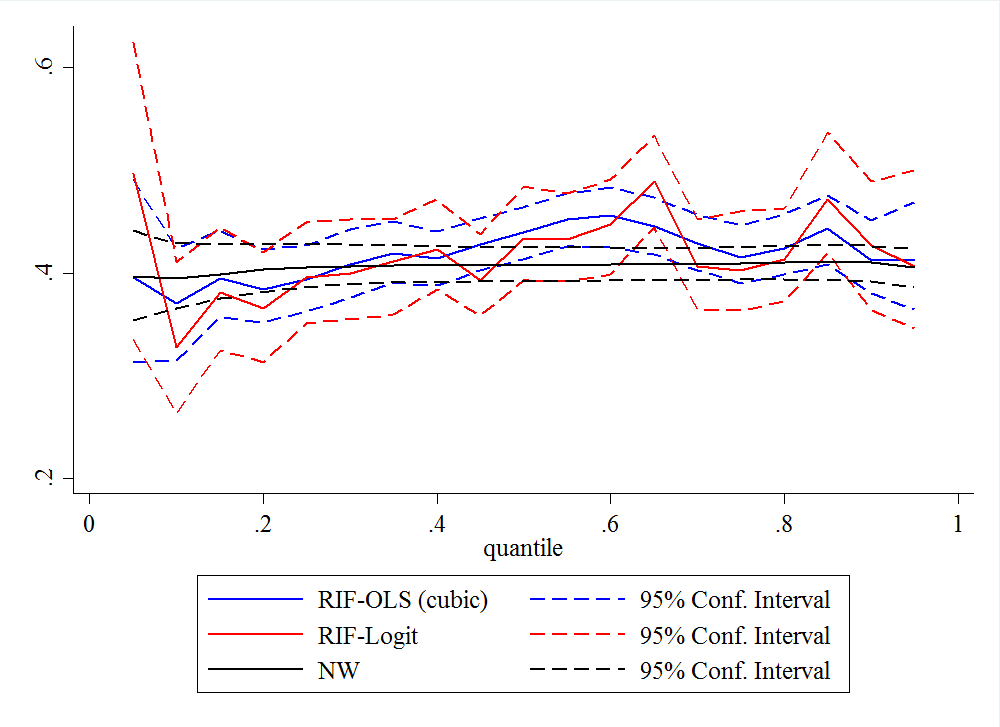}

       \footnotesize Notes: UQPE NW (black), RIF-OLS (cubic polynomial, blue) and RIF-Logit (red) estimates together with 95\% confidence intervals estimated using bootstrap with 200 replications.
\end{figure}

\begin{figure}[htbp]
         \caption{Estimated matching function - Food expenditures}         \label{fig:linkfood}
         \centering
        \includegraphics[width=10cm]{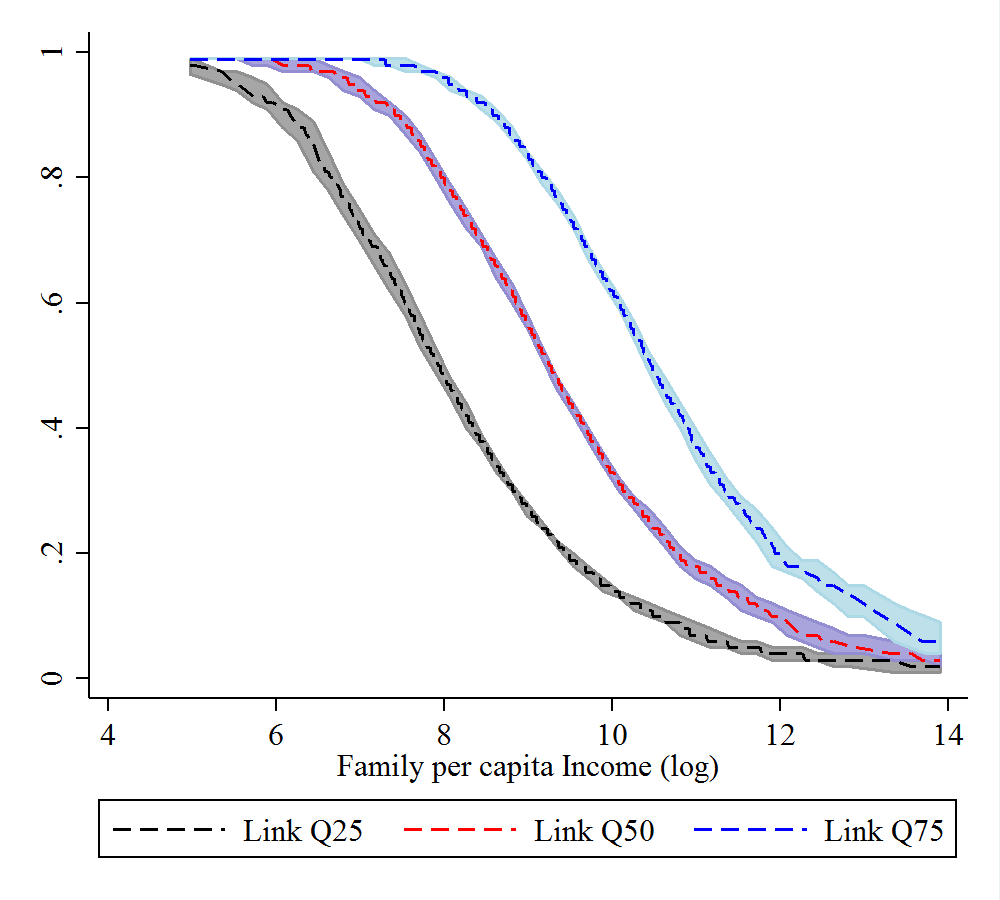}
        
        \footnotesize Notes: Matched coefficients $\hat\beta_1(\hat\xi_\tau(income_i))$ for $\tau\in\{0.25,0.50,0.75\}$ estimates together with 95\% confidence intervals estimated using bootstrap with 200 replications.

\end{figure}

\begin{table}[htbp]
 \centering
  \caption{Engel's curve for total non-durable expenditures}
  \label{tab:nondurables-short}%
     
    \begin{tabular}{rrrrrr}
    \toprule
          & \multicolumn{5}{c}{\textbf{Quantile Partial Effect}} \\
\cmidrule{2-6}          & \multicolumn{1}{c}{\textbf{10}} & \multicolumn{1}{c}{\textbf{25}} & \multicolumn{1}{c}{\textbf{50}} & \multicolumn{1}{c}{\textbf{75}} & \multicolumn{1}{c}{\textbf{90}} \\
    \midrule
    \multicolumn{1}{l}{\textit{Conditional distribution}} &       &       &       &       &  \\
          &       &       &       &       &  \\
    \multicolumn{1}{l}{CQR} & \multicolumn{1}{c}{0.778***} & \multicolumn{1}{c}{0.784***} & \multicolumn{1}{c}{0.776***} & \multicolumn{1}{c}{0.738***} & \multicolumn{1}{c}{0.662***} \\
          & \multicolumn{1}{c}{(0.000359)} & \multicolumn{1}{c}{(0.000265)} & \multicolumn{1}{c}{(0.000211)} & \multicolumn{1}{c}{(0.000283)} & \multicolumn{1}{c}{(0.000295)} \\
    \multicolumn{1}{l}{\textit{Unconditional distribution}} &       &       &       &       &  \\
          &       &       &       &       &  \\
    \multicolumn{1}{l}{RIF-OLS (linear)} & \multicolumn{1}{c}{0.700***} & \multicolumn{1}{c}{0.706***} & \multicolumn{1}{c}{0.761***} & \multicolumn{1}{c}{0.768***} & \multicolumn{1}{c}{0.731***} \\
          & \multicolumn{1}{c}{(0.0295)} & \multicolumn{1}{c}{(0.0194)} & \multicolumn{1}{c}{(0.0216)} & \multicolumn{1}{c}{(0.0229)} & \multicolumn{1}{c}{(0.0292)} \\
    \multicolumn{1}{l}{RIF-OLS (quadratic)} & \multicolumn{1}{c}{0.674***} & \multicolumn{1}{c}{0.693***} & \multicolumn{1}{c}{0.765***} & \multicolumn{1}{c}{0.792***} & \multicolumn{1}{c}{0.769***} \\
          & \multicolumn{1}{c}{(0.0278)} & \multicolumn{1}{c}{(0.0183)} & \multicolumn{1}{c}{(0.0217)} & \multicolumn{1}{c}{(0.0251)} & \multicolumn{1}{c}{(0.0338)} \\
    \multicolumn{1}{l}{RIF-OLS (cubic)} & \multicolumn{1}{c}{0.683***} & \multicolumn{1}{c}{0.719***} & \multicolumn{1}{c}{0.801***} & \multicolumn{1}{c}{0.817***} & \multicolumn{1}{c}{0.774***} \\
          & \multicolumn{1}{c}{(0.0267)} & \multicolumn{1}{c}{(0.0183)} & \multicolumn{1}{c}{(0.0217)} & \multicolumn{1}{c}{(0.0229)} & \multicolumn{1}{c}{(0.0275)} \\
    \multicolumn{1}{l}{RIF-Logit} & \multicolumn{1}{c}{0.622***} & \multicolumn{1}{c}{0.641***} & \multicolumn{1}{c}{0.755***} & \multicolumn{1}{c}{0.794***} & \multicolumn{1}{c}{0.811***} \\
          & \multicolumn{1}{l}{(0.0471)} & \multicolumn{1}{l}{(0.0329)} & \multicolumn{1}{l}{(0.0434)} & \multicolumn{1}{l}{(0.0475)} & \multicolumn{1}{l}{(0.0542)} \\
          &       &       &       &       &  \\
    \multicolumn{1}{l}{NW} & \multicolumn{1}{c}{0.774***} & \multicolumn{1}{c}{0.773***} & \multicolumn{1}{c}{0.760***} & \multicolumn{1}{c}{0.732***} & \multicolumn{1}{c}{0.700***} \\
          & \multicolumn{1}{c}{(0.0140)} & \multicolumn{1}{c}{(0.0102)} & \multicolumn{1}{c}{(0.00839)} & \multicolumn{1}{c}{(0.00863)} & \multicolumn{1}{c}{(0.0103)} \\
          &       &       &       &       &  \\
    \multicolumn{1}{l}{Observations} & \multicolumn{1}{c}{21,461} & \multicolumn{1}{c}{21,461} & \multicolumn{1}{c}{21,461} & \multicolumn{1}{c}{21,461} & \multicolumn{1}{c}{21,461} \\
          &       &       &       &       &  \\
    \bottomrule
    \end{tabular}%
    
\footnotesize
Notes: The CQR analysis corresponds to a regression of log non-durable expenditures on log income. UQPE estimates the effect of a marginal change in log income on the unconditional distribution of log non-durable expenditures.
Standard errors in parentheses (analytical for CQR, bootstrap with 200 replications for RIF and NW). * indicates significance at 10 \%, ** at 5 \% and *** at 1 \%.					\\
\end{table}%

\begin{figure}[htbp]
         \caption{Engel's curves for total non-durable expenditures.}
        \label{fig:nondurables-short}     \centering
       	\includegraphics[width=13cm]{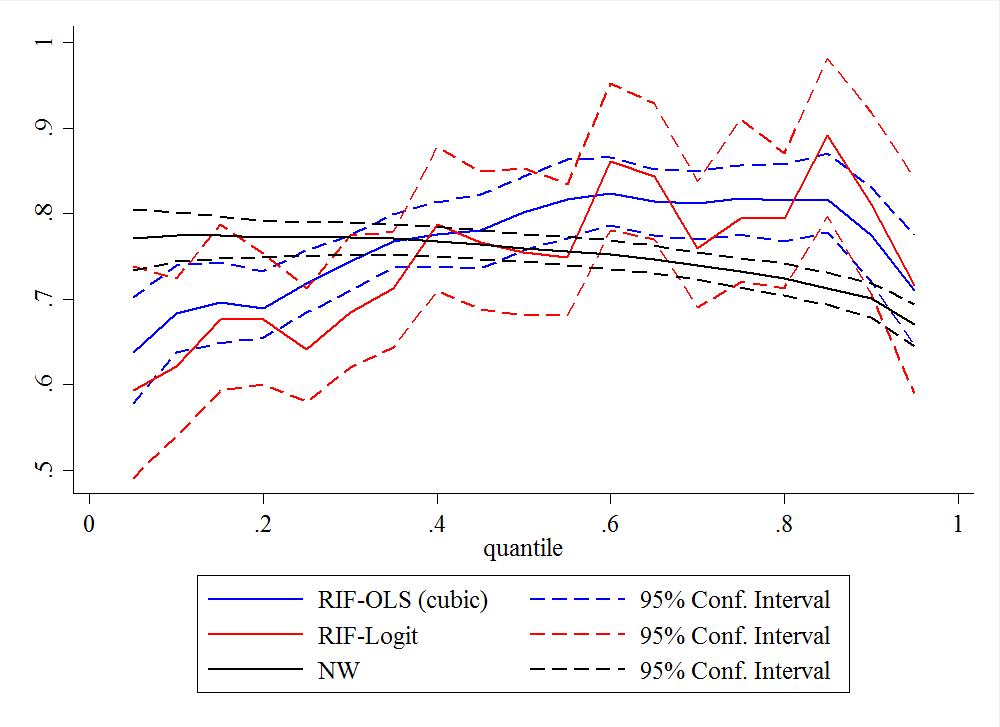}

        \footnotesize Notes: UQPE NW (black), RIF-OLS (cubic polynomial, blue) and RIF-Logit (red) estimates together with 95\% confidence intervals estimated using bootstrap with 200 replications).
\end{figure}

\begin{figure}[htbp]
        \centering
        \caption{Estimated matching function - Non-durable expenditures}         \label{fig:link}
        \includegraphics[width=10cm]{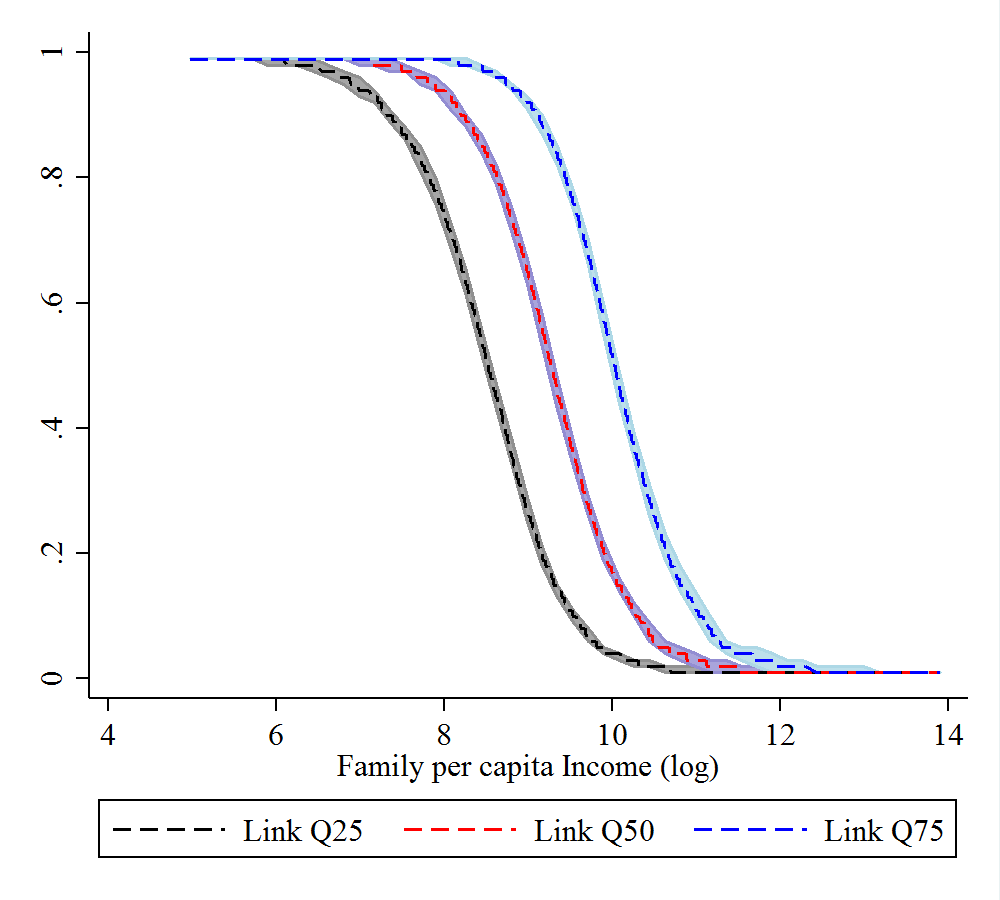}

        \footnotesize Notes: Matched coefficients $\hat\beta_1(\hat\xi_\tau(income_i))$ for $\tau\in\{0.25,0.50,0.75\}$ estimates together with 95\% confidence intervals estimated using bootstrap with 200 replications.
\end{figure}

We estimate these models for different quantiles. The results are collected in Table \ref{tab:food-short} and Figure \ref{fig:food-short} for food expenditures, and Table \ref{tab:nondurables-short} and Figure \ref{fig:nondurables-short} for total non-durable expenditures. 
In order to explore the results in more detail, we also plot the by-product of this analysis that is the matching quantile function. This was implicitly used for the estimation of the UQPE NW estimator. Figures \ref{fig:linkfood} and \ref{fig:link} plot the estimated match for $\tau=\{0.25,0.50,0.75\}$ for different values of log income, for food and non-durable expenditures, respectively. The last figures illustrate that for each $\tau$ there is a full range of variation in the corresponding CQPE model indexed by $\eta$.

The results for food expenditures show evidence that CQR coefficients are roughly constant across $\eta$, although mildly increasing. The proposed UQPE NW estimator is then also roughly constant across $\tau$. The RIF estimates also have this pattern although they are estimated in a less precise manner. In all cases, the estimated effects can be interpreted as elasticities, implying that a 1\% increase in income increase food consumption in less than 1\%, about 0.4\%. Moreover, since the CQR coefficients are mildly increasing, the variation in the UQPE has to be coming from the variation in the density of $X$ given $Y=Q_\tau[Y]$. As $\tau$ increases, for the UQPE to increase, higher CQR coefficients must be getting higher weight. This happens if the density of ``income|food$=Q_\tau[\text{food}]$'' is moving to the right. 

For the case of non-durable expenditures, the RIF estimates are increasing along $\tau$, while the UQPE NW estimator is decreasing.
The fact that RIF estimates have a larger range of variation than CQR and that it gives the counter-intuitive increasing pattern suggest that it might be misspecified. That is, if we assume that richest families have a higher saving rate than the poorer ones, the unconditional estimates should not increase along the quantiles.

A take away note from this example is that the empirical researcher should be aware of different modelling choices. RIF and UQPE NW estimators both rely on different assumptions to estimate the same parameter. As such, it may be a good procedure to report different alternatives and to highlight if differences occur.

\section{Conclusion}\label{conclusion}

This paper considers the use of conditional quantile regression analysis to estimate unconditional quantile partial effects. The proposed methodology is based on a matching and reweighting result to link the unconditional effects to the conditional ones. This method thus benefits from the usual conditional quantile regression estimation techniques, and suggests a two-step estimator for the unconditional effects. In the first step one estimates a structural quantile regression model, and in the second step a nonparametric regression is applied. We establish the asymptotic properties of the estimator. Monte Carlo simulations show evidence that the estimator has good finite sample performance and is robust to the selection of bandwidth and kernel. To illustrate the proposed methods, we study Engel's curves in Argentina.

The current paper can be extended in several directions. First, the proposed model uses a simple linear quantile regression framework and is based on its coefficient estimators. The current framework can be applied to any other $\sqrt{n}$-consistent estimation procedure. In particular, as an example, instrumental variables quantile regression and/or panel data models deliver consistent estimators for the conditional effects in several related statistical models. The current methodology could be extended to evaluate unconditional effects, starting from any initial consistent conditional estimation procedure. Second, the current proposed framework can be used to evaluate any other functional analysis related to the unconditional quantile regression one. In other words, to recover general distributional effects. Third, the Nadaraya-Watson estimator is the first approximation to a larger family of estimators that can be used to estimate the unconditional effects. Local linear regression models is a proposed refinement to obtain possibly better asymptotic properties. Fourth, we leave optimal bandwidth choice for future research.
Finally, the analysis may be extended in  formalizing the resampling procedure and studying to practically construct uniform confidence bands.

\newpage

\section*{Appendix A}

\renewcommand{\thesubsection}{A.\arabic{subsection}}

\setcounter{equation}{0} \renewcommand\theequation{A.\arabic{equation}}

\subsection{Proof of Lemma \ref{lemma:reverse}}

Recall that $X=(X_1,X_2')'$, $x=(x_1,x_2')'$, and $X_1$ and $x_1$ are one-dimensional. First we show that $CQPE_{X_1}(\eta, x)$, defined in \eqref{eq:cqpe0} as
\begin{align*}
CQPE_{X_1}(\eta, x) :&= \frac{\partial Q_Y[\tau|X_1 = z, X_2=x_2]}{\partial z}\bigg|_{z=x_1}
\end{align*}
can be written in the way of \eqref{eq:cqpe} as
\begin{align*}
CQPE_{X_1}(\eta, x) &= -\frac{1}{f_{Y| X}(Q_Y[\tau| X= x]| x)}   \frac{\partial F_{Y| X}(Q_Y[\tau| X= x]|z,x_2 )}{\partial z}\bigg|_{z=x_1}.
\end{align*}
By definition of quantiles, we have that this identity holds for all $x\in \mathcal X$ for a given fixed $\tau$:
\begin{align*}
 F_{Y| X}(Q_Y[\tau| X= x]|x ) = \tau.
\end{align*}
Differentiating both sides with respect to $x_1$, we obtain
\begin{align*}
 f_{Y| X}(Q_Y[\tau| X= x]|x )CQPE_{X_1}(\eta, x)+ \frac{\partial F_{Y| X}(Q_Y[\tau| X= x]|x )}{\partial z}\bigg|_{z=x_1} = 0.
\end{align*}
Since $f_{Y| X}(Q_Y[\tau| X= x]|x )\neq 0$, then the result follows by solving for $CQPE_{X_1}(\eta, x)$. Since the matching is a singleton, then for every $x$, and any $\tau$, we have  $Q_Y[\xi_\tau(x)|X = x]=Q_Y[\tau]$. Thus, we evaluate $CQPE_{X_1}(\eta, x)$ at $\tau = \xi_\tau(x)$ to yield
\begin{align*}
CQPE_{X_1}(\xi_\tau(x),x)
&=-\frac{1}{f_{Y|X}(Q_Y[\tau] |x)}   \frac{\partial F_{Y|X}(Q_Y[\tau] |z,x_2 )}{\partial z}\bigg|_{z=x_1}.
\end{align*}
Given the identification result for $UQPE_{X_1}(\tau)$ in equation \eqref{eq:uqpe}, we have that
\begin{align*}\label{eq:cqpe_uqpe}
UQPE_{X_1}(\tau) = \int CQPE_{X_1}(\xi_\tau(x),x) \frac{f_{Y|X}(Q_Y[\tau] |x)}{f_{Y}(Q_Y[\tau])}dF_{X}(x), 
\end{align*}
which is the result in \eqref{eq:cqpe_uqpe}.
Since $f_Y$ and $f_X$ are non-zero, then 
\begin{align*}
\frac{f_{Y|X}(Q_Y[\tau] |x)}{f_{Y}(Q_Y[\tau])}f_{X}(x) = \frac{f_{Y,X}(Q_Y[\tau] ,x)}{f_{Y}(Q_Y[\tau])f_{X}(x)}f_{X}(x) = f_{X|Y}(x|Q_Y[\tau]).
\end{align*}
Therefore, we obtain \eqref{eq:cqpe_uqpe_2}:
\begin{align*}
UQPE_{X_1}(\tau)
= E \bigl[ CQPE_{X_1}(\xi_\tau(X),X)|Y= Q_Y[\tau]\bigr].
\end{align*}

\subsection{Proof of Theorem \ref{thm:matching}}
Let
\begin{align*}
\Psi_\tau(\eta|x) = Q_Y[\eta|X = x] -  Q_Y[\tau].
\end{align*}
Here, $\tau$ and $x$ are fixed, and the criterion function is the map $[\epsilon,1-\epsilon]\ni \eta\mapsto \Psi_\tau(\eta|x) $ for $0<\epsilon<1/2$. Under Assumption \ref{assumption_matching}.\ref{assumption_matching_positive_density}, $y\mapsto F_{Y|X}(y|x)$ is strictly increasing, and hence $\Psi_\tau(\eta|x) $ has a unique zero given by $\xi_\tau(x) = F_{Y|X}(Q_\tau[Y]|x)$. This shows that $Q_\tau[Y]$ is $\xi_\tau(x)$-conditional quantile of $Y|X=x$. By Assumption \ref{assumption_matching}.\ref{assumption_matching_linearity}, this can be written as $x'\beta(\xi_\tau(x))=Q_\tau[Y]$.

Now we will show consistency: $\hat \xi_\tau(x) \overset{p}{\to} \xi_\tau(x) $. The matching function is defined to be the (approximate) zero of the random criterion function $\Psi_{\tau,n}(\eta|x)$:
\begin{align*}
\Psi_{\tau,n}(\hat \xi_\tau(x) |x) &= x' \hat{ \beta}(\hat \xi_\tau(x))- \hat Q_Y[\tau].
\end{align*}
The computational procedure outlined in equation \eqref{eq:mat_func} implicitly defines $\hat \xi_\tau(x) $ as $\eta_j$ for some $j$ in $\left\{1,2,\ldots,m\right\}$ such that
\begin{align*}
x'\hat{\beta}(\eta_j) \leq \hat Q_Y[\tau] < x'\hat{\beta}(\eta_{j+1}).
\end{align*}
We want to show that this, together with Assumption \ref{assumption_matching}.\ref{assumption_matching_approx_zero} that ensures $\Delta\eta = \eta_{j+1}-\eta_j=o(n^{-1/2})$, imply that
 $\Psi_{\tau,n}(\hat \xi_\tau(x) |x)=o_p(n^{-1/2})$. For a given $n$, let $\eta_j=\hat \xi_\tau(x) $, that is
\begin{align*}
x'\hat{\beta}(\eta_j)& \leq \hat Q_Y[\tau] < x'\hat{\beta}(\eta_{j+1})\\
0&\leq \hat Q_Y[\tau] -x'\hat{\beta}(\eta_j)< x'\hat{\beta}(\eta_{j+1})-x'\hat{\beta}(\eta_j)\\
0&\leq -\Psi_{\tau,n}(\hat \xi_\tau(x) |x) < x'\left (\hat{\beta}(\eta_{j+1})-\hat{\beta}(\eta_j)\right).
\end{align*}
We focus on the difference $\hat{\beta}(\eta_{j+1})-\hat{\beta}(\eta_j)$. We write
\begin{align*}
\hat{\beta}(\eta_{j+1})-\hat{\beta}(\eta_j) &= \hat{\beta}(\eta_{j+1}) -  {\beta}(\eta_{j+1}) - \left(\hat{\beta}(\eta_j) - {\beta}(\eta_j) \right)\\
&+{\beta}(\eta_{j+1})-{\beta}(\eta_j).
\end{align*}
For the last term, we can write ${\beta}(\eta_{j+1})-{\beta}(\eta_j)=\dot \beta(\tilde \eta)(\eta_{j+1}-\eta_{j})=o(n^{-1/2})$ because the derivative is bounded by Assumption \ref{assumption_matching}.\ref{assumption_matching_conditional_sup}. To alleviate notation, define:
\begin{align*}
J(\eta) = E\left[f_{Y|X}(X ' { \beta}(\eta)|X)XX'\right],
\end{align*}
which is differentiable with bounded derivative by \ref{assumption_matching}.\ref{assumption_matching_conditional_sup}. Using Assumption \ref{assumption_matching}.\ref{assumption_matching_conditional_sup} we have that
\begin{align*}
\hat{ \beta}(\eta)-{ \beta}(\eta) &=   J(\eta) ^{-1}\frac{1}{n}\sum_{i=1}^n\left ( \eta- \mathds 1\left\{ y_i\leq x_i'  { \beta}(\eta) \right\} \right)x_i+ o_p(n^{-1/2}),
\end{align*}
so that 
\begin{align*}
\hat{\beta}(\eta_{j+1}) -  {\beta}(\eta_{j+1}) - \left(\hat{\beta}(\eta_j) - {\beta}(\eta_j) \right)&=   J(\eta_{j+1})^{-1}\frac{1}{n}\sum_{i=1}^n\left ( \eta_{j+1}- \mathds 1\left\{ y_i\leq x_i'  { \beta}(\eta_{j+1}) \right\} \right)x_i\\
&-J(\eta_{j})^{-1}\frac{1}{n}\sum_{i=1}^n\left ( \eta_j- \mathds 1\left\{ y_i\leq x_i'  { \beta}(\eta_j) \right\} \right)x_i + o_p(n^{-1/2})\\
&=J(\eta_{j+1})^{-1}\frac{1}{n}\sum_{i=1}^n\bigg[ \left ( \eta_{j+1}- \mathds 1\left\{ y_i\leq x_i'  { \beta}(\eta_{j+1}) \right\} \right)x_i\\
&-\left ( \eta_j- \mathds 1\left\{ y_i\leq x_i'  { \beta}(\eta_j) \right\} \right)x_i \bigg]\\
&+\left( J(\eta_{j+1})^{-1} - J(\eta_{j})^{-1} \right)\frac{1}{n}\sum_{i=1}^n\left ( \eta_j- \mathds 1\left\{ y_i\leq x_i'  { \beta}(\eta_j) \right\} \right)x_i + o_p(n^{-1/2}).
\end{align*}
By Assumption  \ref{assumption_matching}.\ref{assumption_matching_positive_density}, $J(\eta)$ is bounded away from zero, so that $J(\eta)^{-1}$ is bounded.
We focus first on the difference in the sums. 
\begin{align*}
&\frac{1}{n}\sum_{i=1}^n\left ( \eta_{j+1}- \mathds 1\left\{ y_i\leq x_i'  { \beta}(\eta_{j+1}) \right\} - \eta_j+ \mathds 1\left\{ y_i\leq x_i'  { \beta}(\eta_j) \right\} \right)x_i .
\end{align*}
We note first that since $\eta_{j}<\eta_{j+1}$, by definition of quantiles, $x_i'  { \beta}(\eta_{j})<x_i'  { \beta}(\eta_{j+1})$. This means that if $\mathds 1\left\{ y_i\leq x_i'  { \beta}(\eta_{j+1}) \right\} =0$, then $\mathds 1\left\{ y_i\leq x_i'  { \beta}(\eta_j) \right\}=0$, and if $\mathds 1\left\{ y_i\leq x_i'  { \beta}(\eta_{j+1}) \right\} =1$, then either $\mathds 1\left\{ y_i\leq x_i'  { \beta}(\eta_j) \right\}=0$, or $\mathds 1\left\{ y_i\leq x_i'  { \beta}(\eta_j) \right\}=1$. Thus, the difference
\begin{align*}
 \mathds 1\left\{ y_i\leq x_i'  { \beta}(\eta_{j+1}) \right\} - \mathds 1\left\{ y_i\leq x_i'  { \beta}(\eta_j) \right\}
\end{align*}
is either 1 or 0. Using this, we have
\begin{align*}
&\left|\frac{1}{n}\sum_{i=1}^n\left ( \eta_{j+1}- \mathds 1\left\{ y_i\leq x_i'  { \beta}(\eta_{j+1}) \right\} - \eta_j+ \mathds 1\left\{ y_i\leq x_i'  { \beta}(\eta_j) \right\} \right)x_i \right|\\
&\leq \frac{1}{n}\sum_{i=1}^n\left|  \eta_{j+1}- \mathds 1\left\{ y_i\leq x_i'  { \beta}(\eta_{j+1}) \right\} - \eta_j+ \mathds 1\left\{ y_i\leq x_i'  { \beta}(\eta_j) \right\} \right| \left|x_i \right|\\
&\leq \left| \eta_{j+1} - \eta_j \right|  \frac{1}{n}\sum_{i=1}^n \left| x_i \right|\\
&=o(n^{-1/2}) O_p(1)=o_p(n^{-1/2}).
\end{align*}
Now, for the second term, we note that
\begin{align*}
\left |\left( J(\eta_{j+1})^{-1} - J(\eta_{j})^{-1} \right)\frac{1}{n}\sum_{i=1}^n\left ( \eta_j- \mathds 1\left\{ y_i\leq x_i'  { \beta}(\eta_j) \right\} \right)x_i\right| 
&\leq \left | \frac{J(\eta_{j})-J(\eta_{j+1})}{J(\eta_{j})J(\eta_{j+1})} \right| \frac{1}{n}\sum_{i=1}^n|x_i|\\
&=  \left | \frac{J'(\tilde \eta)(\eta_{j+1}-\eta_{j})}{J(\eta_{j})J(\eta_{j+1})} \right| \frac{1}{n}\sum_{i=1}^n|x_i|\\
&=o(n^{-1/2}) O_p(1)=o_p(n^{-1/2}).
\end{align*}
where $\tilde\eta$ is between $\eta_{j}$ and $\eta_{j+1}$. Therefore, we have that 
\begin{align*}
|\Psi_{\tau,n}(\hat \xi_\tau(x) |x)| & < |x'\left (\hat{\beta}(\eta_{j+1})-\hat{\beta}(\eta_j)\right) |\\
& < \sup ||x|| \cdot |\left (\hat{\beta}(\eta_{j+1})-\hat{\beta}(\eta_j)\right) |\\
& = o_p(n^{1/2}) o_p(n^{-1/2})=o_p(1).
\end{align*}

Since $\hat \xi_\tau(x) $ is a Z-estimator, we follow Theorem 5.9 in \cite{vanderVaart98}. We need to show $(i)$ that the criterion function converges uniformly in probability:
\begin{equation}\label{eq:uniform_criterion}
\sup_{\eta\in [\epsilon,1-\epsilon] }|\Psi_{\tau,n}(\eta|x)- \Psi_\tau(\eta|x) | \overset{p}{\to} 0,
\end{equation}
and $(ii)$ that the zero is well-separated: for any $\Delta>0$
\begin{align*}
\inf_{\eta:|\eta-\xi_\tau(x)| >\Delta} |\Psi_{\tau}(\eta|X)|>0.
\end{align*}
To show \eqref{eq:uniform_criterion}, we note that by Assumption \ref{assumption_matching}.\ref{assumption_matching_linearity} 
\begin{align*}
\sup_{\eta\in [\epsilon,1-\epsilon] }|\Psi_{\tau,n}(\eta|x)- \Psi_\tau(\eta|x) |  &=\sup_{\eta\in [\epsilon,1-\epsilon] }| x' \hat{ \beta}(\eta)- \hat Q_Y[\tau]  - x' { \beta}( \eta) +  Q_Y[\tau]|\\
&\leq ||x|| \sup_{\eta\in [\epsilon,1-\epsilon] } |\hat{ \beta}(\eta)-{ \beta}(\eta) | + |\hat Q_Y[\tau]- Q_Y[\tau]|\\
&= ||x|| O_p(n^{-1/2}) + O_p(n^{-1/2})\\
&=o_p(1).
\end{align*}
where $\|\cdot\|$ is the Euclidean norm, and the bounds follow from Assumptions \ref{assumption_matching}.\ref{assumption_matching_conditional_sup} and \ref{assumption_matching}.\ref{assumption_matching_unconditional}. 
To show that the zero is well-separated, we note that by Assumption \ref{assumption_matching}.\ref{assumption_matching_positive_density}, $y\mapsto F_{Y|X}(y|x)$ is strictly increasing, so that for any $\Delta>0$, if $|\eta-\xi_\tau(x)| >\Delta$, then by the mean-value theorem
\begin{align*}
|\eta - F_{Y|X}(Q_\tau[Y]|x)|=|F_{Y|X}(Q_\eta[Y|X=x]|x) - F_{Y|X}(Q_\tau[Y]|x)|= f_{Y|X}(\tilde Q|x) |Q_\eta[Y|X=x]-Q_\tau[Y]|,
\end{align*}
where $\tilde Q$ is between $Q_\eta[Y|X=x]$ and $Q_\tau[Y]$. Now, $f_{Y|X}(\tilde Q|x)>0$. Moreover, if $\eta>\xi_\tau(x)$, then $\eta>\xi_\tau(x)+\Delta$, so that $Q_\eta[Y|X=x]>Q_{\xi_\tau(x)+\Delta}[Y|X=x]>Q_{\xi_\tau(x)}[Y|X=x]=Q_\tau[Y] $, where we take $\Delta$ small enough such that $\eta<1$. The same analysis can be carried out for $\eta<\xi_\tau(x)$, in which case: $Q_\eta[Y|X=x]<Q_{\xi_\tau(x)-\Delta}[Y|X=x]<Q_{\xi_\tau(x)}[Y|X=x]=Q_\tau[Y]$. Therefore, 
\begin{align*}
\inf_{\eta:|\eta-\xi_\tau(x)| >\Delta} |\Psi_{\tau}(\eta|X)|&= \inf_{\eta:|\eta-\xi_\tau(x)| >\Delta} f_{Y|X}(\tilde Q|x) |Q_\eta[Y|X=x]-Q_\tau[Y]|\\
&\geq \inf_{\eta:|\eta-\xi_\tau(x)| >\Delta} f_{Y|X}(\tilde Q|x) \times \inf_{\eta:|\eta-\xi_\tau(x)| >\Delta} |Q_\eta[Y|X=x]-Q_\tau[Y]|\\
&\geq \inf_{y\in \mathbb R} f_{Y|X}(y|x) \times \min\left\{|Q_{\xi_\tau(x)+\Delta}[Y|X=x]-Q_\tau[Y]|, |Q_{\xi_\tau(x)-\Delta}[Y|X=x]-Q_\tau[Y]|\right\} \\
&>0,
\end{align*}
where we have used that $f_{Y|X}(y|x)$ is bounded away from zero. Finally, we can invoke Theorem 5.9 in \cite{vanderVaart98}, since we also showed that $\Psi_{\tau,n}(\hat \xi_\tau(x) |x) = o_p(n^{-1/2})$, therefore, $\hat \xi_\tau(x) \overset{p}{\to} \xi_\tau(x) $.

Having shown consistency, we now prove it is actually $\sqrt n$-consistent. To that end, we use Assumption \ref{assumption_matching}.\ref{assumption_matching_approx_zero}:
\begin{align}\label{eq:exp_zero}
o_p(n^{-1/2}) &= x' \hat{\beta}(\hat \xi_\tau(x))- \hat Q_Y[\tau] \\
&= x' \left ( \hat{\beta}(\hat \xi_\tau(x))- {\beta}(\hat \xi_\tau(x)) \right) + 
x' \left ( {\beta}(\hat \xi_\tau(x))- {\beta}( \xi_\tau(x)) \right) \notag\\
&+ \underbrace{x' {\beta}( \xi_\tau(x))  - Q_Y[\tau] }_{=\Psi_\tau(\xi_\tau(x)|x) =0} - \left ( \hat Q_Y[\tau] - Q_Y[\tau] \right).\notag
\end{align}
By Assumptions \ref{assumption_matching}.\ref{assumption_matching_linearity}, \ref{assumption_matching}.\ref{assumption_matching_positive_density}, we have that $F_{Y|X}(x'\beta(\eta)|x)=\eta$, so that $x'{ \dot\beta}( \eta) = f_{Y|X}(x'\beta(\eta)|x)^{-1}>0$, so that we can do a first order term-by-term Taylor expansion to obtain
\begin{align*}
{\beta}(\hat \xi_\tau(x))- {\beta}( \xi_\tau(x)) =  {\dot\beta}( \xi_\tau(x)) \left(\hat \xi_\tau(x)- \xi_\tau(x) \right) + o_p(|\hat \xi_\tau(x)- \xi_\tau(x)|).
\end{align*}
Here, $ {\dot\beta}( \xi_\tau(X))$ is the Jacobian vector: the derivative of the map $\tau\mapsto {\beta}(\tau)$ and $o_p(|\hat \xi_\tau(x)- \xi_\tau(x)|)$ is a vector of residuals of the expansion. Plugging this into the previous display, we obtain
\begin{align}\label{eq:main_expansion}
o_p(n^{-1/2}) &= x' \left ( \hat{\beta}(\hat \xi_\tau(x))- {\beta}(\hat \xi_\tau(x)) \right) + 
x'  {\dot\beta}( \xi_\tau(x)) \left(\hat \xi_\tau(x)- \xi_\tau(x) \right) \notag\\
&+  o_p(|\hat \xi_\tau(x)- \xi_\tau(x)|)- \left ( \hat Q_Y[\tau] - Q_Y[\tau] \right).
\end{align}
Here the term $o_p(|\hat \xi_\tau(x)- \xi_\tau(x)|)$ is scalar-valued and collects all the terms from $x'o_p(|\hat \xi_\tau(x)- \xi_\tau(x)|)$. 
Now, $\hat Q_Y[\tau] - Q_Y[\tau] = O_p(n^{-1/2})$ by Assumption \ref{assumption_matching}.\ref{assumption_matching_unconditional}. Also, by Assumption \ref{assumption_matching}.\ref{assumption_matching_conditional_sup} $ \hat{\beta}(\hat \xi_\tau(x))- {\beta}(\hat \xi_\tau(x))\leq  \sup_{\eta\in [\epsilon,1-\epsilon] } |\hat{\beta}(\eta)-{\beta}(\eta) |=O_p(n^{-1/2})$. Therefore, since $x'{\dot\beta}( \xi_\tau(x)) \neq 0$, we have that $|\hat \xi_\tau(x)- \xi_\tau(x)|=O_p(n^{-1/2})$. 

Finally to obtain the asymptotic distribution, we go back to \eqref{eq:main_expansion}, and using the stochastic equicontinuity guaranteed by Assumption \ref{assumption_matching}.\ref{assumption_matching_conditional_sup}, we replace $ \hat{\beta}(\hat \xi_\tau(x)) - {\beta}(\hat \xi_\tau(x))$ by $ \hat{\beta}( \xi_\tau(x))-{\beta}( \xi_\tau(x))+o_p(n^{-1/2})$. Therefore, we obtain
\begin{align*}
\hat \xi_\tau(x)- \xi_\tau(x)  &= -\frac{1}{x'  {\dot\beta}( \xi_\tau(x)) } x' \left ( \hat{\beta}( \xi_\tau(x))- {\beta}( \xi_\tau(x)) \right) +\frac{1}{x'  {\dot\beta}( \xi_\tau(x)) }  \left ( \hat Q_Y[\tau] - Q_Y[\tau] \right) + o_p(n^{-1/2})\\
&=-\frac{1}{x'  {\dot\beta}( \xi_\tau(x)) } \frac{1}{n}\sum_{i=1}^n  x'\Psi_{i}(\xi_\tau(x)) + \frac{1}{x'  {\dot\beta}( \xi_\tau(x)) } \frac{1}{n}\sum_{i=1}^n\psi_{i}(\tau)+o_p(n^{-1/2}).
\end{align*}
Now, using Assumptions \ref{assumption_matching}.\ref{assumption_matching_linearity} and  \ref{assumption_matching}.\ref{assumption_matching_positive_density}, which allow us to write $x'{ \dot\beta}( \eta) = f_{Y|X}(x'\beta(\eta)|x)^{-1}>0$, we have
\begin{align*}
\hat \xi_\tau(x)- \xi_\tau(x)  &= f_{Y|X}(x'\beta(\xi_\tau(x))|x) \left[-x' \left ( \hat{\beta}( \xi_\tau(x))- {\beta}( \xi_\tau(x)) \right) +  \left ( \hat Q_Y[\tau] - Q_Y[\tau] \right)\right] + o_p(n^{-1/2}).
\end{align*}
To obtain the main statement of the theorem we write
\begin{align*}
\hat \beta_1(\hat \xi_\tau(x)) - \beta_1( \xi_\tau(x)) &= \hat \beta_1(\hat \xi_\tau(x)) - \beta_1( \hat \xi_\tau(x)) + \beta_1(\hat \xi_\tau(x)) - \beta_1( \xi_\tau(x))\\
&=\hat \beta_1( \xi_\tau(x)) - \beta_1(  \xi_\tau(x)) + {\dot\beta_1}( \xi_\tau(x) ) (\hat \xi_\tau(x)- \xi_\tau(x) ) + o_p(n^{-1/2}).
\end{align*}

\subsection{Proof of Theorem \ref{thm:first_order_equiv}}
To alleviate notation, we write:
\begin{align*}
\hat m_1 (q,b,e) &:=\frac{1}{n} \sum_{i=1}^nK_h(y_i- q)\cdot  b( e(x_i))\\
\hat m_2 (q) &:=\frac{1}{n}\sum_{i=1}^nK_h(y_i- q).
\end{align*}
Thus, our estimator of $UQPE_{X_1}(\tau)$ can be written as
\begin{align*}
\widehat{UQPE_{X_1}}(\tau)=\frac{\hat m_1 ( \hat Q_Y[\tau],\hat \beta_1,\hat \xi_\tau) }{\hat m_2 ( \hat Q_Y[\tau])}.
\end{align*}
The unfeasible version is then
\begin{align*}
\widetilde{UQPE_{X_1}}(\tau)=\frac{\hat m_1 (  Q_Y[\tau], \beta_1, \xi_\tau) }{\hat m_2 (  Q_Y[\tau])}.
\end{align*}
Consider the difference
\begin{align}\label{eq:expansion_1}
\widehat{UQPE_{X_1}}(\tau)-\widetilde{UQPE_{X_1}}(\tau)
&= \frac{\hat m_1 ( \hat Q_Y[\tau],\hat \beta_1,\hat \xi_\tau) }{\hat m_2 ( \hat Q_Y[\tau])} -  \frac{ \hat m_1 (  Q_Y[\tau], \beta_1, \xi_\tau) }{ \hat m_2 (  Q_Y[\tau])}\notag\\
&=
 \frac{\hat m_2 (  Q_Y[\tau])\hat m_1 ( \hat Q_Y[\tau],\hat \beta_1,\hat \xi_\tau)-\hat m_2 ( \hat Q_Y[\tau]) \hat m_1 (  Q_Y[\tau], \beta_1, \xi_\tau)}{\hat m_2 ( \hat Q_Y[\tau])\hat m_2 (  Q_Y[\tau])}\notag\\
 &=\frac{\hat m_1 ( \hat Q_Y[\tau],\hat \beta_1,\hat \xi_\tau)-\hat m_1 (  Q_Y[\tau], \beta_1, \xi_\tau)  }{\hat m_2 ( \hat Q_Y[\tau])}\notag\\
 &-\frac{\hat m_1 (  Q_Y[\tau], \beta_1, \xi_\tau) }{\hat m_2 ( \hat Q_Y[\tau])\hat m_2 (  Q_Y[\tau])}\left (  \hat m_2 (  \hat Q_Y[\tau])- \hat m_2 (  Q_Y[\tau]) \right).
\end{align}

First we focus on the second term of \eqref{eq:expansion_1}. We note that
 \begin{align*}
\hat m_2 ( \hat Q_Y[\tau]) &:=\frac{1}{n}\sum_{i=1}^nK_h(y_i-  \hat Q_Y[\tau]) = \hat f_Y(\hat Q_Y[\tau])
\end{align*}
is an estimator of the density of $Y$ evaluated at $\hat Q_Y[\tau]$, the estimator of $Q_Y[\tau]$; while
\begin{align*}
\hat m_2 (  Q_Y[\tau]) &:=\frac{1}{n}\sum_{i=1}^nK_h(y_i-   Q_Y[\tau])=\hat f_Y( Q_Y[\tau])
\end{align*}
is an estimator of the density of $Y$ evaluated at $Q_Y[\tau]$. The next lemma, proved below, will be useful.\footnote{We thank an anonymous referee for pointing us in this direction.}
\begin{lemma}\label{lemma_density_two_step}
    Under Assumptions \ref{assumption_kernel}, \ref{assumption_bandwitdh}, and \ref{assumption_density}
 \begin{align*}
\hat f_Y(\hat Q_Y[\tau]) - \hat f_Y( Q_Y[\tau])&= f_Y'(Q_Y[\tau])  (\hat Q_Y[\tau]-Q_Y[\tau]) + O_p(h^{r} + \ln (n)^{1/2}n^{-1/2}h^{-3/2})O_p(n^{-1/2})\\
&=  f_Y'(Q_Y[\tau])  (\hat Q_Y[\tau]-Q_Y[\tau]) + o_p(n^{-1/2}h^{-1/2}).
\end{align*}
\end{lemma}
Under the above lemma, we can write
\begin{align}\label{eq:m2}
\hat m_2 (  \hat Q_Y[\tau]) - \hat m_2 (   Q_Y[\tau]) &= \hat f_Y(\hat Q_Y[\tau]) - \hat f_Y( Q_Y[\tau])\notag\\
&= f_Y'(Q_Y[\tau])(\hat Q_Y[\tau]- Q_Y[\tau]) + o_p(n^{-1/2}h^{-1/2}),
\end{align}
which implies that $\hat m_2 (  \hat Q_Y[\tau]) - \hat m_2 (   Q_Y[\tau]) = o_p(n^{-1/2}h^{-1/2})$.

Now we focus on the first term of \eqref{eq:expansion_1}. For the numerator, consider the following decomposition
\begin{align}\label{eq:num_decomp}
\hat m_1 ( \hat Q_Y[\tau],\hat \beta_1,\hat \xi_\tau)-\hat m_1 (  Q_Y[\tau], \beta_1, \xi_\tau) &=\frac{1}{n}\sum_{i=1}^nK_h(y_i-\hat Q_Y[\tau])\cdot \hat \beta_1(\hat \xi_\tau(x_i)) - \frac{1}{n}\sum_{i=1}^nK_h(y_i- Q_Y[\tau])\cdot  \beta_1( \xi_\tau(x_i)) \notag\\
&=\underbrace{\frac{1}{n}\sum_{i=1}^n\left[K_h(y_i-\hat Q_Y[\tau]) - K_h(y_i- Q_Y[\tau])\right] \cdot  \beta_1( \xi_\tau(x_i))}_{:=T_1} \notag\\
&+\underbrace{\frac{1}{n}\sum_{i=1}^nK_h(y_i- Q_Y[\tau])\cdot  \left[\hat\beta_1( \hat\xi_\tau(x_i))  -  \beta_1( \xi_\tau(x_i))\right]}_{:=T_2}\notag\\
&+\underbrace{\frac{1}{n}\sum_{i=1}^n\left[K_h(y_i-\hat Q_Y[\tau]) - K_h(y_i- Q_Y[\tau])\right] \cdot   \left[\hat\beta_1( \hat\xi_\tau(x_i))  -  \beta_1( \xi_\tau(x_i))\right]}_{:=T_3}\notag\\
&=T_1 + T_2 + T_3.
\end{align}

First we will show that $T_1=o_p(n^{-1}h^{-1/2})$. We do a first order Taylor expansion to obtain
\begin{align}\label{eq:expansion_2}
T_1&:=\frac{1}{n}\sum_{i=1}^n\left[K_h(y_i-\hat Q_Y[\tau]) - K_h(y_i- Q_Y[\tau])\right] \cdot  \beta_1( \xi_\tau(x_i))\notag
\\ &= \left( \hat Q_Y[\tau]-Q_Y[\tau]\right)\frac{1}{n}\sum_{i=1}^n \frac{\partial K_h(y_i-q)}{\partial q}\bigg\vert_{q=Q_Y[\tau]} \cdot  \beta_1( \xi_\tau(x_i))\notag\\
& \left( \hat Q_Y[\tau]-Q_Y[\tau]\right)
 \frac{1}{n}\sum_{i=1}^n \left(\frac{\partial K_h(y_i-q)}{\partial q}\bigg\vert_{q=\tilde q} -  \frac{\partial K_h(y_i-q)}{\partial q}\bigg\vert_{q=Q_Y[\tau]} \right) \cdot  \beta_1( \xi_\tau(x_i)).
\end{align}

Consider the first term. Let $f^{(j)}_{Y,X}(y,x)$ denote the $j$-th partial derivative of $f_{Y,X}(y,x)$ with respect to $y$.
The expected value is 
\begin{align*}
    E\left[\frac{1}{n}\sum_{i=1}^n \frac{\partial K_h(Y_i-q)}{\partial q}\bigg\vert_{q=Q_Y[\tau]} \cdot  \beta_1( \xi_\tau(X_i)) \right] &=     E\left[ \frac{\partial K_h(Y-q)}{\partial q}\bigg\vert_{q=Q_Y[\tau]} \cdot  \beta_1( \xi_\tau(X)) \right] \\
    &=-\frac{1}{h^2}E\left[  K'\left( \frac{Y-Q_Y[\tau]}{h} \right)  \cdot  \beta_1( \xi_\tau(X)) \right]\\
    &=-\frac{1}{h^2}\int\int  K'\left( \frac{y-Q_Y[\tau]}{h} \right)    \beta_1( \xi_\tau(x))f_{Y,X}(y,x)dydx\\
    &=-\frac{1}{h}\int\int  K'(u)   \beta_1( \xi_\tau(x))f_{Y,X}(Q_Y[\tau]+hu,x)dudx\\
    &=-\frac{1}{h} \int\int K'(u) \beta_1( \xi_\tau(x)) \bigg[ f_{Y,X}(Q_Y[\tau],x) \\
    &+ \sum_{j=1}^r\frac{h^ju^jf_{Y,X}^{(j)}(Q_Y[\tau],x)}{j!} + \frac{h^{r+1}u^{r+1}f_{Y,X}^{(r+1)}(\tilde Q_Y[\tau],x)}{(r+1)!}  \bigg]dudx,
\end{align*}
where we used Assumption \ref{assumption_density} to expand the joint density. The properties of the kernel of Assumption \ref{assumption_kernel} yield
\begin{align*}
    E\left[\frac{1}{n}\sum_{i=1}^n \frac{\partial K_h(Y_i-q)}{\partial q}\bigg\vert_{q=Q_Y[\tau]} \cdot  \beta_1( \xi_\tau(X_i)) \right] 
    &=-\frac{1}{h} \int\int K'(u) \beta_1( \xi_\tau(x)) \bigg[ f_{Y,X}(Q_Y[\tau],x) \\
    &+ \sum_{j=1}^r\frac{h^ju^jf_{Y,X}^{(j)}(Q_Y[\tau],x)}{j!} + \frac{h^{r+1}u^{r+1}f_{Y,X}^{(r+1)}(\tilde Q_Y[\tau],x)}{(r+1)!}  \bigg]dudx\\
    &= \int  \beta_1( \xi_\tau(x))  f_{Y,X}^{(1)}(Q_Y[\tau],x) dx \\
    &-h^r\int \int K'(u) \beta_1( \xi_\tau(x)) u^{r+1}f_{Y,X}^{(r+1)}(\tilde Q_Y[\tau],x) \beta_1( \xi_\tau(x))  dx du.
\end{align*}
Therefore, the bias is of order $O(h^r)$:
\begin{align*}
    E\left[\frac{1}{n}\sum_{i=1}^n \frac{\partial K_h(Y_i-q)}{\partial q}\bigg\vert_{q=Q_Y[\tau]} \cdot  \beta_1( \xi_\tau(X_i)) \right] 
    &= \int  \beta_1( \xi_\tau(x))  f_{Y,X}^{(1)}(Q_Y[\tau],x) dx + O(h^r).
\end{align*}
For the variance, we have
\begin{align*}
    Var\left[\frac{1}{n}\sum_{i=1}^n \frac{\partial K_h(Y_i-q)}{\partial q}\bigg\vert_{q=Q_Y[\tau]} \cdot  \beta_1( \xi_\tau(X_i)) \right] &=    \frac{1}{nh^4} Var\left [ K'\left( \frac{Y-Q_Y[\tau]}{h} \right)  \beta_1( \xi_\tau(X)) \right]\\
    &=\frac{1}{nh^4} E\left [ K'\left( \frac{Y-Q_Y[\tau]}{h} \right)^2  \beta_1( \xi_\tau(X))^2 \right] \\
    &- \frac{1}{nh^4} E\left [ K'\left( \frac{Y-Q_Y[\tau]}{h} \right)  \beta_1( \xi_\tau(X)) \right]^2.
\end{align*}
We take care of each term at a time.
\begin{align*}
\frac{1}{nh^4} E\left [ K'\left( \frac{Y-Q_Y[\tau]}{h} \right)^2  \beta_1( \xi_\tau(X))^2 \right]
 &=\frac{1}{nh^4}\int\int K'\left(\frac{y-q}{h} \right)^2\beta_1( \xi_\tau(x))^2f_{Y,X}(Q_Y[\tau],x)dydx\\
 &=\frac{1}{nh^3}\int\int K'(u)^2\beta_1( \xi_\tau(x))^2f_{Y,X}(Q_Y[\tau]+hu,x)dudx\\
 &= \frac{1}{nh^3}\int\int K'(u)^2\beta_1( \xi_\tau(x))^2\left[f_{Y,X}(Q_Y[\tau],x) + hu f_{Y,X}^{(1)}(\tilde Q_Y[\tau],x)\right]dudx\\
 &=\frac{1}{nh^3}\int \int K'(u)^2 \beta_1( \xi_\tau(x))^2 f_{Y,X}(Q_Y[\tau],x) dudx + o(n^{-1}h^{-2}).
\end{align*}
For the other term, we have
\begin{align*}
\frac{1}{nh^4} E\left [ K'\left( \frac{Y-Q_Y[\tau]}{h} \right)  \beta_1( \xi_\tau(X)) \right]^2 &= \frac{1}{nh^4} \left [ \int \int K'\left( \frac{y-Q_Y[\tau]}{h} \right)  \beta_1( \xi_\tau(x))f_{Y,X}(y,x)dydx \right]^2 \\
&=\frac{1}{nh^2} \left [ \int \int K'(u) \beta_1( \xi_\tau(x))f_{Y,X}(Q_Y[\tau]+hu,x)dudx \right]^2\\
&= O(n^{-1}h^{-2}).
\end{align*}
Combining the bias and variance results, we obtain
\begin{align*}
    E\left[\frac{1}{n}\sum_{i=1}^n \frac{\partial K_h(Y_i-q)}{\partial q}\bigg\vert_{q=Q_Y[\tau]} \cdot  \beta_1( \xi_\tau(X_i)) \right] 
    - \int  \beta_1( \xi_\tau(x))  f_{Y,X}^{(1)}(Q_Y[\tau],x) dx =  O_p(n^{-1/2}h^{-3/2}+h^r).
\end{align*}

The remaining term in \eqref{eq:expansion_2} is bounded by
\begin{align*}
&\left( \hat Q_Y[\tau]-Q_Y[\tau]\right)
 \frac{1}{n}\sum_{i=1}^n \left(\frac{\partial K_h(y_i-q)}{\partial q}\bigg\vert_{q=\tilde q} -  \frac{\partial K_h(y_i-q)}{\partial q}\bigg\vert_{q=Q_Y[\tau]} \right) \cdot  \beta_1( \xi_\tau(x_i))\\
 &\leq |\hat Q_Y[\tau]-Q_Y[\tau]|\cdot \sup_{\eta\in [\varepsilon, 1-\varepsilon]}\beta_1(\eta) \cdot |\hat f_Y'(\tilde q) - \hat f_Y'(Q_Y[\tau])|\\
 &\leq |\hat Q_Y[\tau]-Q_Y[\tau]|\cdot \sup_{\eta\in [\varepsilon, 1-\varepsilon]}\beta_1(\eta) \cdot \left (|\hat f_Y'(\tilde q) - f_Y'(Q_Y[\tau])| + |f_Y'(Q_Y[\tau]) - \hat f_Y'(Q_Y[\tau])| \right)
\end{align*}
By Lemma \ref{lemma_density_two_step}, this term is of order $O_p(n^{-1/2})(o_p(n^{-1/2}h^{-1/2})+o_p(n^{-1/2}h^{-1/2}))$. Therefore, $T_1 = o_p(n^{-1}h^{-1/2})$. 

Now we show that $T_2$ in \eqref{eq:num_decomp} satisfies
\begin{align*}
T_2:=\frac{1}{n}\sum_{i=1}^nK_h(y_i- Q_Y[\tau])\cdot  \left[\hat\beta_1( \hat\xi_\tau(x_i))  -  \beta_1( \xi_\tau(x_i))\right] = O_p(n^{-1/2}).
\end{align*}
We use the following decomposition, similar to the one in Theorem \ref{thm:matching}:
\begin{align*}
\hat \beta_1(\hat \xi_\tau(x)) - \beta_1( \xi_\tau(x)) &= \hat \beta_1(\hat \xi_\tau(x)) - \beta_1( \hat \xi_\tau(x)) + \beta_1(\hat \xi_\tau(x)) - \beta_1( \xi_\tau(x)).
\end{align*}
We have
\begin{align*}
T_2&:=\frac{1}{n}\sum_{i=1}^nK_h(y_i- Q_Y[\tau])\cdot  \left[\hat\beta_1( \hat\xi_\tau(x_i))  -  \beta_1( \xi_\tau(x_i))\right]\\ 
&= \frac{1}{n}\sum_{i=1}^nK_h(y_i- Q_Y[\tau])\cdot  \left[\hat \beta_1(\hat \xi_\tau(x_i)) - \beta_1( \hat \xi_\tau(x_i))\right]
+\frac{1}{n}\sum_{i=1}^nK_h(y_i- Q_Y[\tau])\cdot \left[ \beta_1(\hat \xi_\tau(x_i)) - \beta_1( \xi_\tau(x_i)) \right]\\
&\leq \sup_{\eta\in [\epsilon,1-\epsilon] } |\hat{\beta}(\eta)-{\beta}(\eta) | \frac{1}{n}\sum_{i=1}^nK_h(y_i- Q_Y[\tau])
+\frac{1}{n}\sum_{i=1}^nK_h(y_i- Q_Y[\tau])\cdot  \beta_1(\hat \xi_\tau(x_i)) - \beta_1( \xi_\tau(x_i)) 
\end{align*}
Here we use $\sup_{\eta\in [\epsilon,1-\epsilon] } |\hat{\beta}(\eta)-{\beta}(\eta) |=O_p(n^{-1/2})$
and $\frac{1}{n}\sum_{i=1}^nK_h(y_i- Q_Y[\tau])= O_p(1),
$ to conclude that the first term is $O_p(n^{-1/2})$. For the second term, we use the mean value theorem and Theorem \ref{thm:matching} to get
\begin{align}\label{eq:t2_hard}
&\frac{1}{n}\sum_{i=1}^nK_h(y_i- Q_Y[\tau])\cdot \left[ \beta_1(\hat \xi_\tau(x_i)) - \beta_1( \xi_\tau(x_i)) \right]\notag\\
&=\frac{1}{n}\sum_{i=1}^nK_h(y_i- Q_Y[\tau])\cdot \dot \beta_1(\eta_i)\cdot ( \hat \xi_\tau(x_i) - \xi_\tau(x_i) )\\
&=-\frac{1}{n}\sum_{i=1}^nK_h(y_i- Q_Y[\tau])\cdot \dot \beta_1(\eta_i)\cdot f_{Y|X}(x_i'\beta(\xi_\tau(x_i))|x_i)\cdot (x_i'  \hat{\beta}( \xi_\tau(x_i))- x_i'{\beta}( \xi_\tau(x_i)))\notag\\
&+ \left ( \hat Q_Y[\tau] - Q_Y[\tau] \right)\frac{1}{n}\sum_{i=1}^nK_h(y_i- Q_Y[\tau])\cdot \dot \beta_1(\eta_i)\cdot f_{Y|X}(x_i'\beta(\xi_\tau(x_i))|x_i)\notag\\
&+\frac{1}{n}\sum_{i=1}^nK_h(y_i- Q_Y[\tau])\cdot \dot \beta_1(\eta_i)\cdot R_i,\notag
\end{align}
where $R_i$ is the remainder of the expansion of $\hat \xi_\tau(x_i) - \xi_\tau(x_i)$ in Theorem \ref{thm:matching}, which, by Assumption \ref{assumption_approx}, is $o_p(n^{-1/2})$ over the support of $X$. Now, by Assumption \ref{assumption_approx}, $\sup_{\eta\in[\epsilon,1-\epsilon]}|\dot \beta_1(\eta)|$, $(y,x)\mapsto f_{Y|X}(y|x)$ are bounded. Moreover, $\sup_{\eta\in [\epsilon,1-\epsilon] } |\hat{\beta}(\eta)-{\beta}(\eta) |=O_p(n^{-1/2})$. All these facts together imply that in the above display, the order of the terms are $O_p(n^{-1/2})$, $O_p(n^{-1/2})$, and $o_p(n^{-1/2})$ respectively. Therefore, we can conclude that $T_2=O_p(n^{-1/2})$.

For $T_3$ in \eqref{eq:num_decomp}, we follow the same strategy as with $T_2$.  We write
\begin{align*}
T_3&:=\frac{1}{n}\sum_{i=1}^n\left[K_h(y_i-\hat Q_Y[\tau]) - K_h(y_i- Q_Y[\tau])\right] \cdot   \left[\hat\beta_1( \hat\xi_\tau(x_i))  -  \beta_1( \xi_\tau(x_i))\right]\\&\leq \sup_{\eta\in [\epsilon,1-\epsilon] } |\hat{\beta}(\eta)-{\beta}(\eta) |\frac{1}{n}\sum_{i=1}^n\left[K_h(y_i-\hat Q_Y[\tau]) - K_h(y_i- Q_Y[\tau])\right]\\
&+\frac{1}{n}\sum_{i=1}^n\left[K_h(y_i-\hat Q_Y[\tau]) - K_h(y_i- Q_Y[\tau])\right] \cdot   \left[\beta_1( \hat\xi_\tau(x_i))  -  \beta_1( \xi_\tau(x_i))\right].
\end{align*}
By the result of Lemma \ref{lemma_density_two_step} we have that 
\begin{align*}
\sup_{\eta\in [\epsilon,1-\epsilon] } |\hat{\beta}(\eta)-{\beta}(\eta) |\frac{1}{n}\sum_{i=1}^n\left[K_h(y_i-\hat Q_Y[\tau]) - K_h(y_i- Q_Y[\tau])\right] &= O_p(n^{-1/2})(O_p(n^{-1/2})+o_p(n^{-1/2}h^{-1/2}))\\
&=O_p(n^{-1}) + o_p(n^{-1}h^{-1/2}).
\end{align*}
For other term, an argument analogous to the one in equation \eqref{eq:t2_hard} implies it is of order $O_p(n^{-1/2})(O_p(n^{-1/2})+o_p(n^{-1/2}h^{-1/2}))$. Therefore, $T_3$ is $O_p(n^{-1}) + o_p(n^{-1}h^{-1/2})$.
This means that
\begin{align}\label{eq:m1}
\hat m_1 ( \hat Q_Y[\tau],\hat \beta_1,\hat \xi_\tau)-\hat m_1 (  Q_Y[\tau], \beta_1, \xi_\tau)&= o_p(n^{-1}h^{-1/2}) + O_p(n^{-1/2}) + O_p(n^{-1}) + o_p(n^{-1}h^{-1/2})\notag\\
&=o_p(n^{-1/2}h^{-1/2}).
\end{align}
Therefore, \eqref{eq:m2} and \eqref{eq:m1} imply that \eqref{eq:expansion_1} is actually
\begin{align*}
\widehat{UQPE_{X_1}}(\tau)-\widetilde{UQPE_{X_1}}(\tau) = o_p(n^{-1/2}h^{-1/2}). 
\end{align*}

\subsection{Proof of Lemma \ref{lemma_density_two_step}}
We have for some random $\tilde Q_Y[\tau]$,
\begin{align}\label{eq:der_decomp}
\hat f_Y(\hat Q_Y[\tau]) - \hat f_Y( Q_Y[\tau]) &= \hat f_Y'(\tilde Q_Y[\tau])  (\hat Q_Y[\tau]-Q_Y[\tau])\notag\\
&=  f_Y'(Q_Y[\tau])  (\hat Q_Y[\tau]-Q_Y[\tau])
+(\hat f_Y'(\tilde Q_Y[\tau])- f_Y'(Q_Y[\tau])) (\hat Q_Y[\tau]-Q_Y[\tau]).
\end{align}
 We focus on the difference $\hat f_Y'(\tilde Q_Y[\tau])- f_Y'(Q_Y[\tau]) $.
\begin{align*}
\hat f_Y'(\tilde Q_Y[\tau])- f_Y'(Q_Y[\tau]) &= \hat f_Y'(\tilde Q_Y[\tau])- f_Y'(\tilde Q_Y[\tau]) + f_Y'(\tilde Q_Y[\tau]) - f_Y'(Q_Y[\tau])\\
&\leq \sup_{q\in\mathcal S}|\hat f_Y'(q)- f_Y'(q)| + o_p(1)
\end{align*}
The $o_p(1)$ terms comes from the continuity of the first derivative which follows from Assumption \ref{assumption_density}. Now we derive the uniform rate of the estimator of the first derivative. We follow the exposition in Section 1.12 of Li and Racine, and modify it to the case of the derivative. We write
\begin{align}\label{eq:uniform_1}
\hat f_Y'(q)- f_Y'(q) = \hat f_Y'(q)- E[\hat f_Y'(q)] + E[\hat f_Y'(q)] - f_Y'(q),
\end{align}
and take care of each term at a time. We start with the second term, the bias term.
The first derivative of the kernel estimator is 
\begin{align*}
 \hat f'_Y(q) = -\frac{1}{nh^2}\sum_{i=1}^nK'\left(\frac{y_i-q}{h} \right).
\end{align*}
We will use the fact that, by Assumption \ref{assumption_kernel}, $\int K'(u)du=0$,  $\int K'(u)udu=-1$, $\int K'(u)u^jdu=0$ for $j=2,...,r$, and $\int K'(u)u^{r+1}du<\infty$. In the following, $f_Y^{(j)}$ denotes the $j$-th derivative. 
\begin{align*}
 E[\hat f'(q)] &= -\frac{1}{h^2} E\left[K'\left(\frac{Y-q}{h} \right)\right]\\
 &=-\frac{1}{h^2} \int_{\mathbb R} K'\left(\frac{y-q}{h}  \right)f_{Y}(y)dy\\
 &=-\frac{1}{h} \int_{\mathbb R} K'(u)f_{Y}(q+hu)du\\
 &=-\frac{1}{h} \int_{\mathbb R} K'(u)\left[ f_{Y}(q) + \sum_{j=1}^r\frac{h^ju^jf_Y^{(j)}(q)}{j!} + \frac{h^{r+1}u^{r+1}f_Y^{(r+1)}(\tilde q)}{(r+1)!}  \right]du\\
 &=f_Y'(q) - \frac{h^r}{(r+1)!}\int_{\mathbb R}K'(u)u^{r+1}f_Y^{(r+1)}( \tilde q)du. 
\end{align*}
Since, by Assumption \ref{assumption_density}, the derivatives are bounded, then in the compact set $\mathcal S$, the bias is
\begin{align*}
 \sup_{q\in\mathcal S}|E[\hat f'(q)] - f_{Y}'(q)|=
 O(h^{r}).
\end{align*}
Now we address the first term in \eqref{eq:uniform_1}. As usual, since $\mathcal S$ is a compact set, we can pick a finite number $L(n)$ of intervals $I_k=I_{k,n}$ centered at $q_{k,n}$ for $k=1,\ldots, L(n)$ and length $\ell_n$ that cover $\mathcal S$.
We write
\begin{align*}
\sup_{q\in\mathcal S}|\hat f_Y'(q)- E[\hat f_Y'(q)]| &= \max_{k\in 1,\ldots, L(n)}\sup_{\mathcal S\cap I_{k,n}} |\hat f_Y'(q)- E[\hat f_Y'(q)]|\\
&\leq \max_{k\in 1,\ldots, L(n)}\sup_{\mathcal S\cap I_{k,n}} |\hat f_Y'(q)- \hat f_Y'(q_{k,n})|\\
&+\max_{k\in 1,\ldots, L(n)} |\hat f_Y'(q_{k,n})- E[\hat f_Y'(q_{k,n})]|\\
&+\max_{k\in 1,\ldots, L(n)}\sup_{\mathcal S\cap I_{k,n}} |E[\hat f_Y'(q_{k,n})- E[\hat f_Y'(q)]|\\
&=Q_1 + Q_2 + Q_3
\end{align*}
We start with $Q_2$. Write
\begin{align*}
Z_{ni}(q) &= -\frac{1}{nh^2} \left[K'\left(\frac{y_i-q}{h}\right)    -E\left[K'\left(\frac{y_i-q}{h}\right)\right]  \right]\\
W_{n}(q) &= \sum_{i=1}^nZ_{ni}(q) = \hat f'_Y(q)-E[\hat f_Y'(q)].
\end{align*}
Then, for some $M$, 
\begin{align}\label{eq:markov_0}
\Pr[Q_2> M]&=\Pr\left[\max_{k\in 1,\ldots, L(n)} |\hat f_Y'(q_{k,n})- E[\hat f_Y'(q_{k,n})]|>M\right]\notag\\
&=\Pr\left[\max_{k\in 1,\ldots, L(n)} |W_{n}(q_{k,n})|>M\right]\notag\\
&=\Pr\left[\cup_{k\in 1,\ldots, L(n)}|W_{n}(q_{k,n})|>M\right]\notag\\
&\leq \sum_{k=1}^{L(n)}\Pr\left[|W_{n}(q_{k,n})|>M\right]\notag\\
&\leq L(n) \sup_{q\in\mathcal S}\Pr\left[|W_{n}(q)|>M\right].
\end{align}
Using Markov's inequality\footnote{$\Pr[X>M]\leq E[\exp(aX)]/\exp(aM)$ for $a>0$.} we aim to bound uniformly $\Pr\left[|W_{n}(q)|>M\right]$. We write
\begin{align}\label{eq:markov_1}
\Pr\left[|W_{n}(q)|>M\right] &= \Pr\left[\left|\sum_{i=1}^n Z_{ni}(q)\right|>M\right]\notag \\
&=\Pr\left[\sum_{i=1}^n Z_{ni}(q)>M\right] + \Pr\left[-\sum_{i=1}^n Z_{ni}(q)>M\right]\notag\\
&\leq \frac{E[\exp(a\sum_{i=1}^n Z_{ni}(q))] + E[\exp(-a\sum_{i=1}^n Z_{ni}(q))]}{\exp(Ma)}
\end{align}
for some $a>0$. By Assumption \ref{assumption_kernel}, $K'$ is a bounded function then, 
\begin{align*}
|Z_{ni}(q)| \leq \frac{2\sup_{u}|K'(u)|}{nh^2} = \frac{2A_1}{nh^2}
\end{align*}
for some $A_1$ that does not depend on $q$. Let $a_n>0$, be such that
\begin{align*}
a_n|Z_{ni}(q)| \leq 1/2
\end{align*}
for all $i=1,\ldots,n$, for $n$ sufficiently large. Since $\exp(x)\leq 1 + x + x^2$ for $|x|\leq 1/2$, then
\begin{align*}
\exp(\pm a_nZ_{ni}(q)) \leq 1  \pm a_nZ_{ni}(q) +  a_n^2Z_{ni}(q)^2
\end{align*}
so that
\begin{align*}
E[\exp(\pm a_nZ_{ni}(q))] &\leq 1  \pm a_nE[Z_{ni}(q)] +  a_n^2E[Z_{ni}(q)^2]\\
&=1 + a_n^2E[Z_{ni}(q)^2]\\
&\leq \exp(E[a_n^2Z_{ni}(q)^2]),
\end{align*}
because $E[Z_{ni}(q)]=0$, and $1+x\leq \exp(x)$ for $x>0$. Thus, going back to \eqref{eq:markov_1}, we have for $a=a_n$, and $n$ sufficiently large, that
\begin{align}\label{eq:markov_2}
\Pr\left[|W_{n}(q)|>M\right]
&\leq \frac{E[\exp(a_n\sum_{i=1}^n Z_{ni}(q))] + E[\exp(-a_n\sum_{i=1}^n Z_{ni}(q))]}{\exp(Ma_n)}\notag\\
&\leq \frac{2\exp(a_n^2 \sum_{i=1}^n E[Z_{ni}(q)^2])}{\exp(Ma_n)}.
\end{align}
Now, the second moment of $Z_{ni}(q)$ is bounded by
\begin{align*}
E\left[Z_{ni}(q)^2\right] &\leq \frac{1}{n^2h^4} E\left[K'\left(\frac{y_i-q}{h}\right)^2\right]\\
&= \frac{1}{n^2h^4} \int K'\left(\frac{u-q}{h}\right)^2f_Y(u)du\\
&=\frac{1}{n^2h^3} \int K'\left(v\right)^2f_Y(q+hv)dv\\
&=\frac{1}{n^2h^3} \int K'\left(v\right)^2\left (f_Y(q) + hvf'(\tilde q)\right) dv\\
&=  \frac{f_Y(q)}{n^2 h^3}\int K'(v)^2dv + O(n^{-2} h^{-2})\\
&\leq \frac{A_2}{n^2 h^3}.
\end{align*}
by doing the change of variable $v = (u-q)/h$, where $A_2$ does not depend on $q$ because by Assumption \ref{assumption_density}, $f_Y$ and $f_Y'$ are bounded. Going back to \eqref{eq:markov_2}, we obtain
\begin{align}\label{eq:markov_3}
\Pr\left[|W_{n}(q)|>M\right]
&\leq \frac{2\exp(a_n^2 \sum_{i=1}^n E[Z_{ni}(q)^2])}{\exp(Ma_n)}\notag\\
&\leq \frac{2\exp(A_2a_n^2 n^{-1}h^{-3})}{\exp(Ma_n)}
\end{align}
and this holds uniformly, so that
\begin{align*}
\sup_{q\in\mathcal S}\Pr\left[|W_{n}(q)|>M\right]&\leq \frac{2\exp(A_2a_n^2 n^{-2}h^{-3})}{\exp(Ma_n)}\\
&=2\exp \left (-Ma_n + \frac{A_2a_n^2}{n h^3}    \right).
\end{align*}
Therefore, going back to \eqref{eq:markov_0}, the bound is
\begin{align*}
\Pr[Q_2>M]\leq 2L(n)\exp \left (-Ma_n + \frac{A_2a_n^2}{n h^3}    \right).
\end{align*}
To utilize the Borel-Cantelli lemma,\footnote{If $\sum_{n=1}^\infty \Pr[|X_n|>a_n]<\infty$, then $X_n = O(a_n)$ almost surely.} we need to choose a sequence $M=M_n\to 0$, such that $M_na_n\to \infty$, and such that the bound is summable. We can take
\begin{align*}
a_n &= (nh^{3}\ln(n))^{1/2}\\
M_n &= A_3 \frac{\ln(n)}{a_n} = A_3\left(\frac{\ln(n)}{nh^3}\right)^{1/2},
\end{align*}
so that
\begin{align*}
-M_na_n + \frac{A_2a_n^2}{n h^3} = \underbrace{(A_2-A_3)}_{:=A_4}\ln(n),
\end{align*}
and the bound is
\begin{align*}
\Pr[Q_2>M_n]\leq 2L(n) n^{A_4}
\end{align*}
Therefore, by choosing the order of $L(n)$ and $A_3$ sufficiently large so that $A_4$ is sufficiently small (negative), then the right-hand side of the above display will be summable. This means that, by the Borel-Cantelli lemma, $Q_2=O((\ln(n))^{1/2}({nh^3})^{-1/2})$ almost surely.

Now, for the term $Q_1$, we have
\begin{align*}
Q_1:=\max_{k\in 1,\ldots, L(n)}\sup_{\mathcal S\cap I_{k,n}} |\hat f_Y'(q)- \hat f_Y'(q_{k,n})|
\end{align*}
Here we use the Lipschitz continuity of $ K'$, which is guaranteed by Assumption \ref{assumption_kernel} by the second derivative being bounded. For some $A_6>0$, 
\begin{align*}
 |\hat f'_Y(q)-\hat f'_Y(q_{k,n})| &\leq  \frac{1}{nh^2}\sum_{i=1}^n\left |K'\left(\frac{y_i-q}{h} \right) - K'\left(\frac{y_i-q_{k,n}}{h} \right)\right|\\
 &\leq \frac{A_6}{nh^2}\sum_{i=1}^n \left |\left(\frac{y_i-q}{h} \right) - \left(\frac{y_i-q_{k,n}}{h} \right)\right|\\
 &=\frac{A_6}{h^3}|q-q_{k,n}|\\
 &\leq \frac{A_6}{h^3}\ell_{n},
\end{align*}
where $\ell_n$ is the common length of the $I_{k,n}$ intervals.
Since the bound is independent of $q$, then
\begin{align*}
Q_1\leq \frac{A_6}{h^3}\ell_{n}.
\end{align*}
Thus, we choose $\ell_{n}h^{-3}$ to be of the same order as $Q_2$:
\begin{align*}
\frac{A_6}{h^3}\ell_{n} = O((\ln(n))^{1/2}({nh^3})^{-1/2}),
\end{align*}
which implies we need $\ell_{n}=O(\ln(n)^{1/2}h^{3/2}{n}^{-1/2})$. Therefore, 
\begin{align*}
|Q_1|=O((\ln(n))^{1/2}({nh^3})^{-1/2}).
\end{align*}
Finally, for the term $Q_3$, we have
\begin{align*}
Q_3 := \max_{k\in 1,\ldots, L(n)}\sup_{\mathcal S\cap I_{k,n}} |E[\hat f_Y'(q_{k,n})- E[\hat f_Y'(q)]|
\end{align*}
Here we have
\begin{align*}
|E[\hat f_Y'(q_{k,n})- E[\hat f_Y'(q)]| &= \left | \int \frac{1}{h^2}K'\left(\frac{y-q_{k,n}}{h} \right)f_Y(y)dy -  \int \frac{1}{h^2}K'\left(\frac{y-q}{h} \right) f_Y(y)dy \right|\\
&\leq \frac{A_6}{h^3}\ell_n,
\end{align*}
so that, as with $Q_1$, we have
\begin{align*}
|Q_3|=O((\ln(n))^{1/2}({nh^3})^{-1/2}).
\end{align*}    

Therefore, the order of $\eqref{eq:uniform_1}$ is
\begin{align*}
\hat f_Y'(q)- f_Y'(q) &= \hat f_Y'(q)- E[\hat f_Y'(q)] + E[\hat f_Y'(q)] - f_Y'(q)\\
&=O_p(h^{r} + \ln (n)^{1/2}n^{-1/2}h^{-3/2}).
\end{align*}
Going back to \eqref{eq:der_decomp}, we have
\begin{align*}
\hat f_Y(\hat Q_Y[\tau]) - \hat f_Y( Q_Y[\tau])
&=  f_Y'(Q_Y[\tau])  (\hat Q_Y[\tau]-Q_Y[\tau])
+(\hat f_Y'(\tilde Q_Y[\tau])- f_Y'(Q_Y[\tau])) (\hat Q_Y[\tau]-Q_Y[\tau])\\
&=  f_Y'(Q_Y[\tau])  (\hat Q_Y[\tau]-Q_Y[\tau]) + O_p(h^{r} + \ln (n)^{1/2}n^{-1/2}h^{-3/2})O_p(n^{-1/2}).
\end{align*}
Now we show that under Assumption \ref{assumption_bandwitdh}, the order of the remainder is $o_p(n^{-1/2}h^{-1/2})$. We need the bandwidth to satisfy
\begin{align*}
n^{1/2}h^{1/2}(n^{-1/2}h^r + \log (n)^{1/2}n^{-1}h^{-3/2})\to 0
\end{align*}
which means we need
\begin{align*}
\frac{\log (n)^{1/2}}{n^{1/2}h}\to 0.
\end{align*}
So we have that we need $h\propto n^{-a}$ with $0<a<1/2$. This is also ensures $nh\to \infty$, and $h\to 0$. Now, the bias in the estimation is of order $O(h^{r})$, so that we need the following undersmoothing rate
\begin{align*}
n^{1/2} h^{1/2} h^{r}\to 0
\end{align*}
which means we need $1/2-a(1/2+r)\leq0$, or $1-a(1+2r)\leq 0$, which implies $1/(1+2r)\leq a$. For example, for $r=2$, we have $1/5\leq a< 1/2$. Therefore,
\begin{align*}
\hat f_Y(\hat Q_Y[\tau]) - \hat f_Y( Q_Y[\tau])
&=  f_Y'(Q_Y[\tau])  (\hat Q_Y[\tau]-Q_Y[\tau]) + o_p(n^{-1/2}h^{-1/2}).
\end{align*}

\subsection{Proof of Corollary \ref{cor:clt}}

Recall that by equation \eqref{eq:cqpe_uqpe_3}, we have that
\begin{align*}
UQPE_{X_1}(\tau)=E\left [ \beta_1(\xi_\tau(X))|Y= Q_Y[\tau]\right],
\end{align*}
and that we defined
\begin{align*}
U_{\tau}:= \beta_1(\xi_\tau(X)) - E\left [ \beta_1(\xi_\tau(X))|Y\right]
\end{align*}
and, by construction, $E[U_\tau|Y]=0$ a.s. and, by Assumption \ref{assumption_clt}, $E[U^2_\tau|Y=y]=\sigma^2_\tau(y)<\infty$ for every $y$ in the support of $Y$. Moreover, for $(x_i,y_i)$, we have
\begin{align}\label{eq:u_definition}
u_{\tau,i}:= \beta_1(\xi_\tau(x_i)) - E\left [ \beta_1(\xi_\tau(X))|Y=y_i\right]. 
\end{align}
We focus on 
\begin{align}\label{eq:bias_variance}
\sqrt{nh} \left (\widetilde{UQPE_{X_1}}(\tau) -  {UQPE_{X_1}}(\tau)\right) &= \sqrt{nh} \left ( \frac{\sum_{i=1}^{n}K_h(y_i-Q_Y[\tau])\cdot \beta_1(\xi_\tau(x_i))}{\sum_{i=1}^{n}K_h(y_i-Q_Y[\tau])} - {UQPE_{X_1}}(\tau) \right)\notag\\
&=\sqrt{nh} \left ( \frac{\sum_{i=1}^{n}K_h(y_i-Q_Y[\tau])\cdot \left[\beta_1(\xi_\tau(x_i))-{UQPE_{X_1}}(\tau)\right]}{\sum_{i=1}^{n}K_h(y_i-Q_Y[\tau])} \right)\notag\\
&=\sqrt{nh} \left ( \frac{\sum_{i=1}^{n}K_h(y_i-Q_Y[\tau])\cdot u_{\tau,i}}{\sum_{i=1}^{n}K_h(y_i-Q_Y[\tau])} \right)\notag\\
&+\sqrt{nh} \left ( \frac{\frac{1}{n}\sum_{i=1}^{n}K_h(y_i-Q_Y[\tau])\cdot \left[E\left [ \beta_1(\xi_\tau(X))|Y=y_i\right]-{UQPE_{X_1}}(\tau)\right]}{\frac{1}{n}\sum_{i=1}^{n}K_h(y_i-Q_Y[\tau])} \right).
\end{align}

Consider the first term:
\begin{align*}
\sqrt{nh} \left ( \frac{\sum_{i=1}^{n}K_h(y_i-Q_Y[\tau])\cdot u_{\tau,i}}{\sum_{i=1}^{n}K_h(y_i-Q_Y[\tau])} \right)= \frac{1}{\hat f_Y(Q_Y[\tau])} \sum_{i=1}^{n} \frac{1}{\sqrt{nh} } K\left(\frac{y_i-Q_Y[\tau]}{h} \right)\cdot u_{\tau,i}.
\end{align*}
For this term we use the Lindberg-Feller CLT. First we compute the variance of the sum.
\begin{align*}
Var\left[ \sum_{i=1}^{n} \frac{1}{\sqrt{nh} } K\left(\frac{Y_i-Q_Y[\tau]}{h} \right)\cdot u_{\tau,i}\right] &= \frac{1}{h}E\left[ K\left(\frac{Y-Q_Y[\tau]}{h} \right)^2 \sigma^2_\tau(Y)\right]  \\
&=\frac{1}{h}\int K\left(\frac{y-Q_Y[\tau]}{h} \right)^2 \sigma^2_\tau(y) f_Y(y)dy\\
&=\int K(u)^2 \sigma^2(Q_Y[\tau] + hu) f_Y(Q_Y[\tau]+hu)du\\
&\to  \sigma^2_\tau(Q_Y[\tau]) f_Y(Q_Y[\tau]) \int K(u)^2du.
\end{align*}
because $\sigma^2_\tau(y)$ and $f_Y(y)$ are continuous, and are bounded. The conclusion follows from the dominated convergence theorem. We write
\begin{align*}
\sigma_{\tau,n}^2:=Var\left[ \sum_{i=1}^{n} \frac{1}{\sqrt{nh} } K\left(\frac{Y_i-Q_Y[\tau]}{h} \right)\cdot u_{\tau,i}\right] 
\to  \sigma_{\tau,0}^2.
\end{align*}
To apply the Lindberg-Feller CLT, we define
\begin{align*}
\omega_{in}:=\frac{1}{\sqrt{nh} } K\left(\frac{Y_i-Q_Y[\tau]}{h} \right).
\end{align*}
We need to show that, for some $\delta>0$,
\begin{align*}
\lim_{n\to \infty}\sum_{i=1}^n E\left |\frac{\omega_{in}U_{\tau,i}}{\sigma_{\tau,n}} \right|^{2+\delta} =0.
\end{align*}
We have
\begin{align*}
\sum_{i=1}^n E\left |\frac{\omega_{in}U_{\tau,i}}{\sigma_{\tau,n}} \right|^{2+\delta} = \sum_{i=1}^n \left|\frac{\sigma_{\tau,0}}{\sigma_{\tau,n}} \right|^{2+\delta} E\left |\frac{\omega_{in}u_{\tau,i}}{\sigma_{\tau,0}} \right|^{2+\delta}.
\end{align*}
It will be sufficient to focus on 
\begin{align*}
 \sum_{i=1}^n E\left |\frac{\omega_{in}U_{\tau,i}}{\sigma_{0,\tau}} \right|^{2+\delta} &= \frac{n}{|\sigma_{0,\tau}|^{2+\delta}}E\left |\omega_{in}U_{\tau,i} \right|^{2+\delta}\\
 &=\frac{n}{|\sigma_{0,\tau}|^{2+\delta}}E\left[ |\omega_{in}|^{2+\delta}E[|U_{\tau,i}|^{2+\delta}|Y_i] \right]\\
 &\leq C \frac{n}{|\sigma_{0,\tau}|^{2+\delta}}E\left[ |\omega_{in}|^{2+\delta} \right]\\
 &= \frac{Cn}{|\sigma_{0,\tau}|^{2+\delta}} (nh)^{-1-\delta/2} \int K \left(\frac{y-Q_Y[\tau]}{h} \right)^{2+\delta} f_Y(y)dy\\
 &= \frac{Cn}{|\sigma_{0,\tau}|^{2+\delta}} (nh)^{-1-\delta/2} h\int K (u)^{2+\delta} f_Y(Q_Y[\tau]+hu)du\\
 &= \frac{c}{(nh)^{\delta/2}} \int K (u)^{2+\delta} f_Y(Q_Y[\tau]+hu)du,
\end{align*}
which goes to 0 since $(nh)^{1/2}\to \infty$ and $f_Y(y)$ is continuous at $y=Q_Y[\tau]$. We have used the bound of the higher order conditional expectation of $U_{\tau}$: $E[|U_{\tau,i}|^{2+\delta}|Y_i]<C$ a.s., and that $\int |K(u)|^{2+\delta}du<\infty$.
Therefore,
\begin{align}\label{eq:clt_1}
\frac{1}{\hat f_Y(Q_Y[\tau])} \sum_{i=1}^{n} \frac{1}{\sqrt{nh} } K\left(\frac{y_i-Q_Y[\tau]}{h} \right)\cdot u_{\tau,i}\overset{d}{\to}N\left (0,  \sigma^2_\tau(Q_Y[\tau]) f_Y(Q_Y[\tau])^{-1} \int K(u)^2du \right)
\end{align}
since $\hat f_Y(Q_Y[\tau]) =  f_Y(Q_Y[\tau])+o_p(1)$.

The second term in the expansion of $\sqrt{nh} \left (\widetilde{UQPE_{X_1}}(\tau) -  {UQPE_{X_1}}(\tau)\right)$ is a bias term. We now find its rate of convergence. We start with the numerator. 
\begin{align*}
&E\left[\frac{1}{n}\sum_{i=1}^{n}K_h(Y_i-Q_Y[\tau])\cdot \left[E\left [ \beta_1(\xi_\tau(X))|Y=Y_i\right]-{UQPE_{X_1}}(\tau)\right]\right] \\
&=
\frac{1}{h}\int_{\mathcal Y} K\left( \frac{y-Q_Y[\tau]}{h} \right)
\left[E\left [ \beta_1(\xi_\tau(X))|Y=y\right]-{UQPE_{X_1}}(\tau)\right]
f_{Y}(y)dy\\
&=
\int K(u)
\left[E\left [ \beta_1(\xi_\tau(X))|Y=Q_Y[\tau]+hu\right]-{UQPE_{X_1}}(\tau)\right] f_{Y}(Q_Y[\tau]+hu)du\\
&=
\int K(u)
\left[E\left [ \beta_1(\xi_\tau(X))|Y=Q_Y[\tau]+hu\right]-{UQPE_{X_1}}(\tau)\right] f_{Y}(Q_Y[\tau]+hu)du\\
&= \int K(u) E\left [ \beta_1(\xi_\tau(X))|Y=Q_Y[\tau]+hu\right]f_{Y}(Q_Y[\tau]+hu)du\\
&-{UQPE_{X_1}}(\tau) \int_{\mathcal Y} K(u) f_{Y}(Q_Y[\tau]+hu)du.
\end{align*}

We do a Taylor expansion on the density and the conditional expectation and we use the fact that $\int K(u)du=1$, $\int u^j K(u)du = 0$ when $j=1,\ldots,r-1$,
and $\int u^r K(u)du < \infty$. Let $(Ef_Y)^{(j)}(q)$ denote the $j$-derivative with respect to $y$ of the product $E\left [\beta_1(\xi_\tau(X))|Y=y\right]f_{Y}(y)$ evaluated at $y=q$.
The first term is
\begin{align*}
& \int K(u) E\left [ \beta_1(\xi_\tau(X))|Y=Q_Y[\tau]+hu\right]f_{Y}(Q_Y[\tau]+hu)du\\
&={UQPE_{X_1}}(\tau) +  \int K(u) \sum_{j=1}^{r-1}\frac{h^ju^j (Ef_Y)^{(j)}(Q_Y[\tau]) }{j!} du\\
&+ \frac{h^r}{r!} \int K(u)  u^r (Ef_Y)^{(j)}(\tilde Q_Y[\tau]) du\\
&={UQPE_{X_1}}(\tau) + O(h^{r}),
\end{align*}
since the derivatives are uniformly bounded. Now, for the other term we do a similar expansion of the density.
\begin{align*}
& {UQPE_{X_1}}(\tau) \int_{\mathcal Y} K(u) f_{Y}(Q_Y[\tau]+hu)du\\
&= {UQPE_{X_1}}  + \int_{\mathcal Y} K(u)\sum_{j=1}^{r-1}\frac{h^ju^j f^{(j)}_{Y}(Q_Y[\tau]) }{j!} du +  \frac{h^r}{r!} \int K(u)  u^r f^{(r)}_{Y}(\tilde Q_Y[\tau]) du\\
&={UQPE_{X_1}}(\tau) + O(h^{r}).
\end{align*}
Therefore, we obtain that the bias is of order $O(h^{r})$:
\begin{align*}
E\left[\frac{1}{n}\sum_{i=1}^{n}K_h(Y_i-Q_Y[\tau])\cdot \left[E\left [ \beta_1(\xi_\tau(X))|Y=Y_i\right]-{UQPE_{X_1}}(\tau)\right]\right]&= O(h^{r}).
\end{align*}

Now, for the variance we have
\begin{align*}
&Var\left[\frac{1}{n}\sum_{i=1}^{n}K_h(Y_i-Q_Y[\tau])\cdot \left[E\left [ \beta_1(\xi_\tau(X))|Y=Y_i\right]-{UQPE_{X_1}}(\tau)\right]\right]\\
&=\frac{1}{nh^{2}} Var\left[K\left(\frac{Y_i-Q_Y[\tau]}{h}\right) \left[E\left [ \beta_1(\xi_\tau(X))|Y=Y_i\right]-{UQPE_{X_1}}(\tau)\right]\right]\\
&\leq \frac{1}{nh^{2}} E\left [   K\left(\frac{Y_i-Q_Y[\tau]}{h}\right)^2 \left[E\left [ \beta_1(\xi_\tau(X))|Y=Y_i\right]-{UQPE_{X_1}}(\tau)\right] ^2 \right ]\\
&=\frac{1}{nh^{2}}\int K\left(\frac{y-Q_Y[\tau]}{h}\right)^2 \left[E\left [ \beta_1(\xi_\tau(X))|Y=y\right]-{UQPE_{X_1}}(\tau)\right] ^2 f_Y(y)dy\\
&=\frac{1}{nh}\int K(u)^2 \left[E\left [ \beta_1(\xi_\tau(X))|Y=Q_Y[\tau] + hu\right]-{UQPE_{X_1}}(\tau)\right] ^2 f_Y(Q_Y[\tau] + hu)du\\
&=\frac{1}{nh}\int K(u)^2 [huE^{(1)}\left [ \beta_1(\xi_\tau(X))|Y=\tilde Q_Y[\tau] \right]] ^2 \left[f_Y(Q_Y[\tau]) + hu f^{(1)}_Y(\tilde Q_Y[\tau]) \right]du.
\end{align*}
This implies that
\begin{align*}
&Var\left[\frac{1}{n}\sum_{i=1}^{n}K_h(Y_i-Q_Y[\tau])\cdot \left[E\left [ \beta_1(\xi_\tau(X))|Y=Y_i\right]-{UQPE_{X_1}}(\tau)\right]\right] =O(n^{-1}h).
\end{align*}
Therefore, we obtain that
\begin{align*}
\frac{1}{n}\sum_{i=1}^{n}K_h(Y_i-Q_Y[\tau])\cdot \left[E\left [ \beta_1(\xi_\tau(X))|Y=Y_i\right]-{UQPE_{X_1}}(\tau)\right] = O_p(h^r + n^{-1/2}h^{1/2}).
\end{align*}
Under Assumption \ref{assumption_bandwitdh}, this term is $o_p(n^{-1/2}h^{-1/2})$, since
\begin{align*}
(nh)^{1/2}O_p(h^r + n^{-1/2}h^{1/2}) = O_p((nh)^{1/2}h^r + h^{1/4}) = o_p(1),
\end{align*}
since $(nh)^{1/2}h^r\to 0$, and $h\to 0$ as $n\to\infty.$ Therefore, the bias term is
\begin{align*}
\sqrt{nh} \left ( \frac{\frac{1}{n}\sum_{i=1}^{n}K_h(y_i-Q_Y[\tau])\cdot \left[E\left [ \beta_1(\xi_\tau(X))|Y=y_i\right]-{UQPE_{X_1}}(\tau)\right]}{\frac{1}{n}\sum_{i=1}^{n}K_h(y_i-Q_Y[\tau])} \right) = \frac{o_p(1)}{f_Y(Q_Y[\tau])+o_p(1)} = o_p(1).
\end{align*}

Combining this fact with \eqref{eq:clt_1}, we obtain
\begin{align*}
\sqrt{nh} \left (\widetilde{UQPE_{X_1}}(\tau) -  {UQPE_{X_1}}(\tau)\right)\overset{d}{\to}N\left (0,  \sigma^2_\tau(Q_Y[\tau]) f_Y(Q_Y[\tau])^{-1} \int K(u)^2du \right).
\end{align*}
In view of the result in Theorem \ref{thm:first_order_equiv}, we have
\begin{align*}
\sqrt{nh} \left (\widehat{UQPE_{X_1}}(\tau) -  {UQPE_{X_1}}(\tau)\right)\overset{d}{\to}N\left (0,  \sigma^2_\tau(Q_Y[\tau]) f_Y(Q_Y[\tau])^{-1} \int K(u)^2du \right).
\end{align*}

Now we focus on the term $x'\Psi_{i}(\xi_\tau(x))$, where
\begin{align*}
\Psi_{i}(\eta)=E\left[f_{Y|X}(X ' { \beta}(\eta)|X)XX'\right]^{-1}\left ( \eta- \mathds 1\left\{ y_i\leq x_i'  { \beta}(\eta) \right\} \right)x_i.
\end{align*}
By \cite{Powell91}, it can be estimated by
\begin{align*}
\widehat{x'\Psi_{i}(\xi_\tau(x)) } &= x'\left [\frac{1}{2nh_P}\sum_{j=1}^n\mathds 1\left\{ |y_j-x_j'\hat \beta (\hat \xi_\tau(x))|\leq h_P  \right\}x_jx_j' \right]^{-1} \left ( \hat \xi_\tau(x)- \mathds 1\left\{ y_i\leq x_i'  { \hat \beta}(\hat \xi_\tau(x)) \right\} \right)x_i.
\end{align*}
where $h_P$ is given in Section 3.4.2 in \cite{Koenker05}. Finally, 
\begin{align*}
\psi_{i}(\tau) = f_Y(Q_Y[\tau])^{-1} \left (\tau -\mathds 1\left\{ y_i\leq Q_Y[\tau]  \right\}  \right)
\end{align*}
can be estimated by
\begin{align*}
\widehat{\psi_{i}(\tau)}
= \hat f_Y(\hat Q_Y[\tau])^{-1} \left (\tau -\mathds 1\left\{ y_i\leq \hat Q_Y[\tau]  \right\}  \right).
\end{align*}
So, the estimator of the asymptotic variance is
\begin{align*}
\frac{1}{n}\frac{1}{\widehat{x'  { \dot\beta}( \xi_\tau(x)) }^2} \sum_{i=1}^n
\left (-\widehat{x'\Psi_{i}(\xi_\tau(x)) }  + \widehat{\psi_{i}(\tau)}\right )^2.
\end{align*}

\bibliographystyle{econometrica}
\bibliography{CQ_UQ.bib}

\end{document}